\documentclass[rmp,aps,twocolumn,groupedaddress,showpacs]{revtex4}

\usepackage{amsmath}
\usepackage{amsfonts}
\usepackage{amssymb}
\usepackage{bbm}
\usepackage{graphicx}
\usepackage{hyperref}

\newcommand{\beq}{\begin{equation}}
\newcommand{\eeq}{\end{equation}}
\newcommand{\beqa}{\begin{eqnarray}}
\newcommand{\eeqa}{\end{eqnarray}}

\newcommand{\1}{\protect\ensuremath{\mathbbm{1}}}

\newcommand{\Z}{\ensuremath{\mathbb{Z}}}

\newcommand{\id}{\ensuremath{{\rm id} \, }}

\newcommand{\A}{\ensuremath{\mathcal{A}}}
\newcommand{\B}{\ensuremath{\mathcal{B}}}

\newcommand{\Tr}{\ensuremath{\text{Tr}}}
\newcommand{\tr}{\ensuremath{\text{Tr}}}

\def\<{\langle}
\def\>{\rangle}

\newcommand{\hr}{{\cal H}}
\newcommand{\cc}{{\mathbb C}}

\newcommand{\zz}{{\mathbb Z}}

\begin{document}

\title{Quantum channels and memory effects}

\author{Filippo Caruso}\email{filippo.caruso@lens.unifi.it}
\affiliation{QSTAR, Largo Enrico Fermi 2, I-50125 Firenze, Italy,
LENS and Universit\`a di Firenze, via Carrara 1, I-50019 Sesto Fiorentino, Italy}
\affiliation{Dipartimento di Fisica e Astronomia, Universit\`a di Firenze, via Sansone 1, I-50019 Sesto Fiorentino, Italy}
\affiliation{Institut f\"{u}r Theoretische Physik, Universit\"{a}t Ulm, Albert-Einstein-Allee 11, D-89069 Ulm, Germany}

\author{Vittorio Giovannetti}\email{v.giovannetti@sns.it}
\affiliation{NEST, Scuola Normale Superiore and Istituto Nanoscienze-CNR, Piazza dei Cavalieri 7, I-56126 Pisa, Italy}

\author{Cosmo Lupo}\email{cosmo.lupo@unicam.it}
\affiliation{MIT, Research Laboratory of Electronics, 77 Massachusetts Avenue, Cambridge MA 02139, USA,}
\affiliation{School of Science and Technology, University of Camerino, I-62032 Camerino, Italy}

\author{Stefano Mancini}\email{stefano.mancini@unicam.it}
\affiliation{School of Science and Technology, University of Camerino, I-62032 Camerino, Italy}
\affiliation{INFN-Sezione di Perugia, Via A. Pascoli, I-06123 Perugia, Italy}

\begin{abstract}
Any physical process can be represented as a quantum channel mapping an initial state to a final state. Hence it can be characterized from the point of view of communication theory, i.e., in terms of its ability to transfer information. Quantum information provides a theoretical framework and the proper mathematical tools to accomplish this. In this context the notion of codes and communication capacities have been introduced by generalizing them from the classical Shannon theory of information transmission and error correction. The underlying assumption of this approach is to consider the channel not as acting on a single system, but on sequences of systems, which, when properly initialized allow one to overcome the noisy effects induced by the physical process under consideration. While most of the work produced so far has been focused on the case in which a given channel transformation acts identically and independently on the various elements of the sequence (memoryless configuration in jargon), correlated error models appear to be a more realistic way to approach the problem. A slightly different, yet conceptually related, notion of correlated errors applies to a single quantum system which evolves continuously in time under the influence of an external disturbance which acts on it in a non-Markovian fashion. This leads to the study of memory effects in quantum channels: a fertile ground where interesting novel phenomena emerge at the intersection of quantum information theory and other branches of physics. A survey is taken of the field of quantum channels theory while also embracing these specific and complex settings.\end{abstract}

\pacs{03.67.-a, 89.70.-a, 02.50.-r}

\maketitle

\tableofcontents


\section{Introduction}\label{sec:intro}

In his seminal 1948 work ``A Mathematical Theory of Communication",
C. E. Shannon established the basis of modern  communication
technology~\cite{Ver98}.
 Neglecting all the semantic aspects  (which are irrelevant at the level of engineering)
 he stressed that ``the fundamental problem of communication
is that of reproducing at one point either exactly or approximately a message
selected at another point."
In particular ``the system must be designed to operate for each possible selection,
not just the one which will actually be chosen since this is unknown at the time of design."~\cite{Sh48}.
At the heart of this view is  what one may call the ``channel" formalism, where
any noisy communication line is depicted as a stochastic map connecting input signals selected by the sender of the message (Alice),  who is operating at one end of the line, to their corresponding
output counterparts accessible to the receiver of the messages (Bob), who is operating at the other end.
In the same article Shannon  also proved  that
the performance of a  transmission line can be gauged by a single quantity, the {\it  capacity} of the channel,
which measures
the maximum rate at which information
can be reliably  transferred when Alice and Bob, operating on long  sequences of transmitted signals, follow a pre-established protocol (error correcting code procedure) aimed to nullify the detrimental effects of the communication noise. 
The rational behind this approach (which is typical to communication theory) is that communication is {\it expensive} while
local operations are somehow {\it free} (unless external constraints are explicitly imposed by the selected implementation).

Rolf Landauer was the first to put on firm ground the fact that information
is not  just an abstract, mathematical notion
 but has instead an intrinsic physical nature which poses
 limits on the possibility of  processing and transferring it~\cite{LAND61}.
That is why quantum mechanics, the most advanced physical theory, comes into play in
the study of communication processes~\cite{BS98}.  In this context ({\it quantum information
theory})  it is recognized that any message two parties wish
to exchange  must be
{\it written} into the states of  some quantum system, say a photonic pulse
propagating along an optical fiber~\cite{CD94},
and that the
processing,  the transmission, and the reading of such data must be carried out  following the
rather unconventional  prescriptions established by quantum mechanics.
\begin{figure}[t]
\centering
\includegraphics[width=0.48\textwidth]{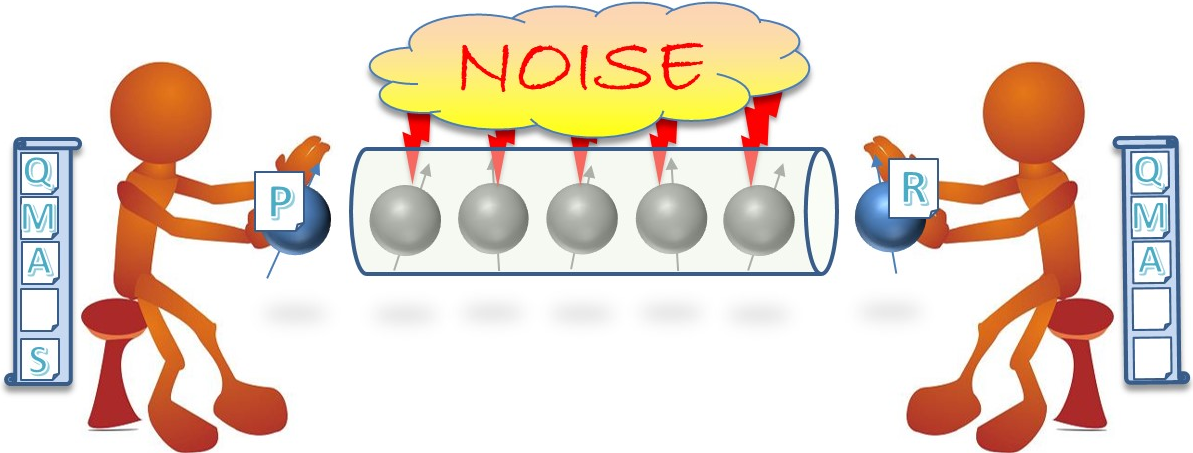}
\caption{Typical communication scenario of two users communicating by encoding (decoding) the letters of a message onto (from) physical systems that are transmitted through a channel. They are subjected to the unavoidable presence of noise in the channel that does introduce errors (eventually correlated) in the transmitted message.}
\label{fig0}
\end{figure}
As in the classical setting  this scenario is properly formalized by introducing
  the notion of  {\it quantum channels}  as those
 mappings which, generalizing the notion of a channel in Shannon theory,
link the initial states of the quantum information carriers
(controlled by Alice)  to the their output states (controlled by Bob), see Fig.~\ref{fig0}.
Interestingly enough, to evaluate the {\it quality} of these exotic communication lines
several non-equivalent notions of coding procedures, as well as of corresponding capacities, must be introduced.
Indeed while Alice and Bob might still be willing to use a quantum channel to
exchange purely classical messages, new forms of communication can now be envisioned.
For instance Alice can be interested in transferring to Bob purely quantum messages, e.g., 
the {\it unknown quantum state} of a quantum memory that is located in her lab, 
or half of a {\it maximally  entangled} state~\cite{GHUNE09,HHHH09} that she has locally produced.
The ability to sustain this special kind of transmission defines what is called
the {\it quantum capacity} of a quantum communication line. 
This in general differs from the {\it classical capacity} that instead, as in the
Shannon setting, measures the ability to transfer purely classical messages.
But there is more. Quantum teleportation~\cite{BetAL93} and the super-dense coding protocol~\cite{BW92}
have shown that quantum {\it entanglement}~\cite{HHHH09,GHUNE09} is a catalytic 
resource for communication. 
Indeed, even though entanglement alone does not constitute a communication link 
between distant parties~\cite{WERNER00}, allowing Alice and Bob to use pre-shared 
entanglement in the design of their communication protocols can boost the performance 
of basically any communication line they have access to (even in terms of the 
quantum capacity). 
This fact naturally brings in the notion of {\it entanglement assisted} capacities of a
quantum channel~\cite{B99}, which  is yet a different way of gauging the performance
of a communication line.

\begin{figure}[t]
\centering
\includegraphics[width=0.48\textwidth]{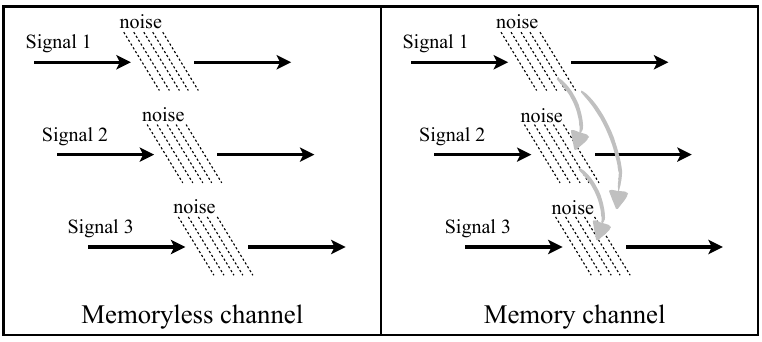}
\caption{Pictorial representation of memoryless (left panel) and memory (right panel) channels. In the former case the individual carriers which compose a sequence of transmitted signals experience the {\it same} noisy transformation. In the latter case  instead cross-talking among the various transmission events can happen and, as indicated by the gray arrows in the figure, the noise model which describes the transformation of the $n$-th carrier depends in principle upon the previous communication exchanges.}
\label{fig000}
\end{figure}

The vast majority of the work on quantum channels has been
concerned with the study of memoryless configurations where sequences of exchanged 
quantum carriers are supposed to undergo the action of noisy transformations which 
affect them {\it independently} and {\it identically}, see left panel of Fig.~\ref{fig000}.
In this scenario coding theorems have been derived which allow one to
express the various capacities of the communication line in terms of rather compact
entropic  formulas. 
For instance the classical (resp.~quantum) capacity of a memoryless quantum channel 
is characterized in terms of the Holevo (resp.~coherent) information~\cite{Holevo98,SN96,SW97,lloyd97,BNS98,BST98,D05,Shorlecture2002}.
The memoryless assumption is indeed a useful hypothesis which permits to simplify 
the input-output mapping induced by the noise. 
It also provides a realistic description for those communication schemes where 
the temporal rates at which signals are fed into the communication line are sufficiently 
low to allow for a {\it resetting} of the channel environment and to prevent signal cross-talking.    
Nonetheless this is  not always justified. For instance, with  increasing signal feeding rates, successive transmissions happen
so rapidly that the environment may retain a ``memory" of past events, see right panel of Fig.~\ref{fig000}.
Optical fibers are an example in which such effects can occur and have been explored experimentally~\cite{BDW+04,BDB04}. 
Similarly, in quantum information processors, especially in solid state
implementations, qubits may be so closely spaced that the same environmental degree of 
freedom will interact jointly with several of them (even if they are not nearest neighbors) 
leading to cross-talks and correlations in the noise~\cite{H07,DG98}.
The consideration of spatial and temporal memory effects is
therefore becoming increasingly pressing with the continuing
miniaturization of information processing devices and with
increasing communication rates through channels.
Moreover, from a fundamental point of view, quantum memory channels
provide a general framework which encompasses the memoryless ones as
a special case.

Apparently, the interest towards information
transmission through quantum channels with memory spread after
a model introduced by \onlinecite{MP02}.
Here an example of a qubit channel with Markovian
correlated noise was analyzed  in which the encoding of information by means of entangled
input states may increase the transmission rate of classical information.
Subsequently, the study of quantum channels with memory has largely
been confined to channels with Markovian correlated noise
with the aim of deriving bounds on the classical capacity, see, e.g.,~\cite{Ham02,BM04,BDM05}.
Then, coding theorems have been devised for a class of quantum memory channels having
structural properties that guarantee ``regular" asymptotic behavior~\cite{KW05, BB08, DD07}.
This approach can be traced back to the work of~\cite{Dob63}, who considered
a wide class of channels exhibiting stationary or ergodic behavior.
A completely different path was taken by Hayashi and
Nagaoka~\cite{HN03}, who applied the ``information-spectrum"
method to obtain a coding theorem for the classical capacity,
following the work by \onlinecite{VH94} on classical channels with memory.
In the  context of continuous variable systems~\cite{BvL05,WPGCRSL11},
generalizing results obtained in the classical setting~\cite{Sh49}
for the capacity of power-constrained Gaussian channels,
quantum ``water-filling" formulae have been recently derived~\cite{SKC09,PLM09}.
Exact expressions for classical and quantum capacities have been computed~\cite{LGM10}
using an ``unravelling" technique based on the Toeplitz distribution theorem~\cite{Gray72},
which allows one to map memory correlations into effective memoryless models.
All these results for continuous variable memory channels have been derived from 
fundamental results in the memoryless setting \cite{GiovannettiPRL92,WPG07} (see also
recent developments in \cite{Giova2014a,Giova2014b}).

Beyond quantum communication, a detailed study of quantum channels, and of the mechanisms responsible
of memory correlations, has implications in the broader research field of quantum open system dynamics~\cite{BPbook}.
As a matter of fact the input-output scheme that underlines the channel formalism
is reminding us of what in physics is conventionally described as a series of scattering events,
the scattered particles playing the role  of the
messages, while the scattering matrix playing the role of  the communication line.
More generally quantum channels  can be used to mimic
 all those  physical processes (temporal evolution, data-processing, etc.) which
  imply  a state change of a system of interest from an initial to a final configuration under the
  influence of an external agent (the system environment).
Exploiting this connection, insight on the system
evolution can then be gained by
analyzing its {\it quality}  as a communication line.
 Along this direction  models have been introduced
in which memory effects of a communication line are described as arising from the interaction with a multi-partite environment
initialized in a correlated state~\cite{GM05}, allowing remarkable links
between information theoretical quantities, like capacities, and statistical properties, like phase-transitions, of the underlining  many-body
environment~\cite{PV07,PV08}.

\begin{figure}[t]
\centering
\includegraphics[width=0.48\textwidth]{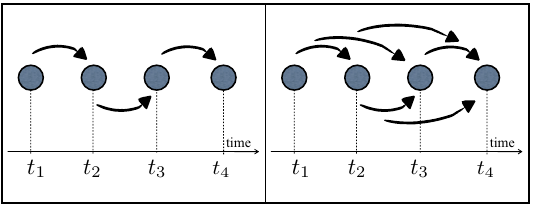}
\caption{Temporal correlations. Left panel: pictorial representation of a trajectory which describes the evolution of a system under dynamical semigroup approximation (discrete evolution steps have been assumed for the sake of simplicity). In this case, as indicated by the black arrows in the figure, the state of the system of interest (gray dot) at time $t_{j+1}$ only depends upon the state it had the time step which
comes immediately before, i.e. $t_{j}$. Right panel: representation of the evolution when the dynamical semigroup approximation fails: in this case the state at time $t_{j+1}$ depends in general by the states the system assumed along the whole trajectory.}
\label{fig001}
\end{figure}

When studying open system dynamics one shall not only deal with the problem of the input-output evolution
of a sequence of otherwise independent carriers. Indeed it is also interesting to address the
problem of the evolution of a single carrier in time, to see wether the associated trajectory
can be described as a collection of quantum channels which are applied sequentially on that system, 
see Fig.~\ref{fig001}. When this is not the case one can talk of memory effects induced by back-action mechanisms arising from the interaction of the system of interest and its own environment. 
At variance from those described in the previous paragraphs these effects have a clear dynamical character which is absent in the scheme of Fig.~\ref{fig000} where the temporal evolution is fixed.  
The study of this topic is intimately related to the semigroup structure of the set of quantum channels,
hence with dynamical maps and master equations~\cite{AL87}. It turns out that dynamical evolutions
which can be split into infinitesimal pieces correspond to the set of solutions of (possibly
time dependent) master equations standardly used to describe open systems dynamics~\cite{WC08}.
A more general approach to open quantum systems uses the
Nakajima-Zwanzig projection operator technique~\cite{N58,Z60} which shows
that, under fairly general conditions, the master equation for the reduced density
operator takes the form of a nonlocal equation in which memory effects are taken
into account through the introduction of a memory kernel.
Then, the problem [first put forward in~\cite{DWM03,DWC+04}] becomes to find those
conditions on the memory kernel ensuring that the time evolution map is a bona fide
quantum channel~\cite{CK12}. While our review is mostly devoted to analyze the memory effects which arise in the input-output paradigm schematized in Fig.~\ref{fig000}, for completeness we shall also briefly report on the most recent results which have being produced in the study of dynamical maps.


This work aims at providing an overview of the field of quantum
channels in a broad framework that includes also memory effects. 
As such it does not pretend to be omni-comprehensive, but it rather touches quantum communication 
subjects for which it has been already possible to venture beyond memoryless assumptions. 
We start considering in Sec. \ref{sec:qcbo} quantum channel maps 
and parallel them to physical processes transforming input states into output ones.
This presumes basic knowledge 
on the structure of quantum states, entanglement, and measurement
that are not reviewed here;  a rather complete report on these topics
can indeed be found elsewhere, e.g. in the books~\cite{NC00,PETZBOOK,SCHUMACHERBOOK,PRESKILL,ZYCBOOK,HOLEVObook}
or in the review articles~\cite{HHHH09,GHUNE09}.
Here the focus is on the representation of quantum channel maps and their properties related to the way they act on quantum states, in particular their composability. 
For pedagogical reasons we present several examples of quantum channels.
Still in Sec. \ref{sec:qcbo} we provide tools to study quantum channels like fidelities, distances, and entropies that will be used throughout the paper.


We then move on discussing in Section \ref{sec:fmmqc} the transition from the
memoryless setting to a more general scenario which allows
for correlations in quantum communication. 
Here much attention will be devoted to the structural properties of quantum channels.
Various classes of memory channels are then reviewed in Sec.\ \ref{sec:tqmc}.


In order to use a quantum channel for information transmission 
one has to cope with the problem of noise altering the transmitted information.
For this reason we next present in Section \ref{sec:codes} the subject of quantum error correction 
and discuss about achievable information transmission rates.
Actually, this Section briefly reviews basic notions of standard quantum error correction 
(mostly suitable for uncorrelated errors) and decoherence free subspaces (mostly suitable 
for completely correlated errors). For more details the reader may refer to \cite{Lidar-book-2013}. 
We then present avenues, not yet fully explored, for correcting partially correlated errors and discuss convolutional 
codes that work with a structure much similar to that of memory channels.


After having introduced the notion of transmission rates, it is natural to ask what are their maximum rates 
that can be achieved in a quantum channel. Thus we address the issue of quantum channel capacities in Sec.\ \ref{sec:ctc}.
A series of papers, at various level, dealt with capacities of memoryless quantum channels. 
For instance, the article~\cite{BS98} can be seen as a sort  of {\it manifesto} for quantum information theory.
The review~\cite{CD94} presents instead a rather detailed account of the mathematical and technological issues one faces when dealing with quantum communication
with photonic sources (even if some of the open problems  discussed there were solved  in more recent years,  this article remains a useful guidance to the field).
Reference~\cite{GALINDO} provides a rather compact overview on quantum information theory
and discusses in a simple but clear  form the basic aspects of the Shannon approach.
A more mathematically oriented point of view is presented in Ref.~\cite{Keyl2002}. Reference~\cite{reviewHOLEVO}  focuses on channel capacities
and their entropic characterization, while  finally~\cite{WPGCRSL11} is a detailed introduction to the field of Gaussian bosonic channels.
In Sec.~\ref{sec:ctc} coding theorems, that allow one to express capacities in a closed form by means of entropic quantities, are succinctly reviewed for memoryless quantum channels. 
We then indulge on the possibility of using them in the memory setting (revisiting what kind of memory permits it) and on their generalization.


For practical purposes we subsequently present in Section \ref{sec:solv} quantum channel models 
that are exactly solvable in terms of capacities. 
Already in the memoryless case these examples are few and 
they are even less in the memory case.
Anyway it is much instructive to see techniques used to solve optimization problems imposed by capacity evaluation.


Finally 
 Sec.\ \ref{sec:memdyn} is devoted to the characterization of the temporal
 correlations  which may arise in the description of the trajectory of quantum system 
 (see Fig.~\ref{fig001}).  In particular we shall review
the divisibility property of quantum channels  
and relate it to properties of 
dynamical maps and master equations.  


A summary of the main results and an outlook on physical realizations
are given in Sec.\ \ref{sec:so}.
Appendixes \ref{sec:dist} and \ref{app:algebras} provide elementary material
about distance measures for states and quasi-local algebras, respectively. 
In Appendix \ref{sec:appendix} an alternative proof of the structure decomposition 
theorem for non-anticipatory channels is presented, while in Appendix~\ref{a:DERIVO} 
an explicit derivation of capacity upper bounds is provided.


\section{Quantum channels: basic definitions and properties}\label{sec:qcbo}

In a typical communication scenario two parties (Alice the sender of the
message and Bob the receiver) aim to exchange (classical or quantum)
information by encoding it into (possibly arbitrarily long) sequences of
signals which propagate through the medium that separate them, see Fig.~\ref{fig0}.
A train of transmitted signals defines a sequence of independent uses of
the communication line ({\em channel uses}), and their input-output
evolution from Alice to Bob is determined by the noise which tampers with
the transmission process.
In classical information theory~\cite{Gbook} this is schematized by assigning an input alphabet $\mathcal{X}$ and an output alphabet $\mathcal{Y}$ 
whose elements $x$ and $y$ represent respectively the individual signals at the input and at the output of the transmission line. The noise instead is
assigned in terms of a stochastic process
characterized by conditional probabilities
that, given an  input sequence $(x_1,x_2,\cdots)$ of elements of $\mathcal{X}$ transmitted by Alice,
Bob will receive the sequence $(y_1, y_2, \cdots)$ of elements of $\mathcal{Y}$.

\begin{figure}[t]
\centering
\includegraphics[width=0.48\textwidth]{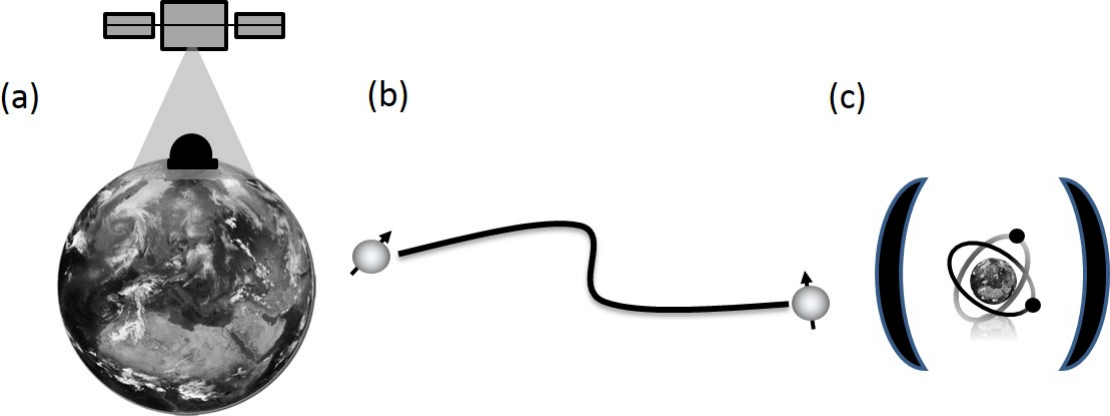}
\caption{Some examples of communication lines which can be described by the quantum channel formalism: (a) quantum communication with Earth- and space-based trasmitter
terminals, (b) transmission of photons along optical fibers, (c) storing information on atoms or molecules trapped in optical cavities. 
The evolution, in space or in time, of a quantum state can be always described by the formalism of quantum channels.
The potential presence of temporally or spatially correlated noise may lead, then, to dealing with memory quantum channels.}
\label{fig1phys}
\end{figure}

In quantum information the channel uses are represented by the degrees of
freedom (e.g., polarization, spins) of a collection $\{ q_1, q_2, \cdots \}$
of identical information carrying objects (e.g., optical pulses, flying atoms or ions)
which are locally produced by Alice and organized in a time-ordered sequence.
In this setting the noise  can then be described by assigning a
proper \emph{mapping} which acts on the (global) input states of the information carriers
to produce the associated (global) output states received by Bob.
The formalism is rather general and provides the proper mathematical tools apt to 
describe all those physical processes that involve the transformation of a  quantum system, induced either by the
direct temporal evolution of  its density matrix on a fixed time interval, or by the transmission through a medium (see  e.g.
Fig. \ref{fig1phys}). 
Concrete examples of these mappings can be encountered  for instance when studying  long-distance quantum communication and cryptography (as quantum key distribution)~\cite{GRTZ02,SBCDLP09}. In this case 
applications are often experimentally realized by identifying the information carriers 
with single-photon pulses which travel
 in free-space or over optical fibers where
air turbulence and absorption losses 
effectively limit the covered
distance
 from tens to hundreds of kilometers 
with the currently available technologies \cite{H02,UJAKLWZ04,GISIN2007,U07,SWFUTSP07,V08,Ma12,Y13}. 
All these effects  can be faithfully described in terms of  a combined action of  amplitude damping 
channels (dissipation and absorption) and
  phase--flip channels (dephasing phenomena) (see Sec.~\ref{sec:examples}).
Similarly, further examples of input-output mapping which admit a proper characterization in terms of quantum channels can be found 
when analyzing the effectiveness of atomic or molecular systems trapped in an optical cavity \cite{LBMW03}
as  quantum memory elements useful for information storage. 
In all these physical implementations, the noise processes may sometimes show temporal or spatial correlations, leading unavoidably to the additional presence of memory effects in the corresponding quantum representation.
Section~\ref{sec:fmmqc} discusses in details how such effects can be characterized.
Before doing so, however, it is useful to recall that
quantum mechanics imposes some fundamental structural constraints
on the transformations describing the evolution of quantum systems,
which must apply  independently from the underlying physical mechanisms that govern
the process and independently from the composite nature of  the input system.

\subsection{CPTP transformations}

\begin{figure}[t]
\centering
\includegraphics[width=0.48\textwidth]{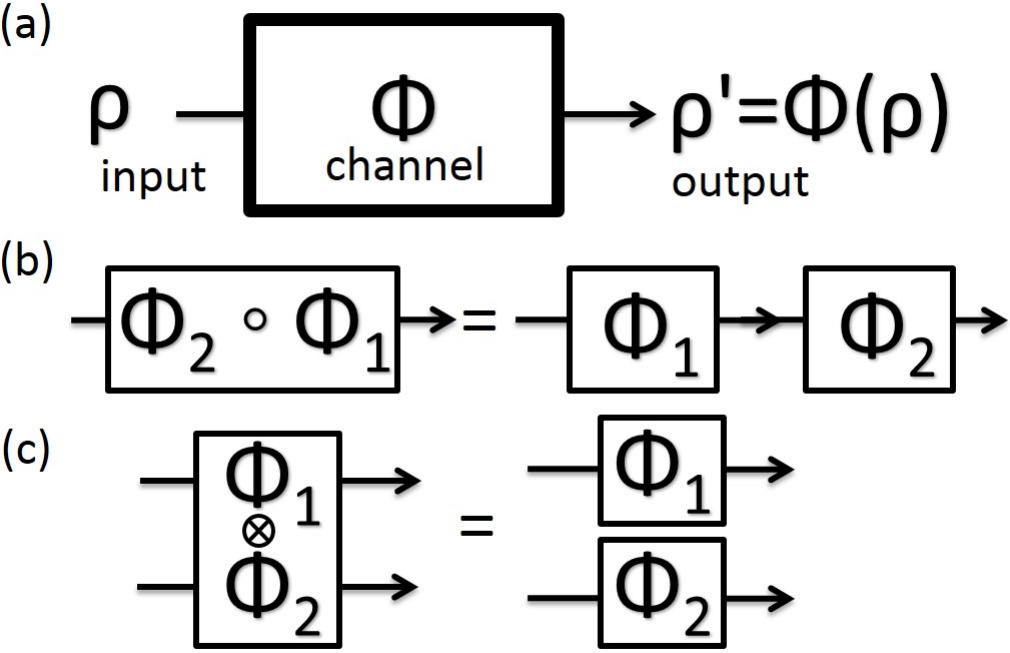}
\caption{(a) Quantum channel $\Phi$ mapping a state $\rho$ into a new one $\rho'$; 
(b) Concatenation $\Phi_1  \circ \Phi_2$ of two channels $\Phi_1$ and $\Phi_2$; 
(c) Tensor product $\Phi_1  \otimes \Phi_2$  of two channels $\Phi_1$ and $\Phi_2$.}
\label{fig1abc}
\end{figure}

Let $\Phi$ be a mapping (see Fig. \ref{fig1abc}a) describing the input-output relations of a generic quantum system
$Q$ (e.g., the carriers $\{ q_1, q_2, \cdots \}$ introduced in the previous
paragraph) evolving under the action of some physical process
\begin{equation} \label{qchmap}
\rho_Q \in \mathfrak{S} ({\cal H}_{Q}) \;\;
{\longmapsto}\;\;  \rho_{Q^{\prime}}=\Phi(\rho_{Q}) \in  \mathfrak{S} ({\cal H}_{Q'}) \,.
\end{equation}
Here $\mathfrak{S} ({\cal H}_{Q})$ and $\mathfrak{S} ({\cal H}_{Q'})$ stand for the sets of  density operators
 (non-negative operators with unit trace) defined on the Hilbert spaces ${\cal H}_{Q}$, ${\cal H}_{Q'}$
(the latter may be different in general) associated, respectively,  to the input and output system 
(unless explicitly stated in what follows it is assumed that
these spaces are finite dimensional). 
Since $\rho_{Q^{\prime}}$ must be a valid density operator it results natural to require 
 the map $\Phi$ to be:
\begin{itemize}

\item[i)] \emph{linear} when extended to the set ${\cal T}({\cal H}_Q)$ of  {\it trace-class}
 linear  operators   of ${\cal H}_{Q}$.
As matter of fact $\Phi$ must transform mixtures of input density operators into 
a mixture of the associated outputs, i.e., $\sum_ip_i \rho_Q(i) \mapsto \sum_ip_i \Phi(\rho_Q(i))$, with $p_i$ the probability associated with the input state $\rho_Q(i)$\footnote{The linearity requirement ensures that the extension of $\Phi$ from
$\mathfrak{S}({\cal H}_Q)$ to ${\cal T}({\cal H}_Q)$ is unique.};

\item[ii)] \emph{trace-preserving} (i.e., it must preserve the normalization of all input states);

\item[iii)] \emph{positive} (i.e., when acting on $Q$ it must preserve the positivity of 
density operators);

\end{itemize}

Actually the latter condition turns out to not be enough to guarantee the positivity of $\Phi(\rho_Q)$ when considering  
$\rho_Q$ as coming from a joint state $\rho_{QA}$ of system $Q$ and system $A$ by tracing out the latter. This is due to possible quantum correlations (entanglement) existing between systems $Q$ and $A$. 
Hence, condition iii) is made tighter as follows:

\begin{itemize}
\item[iii')] \emph{completely  positive} (i.e., when acting on $Q$ the map $\Phi$ must preserve the positivity of any
density operator, including those describing a joint state
$\rho_{QA}$ of $Q$ and an arbitrary ancillary system $A$);
\end{itemize}

A violation of any of conditions i), ii), iii') implies the impossibility of
maintaining the statistical interpretation of the theory ~\cite{ECG61,Kraus83,L75,Holevo72,Stine55}.

Finally any transformation fulfilling all these conditions is said to be CPTP 
(completely positive and trace preserving) linear map and is a \emph{quantum channel}.
Special instances of these last maps are provided by the isometric channels
\begin{eqnarray} \label{UNITARYCH}
\rho_Q \mapsto  {\cal U}(\rho_Q) : = U \rho_Q U^\dag \;,
\end{eqnarray}
which are induced by the action of an isometric  transformation
$U$  connecting   ${\cal H}_Q$ to ${\cal H}_{Q'}$, i.e., $U^\dag U = \openone$ with $U^\dag$ being
the adjoint of $U$ and $\openone$ being the identity on ${\cal H}_Q$.
In particular when $Q=Q'$ and $U$ corresponds to a unitary operator, Eq.~(\ref{UNITARYCH}) 
defines a {\it unitary channel} on $Q$ which admits the channel  ${\cal U}^{-1} (\cdots) := U^\dag (\cdots) U$
as CPTP inverse. Furthermore if $U$ is the identity operator $\openone$ on ${\cal H}_Q$, 
the resulting transformation is the identity channel, denoted $\id$, which
maps any state into itself, i.e., $\id(\rho_Q) = \rho_Q$ for all $\rho_Q$.

\subsection{Composition rules and structural properties} \label{sec:comp}
While referring the reader to~\cite{Keyl2002,PETZBOOK,ZYCBOOK,BPbook,HOLEVObook,
HOLEVObook12} for an exhaustive characterization, here  
the most relevant structural properties of the set
$\mathfrak{P}(Q\mapsto Q')$, formed by the CPTP maps connecting
system $Q$ to system $Q'$, are reviewed.

{\it Convexity:--}
 given $\Phi$, $\Lambda \in \mathfrak{P}(Q\mapsto Q')$ and $p \in[0,1]$,
the transformation
\begin{eqnarray} \label{convexrule}
\rho_Q \mapsto p \; \Phi(\rho_Q) + (1-p)\;  \Lambda(\rho_Q)\;,
\end{eqnarray}
is still an element of $\mathfrak{P}(Q\mapsto Q')$.
For instance convex combinations of unitary channels on $Q$ define the   {\it random unitary channel}
 subset of $\mathfrak{P}(Q\mapsto Q)$.

{\it Concatenation of channels:--} given two CPTP channels, $\Phi$ from $Q$ to $Q'$  and $\Omega$ from $Q'$ to $Q''$, 
one can define (see Fig. \ref{fig1abc}b) their concatenation
$\Omega\circ \Phi$ as the following CPTP transformation from $Q$ to $Q''$,
\begin{eqnarray} \label{compositionNEW}
\rho_Q  \mapsto   (\Omega\circ \Phi)(\rho_Q) := \Omega( \Phi(\rho_Q))\;.
\end{eqnarray}
Channels concatenation allows one to introduce a relation of equivalence between CPTP maps.
In particular two maps $\Phi$, $\Lambda \in \mathfrak{P}(Q\mapsto Q')$  are said to be \emph{unitarily equivalent} 
if there exist unitary channels ${\cal U}\in \mathfrak{P}(Q \mapsto Q)$ and
 ${\cal U}'\in \mathfrak{P}(Q' \mapsto Q')$
  such that
\begin{eqnarray}
\Phi = {\cal U}{'} \circ \Lambda \circ {\cal U} \;,\label{UNIEQ}
\end{eqnarray}
the relation  being reversible in the from $\Lambda = {\cal U}^{-1} \circ \Phi \circ {\cal U'}^{-1}$.
From the above properties it  follows also that  $\mathfrak{P}(Q \mapsto Q)$ equipped with  the
concatenation rule~(\ref{compositionNEW})
possesses a non-Abelian semigroup structure,
 the channel $\id$ being the identity element of the set and the unitary channels  ${\cal U}$ being the only invertible elements.

{\it Tensor product of channels: --}
given two CPTP channels, $\Phi$ from $Q$ to $Q'$  and $\Omega$ from $R$ to $R'$, 
one can define (see Fig. \ref{fig1abc}c) their tensor product  $\Phi \otimes \Omega$ as the CPTP transformation 
from the composite system $QR$ to the composite system  $Q'R'$, which given an arbitrary 
tensor operator $N_Q\otimes M_R\in {\cal T}({\cal H}_Q\otimes {\cal H}_R)$ transforms it into
\begin{eqnarray} \label{compositionNEW1}
  (\Phi\otimes \Omega)(N_{Q}\otimes M_R ) :=
  \Phi(N_{Q})\otimes \Omega(M_R)\;.
\end{eqnarray}
Special instances are provided by the transformation $\Phi \otimes \id$ obtained by
tensoring  $\Phi$ with the identity channel acting on an external system:
such a map is called an {\it extension} of $\Phi$ and represents
the action of such channel when the system $Q$ (where $\Phi$ was originally defined), is described as part of an enlarged composite system  -- notice that  this structure was implicitly assumed when stating  point iii') of the previous section.
\newline

Concatenations and tensor products of quantum channels represent two alternative ways of
composing CPTP maps which, to some extent,  mimic respectively the {\it in-series} and {\it in-parallel}
composition  rules of electrical circuit elements.
In particular, as discussed in Sec.~\ref{sec:memdyn}, channel concatenation
is naturally suited to characterize the {\it temporal} correlations of a single
quantum system schematized in the left panel of Fig.~\ref{fig001} (the sequential  applications of  CPTP maps corresponding to different stages of the system evolution).
  On the contrary the tensor product~(\ref{compositionNEW1}) allows to
  describe {\it spatial} correlations which might be present in the evolution of composite
  quantum systems. Also, as discussed in Sec.~\ref{sec:fmmqc}, tensor products
 can be employed to describe the transformations that a sequence of information carriers
 encounters when transmitted through a communication line, see Fig.~\ref{fig000}.

\subsection{Stinespring representation, Kraus representation, and Choi-Jamiolkowski isomorphism}

It can be shown~\cite{Stine55} that a mapping~(\ref{qchmap})
satisfies the CPTP conditions detailed in the previous section, if and only if it
admits  dilations  that allow one to represent it in terms of
a unitary coupling with an external environment $E$ (which is possibly fictitious 
and may not correspond to the actual physical environment responsible for the system evolution).
For instance, taking for simplicity ${Q}={Q'}$, one can write
 \begin{equation}\label{dia}
\Phi(\rho_Q) = \mathrm{Tr}_E \Big[ U_{QE} \big( \rho_Q \otimes \omega_E \big) U_{QE}^{\dag} \Big] \;,
\end{equation}
where $\omega_E$ is a fixed state of $E$, $U_{QE}$ is the unitary transformation
coupling the latter to the input system $Q$, and $\mbox{Tr}_E$ denotes the
partial trace over the environment\footnote{Note that complete positivity can be violated if
the initial state of $Q$ is entangled with the channel environment, in which case, however, the mapping
represented by Eq.~(\ref{kraus1}) is only defined on a proper subset of $\mathfrak{S}({\cal H}_{Q})$ \cite{Jordan04,Shaji05}.}.
The representation~(\ref{dia}) is not unique.
Nonetheless by enlarging the environment $E$ to describe the environment state 
as a pure state, $\omega_E = |\omega\rangle_E\langle\omega|$ (see Sec.~\ref{sec:fidecha} for a proper definition of this \emph{purification} mechanism), the
choice of $U_{QE}$ can be shown to be unique up to a local isometric transformation on $E$.
Under this condition the dilation~(\ref{dia}) provides what is generally known as  the Stinespring representation for $\Phi$.

The CPTP conditions are also equivalent to the possibility of expressing $\Phi$ in {\em operator sum} 
(or {\em Kraus}) representation~\cite{ECG61,KRAUS,CHOI},
\begin{equation} \label{kraus1}
\Phi(\rho_Q) = \sum_{j} \; K_j \rho_Q K_j^\dag \;,
\end{equation}
with $\{K_j\}$ being operators on ${\cal H}_Q$ satisfying the normalization condition $\sum_j K_j^\dag K_j =\openone$.
The number of non-zero operators in the representation (\ref{kraus1}) is called the Kraus rank.
As in the case of the unitary dilation~(\ref{dia}), their choice is in general not unique. 
One can however guarantee that a Kraus representation exists with no more than $d^2$ elements 
($d$ being the dimension of ${\cal H}_Q$). 
Kraus and the Stinespring representations can also be put in mutual correspondence by 
identifying the operator $K_j$ with the linear operator
${_E\langle} e_j|
U_{QE} |\omega\rangle_E$ of ${\cal H}_Q$, where  $\{ |e_j\rangle_E\}$
is an orthonormal basis of $E$.

\begin{figure}[t]
\centering
\includegraphics[width=0.48\textwidth]{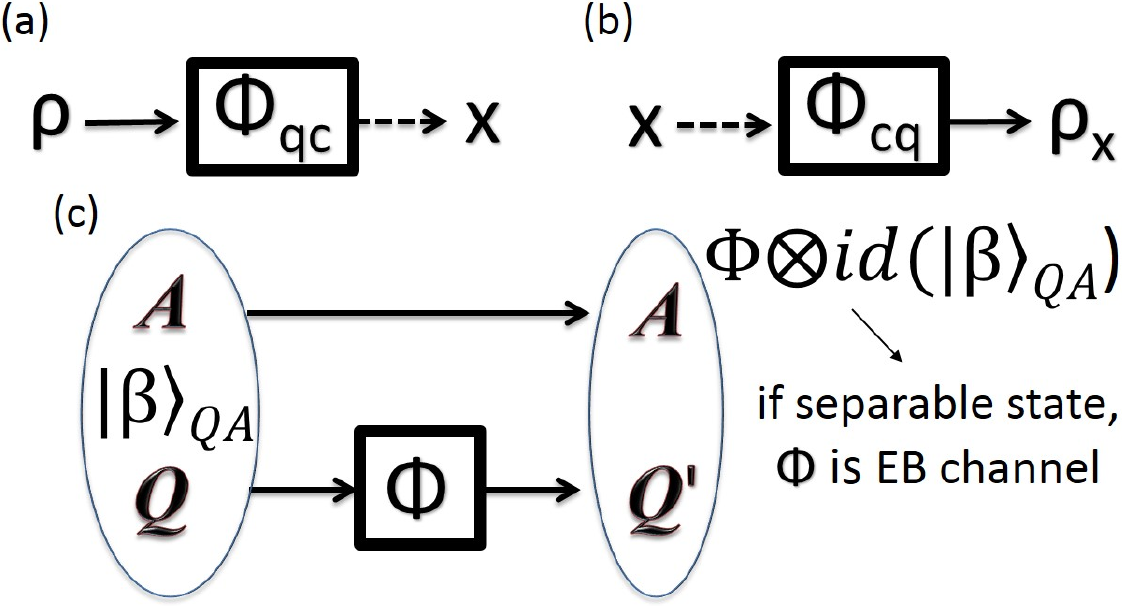}
\caption{(a) qc-channels mapping a quantum state $\rho$ into a classical symbol $x$, while the 
reversed mapping is represented by cq-channels (b); (c) Definition of the Choi-Jamiolkowski (CJ) state
in terms of a channel $\Phi$ and a maximally entangled state $|\beta \rangle_{QA}$ of the system $Q$ and an auxiliary
system $A$ subjected to the identity channel. When CJ state is separable, the corresponding map $\Phi$
is called entanglement-breaking (EB).}\label{fig4abc}
\end{figure}

It is finally worth recalling that there exists a fundamental relation, known as the Choi-Jamiolkowski (CJ) isomorphism \cite{CHOI,J72},
which permits to describe any CPTP $\Phi$
as a density operator of a composite system $QA$ with
$A$ being an auxiliary system having the same dimension $d$ as~$Q$ -- see Fig. \ref{fig4abc}c.
The explicit connection is obtained by applying the map $\Phi$ to half
of a maximally entangled state~\cite{HHHH09,GHUNE09} 
$|\beta\rangle_{QA} = \sum_j |e_j\rangle_Q \otimes |e_j\rangle_A/\sqrt{d}$ of $QA$
to create the so called CJ state of the channel
\begin{eqnarray}
\label{CJiso}
\rho_{QA}^{(\Phi)}:= (\Phi \otimes \id) ( |\beta\rangle_{QA}\langle \beta|)\;,
\end{eqnarray}
where $\id$ stands for the identity map on $A$ and
$\{|e_j\rangle_{Q}\}$ and $\{ |e_j\rangle_A\}$ represent orthonormal basis of $Q$ and $A$ respectively.

\subsection{Heisenberg picture: the dual channel} \label{sec:HEIS}

Equation~(\ref{qchmap}) implicitly assumes the Schr\"odinger picture in which
the states of the system are evolved while the observables are kept fixed.
In the Heisenberg picture, in which instead the states are fixed and the observables
evolve in time, the CPTP transformation $\Phi\in \mathfrak{P}(Q\mapsto Q')$ is replaced by its dual map
\begin{eqnarray}\label{equazioneHEIS}
O \in {\cal B}({\cal H}_{Q'}) \mapsto\Phi^*(O) \in {\cal B}({\cal H}_Q)\;,
\end{eqnarray}
operating on the {\it bounded} operator algebra ${\cal B}({\cal H}_{Q'})$ 
of the receiver observable and defined through the identity
\begin{equation}\label{defduale}
\Tr \left[ \Phi(\rho_Q) O \right] = \Tr \left[ \rho_Q\Phi^*(O) \right] \, ,
\end{equation}
which holds for all $O \in {\cal B}({\cal H}_{Q'})$  and for all $\rho_Q\in \mathfrak{S} ({\cal H}_{Q})$.
The Heisenberg-picture transformation $\Phi^*$ is linear and completely positive,
but in general it is not trace preserving. On the other hand it is always {\em unital},
i.e., it maps the identity operator into itself. Operator sum representations for $\Phi^*$ can
be easily constructed from those of $\Phi$~(\ref{kraus1}),
yielding
\begin{equation}
\Phi^*(O) = \sum_{j} K_j^\dag O K_j \;.
\end{equation}
Notice also  that in the dual picture the concatenation of channels goes in reverse order with respect to the
Schr\"odinger picture, i.e., given $\Phi$ and $\Omega$ CPTP maps,
\begin{eqnarray}
({\Omega}\circ {\Phi})^* = \Phi^* \circ \Omega^*\;.
\end{eqnarray}

\subsection{cq- and qc-channels} \label{CQQC}

Besides considering  physical transformations which represent the evolution 
of quantum carriers, in quantum information it is useful to describe processes 
which map classical inputs into quantum states (cq-channels) or, vice-versa, 
quantum states into classical outputs (qc-channels) -- see Fig. \ref{fig4abc}a-b.
Specifically the former define state preparation procedures where a symbol $x$
extracted from a classical alphabet $\mathcal{X}$ with probability $p_x$ is {\it encoded} 
into a state $\rho_Q^{(x)}$ of the quantum system $Q$, thus producing an average 
density operator $\sum_{x\in \mathcal{X}} p_x \rho_Q^{(x)}$.
qc-channels instead correspond to measurement procedures which, given $\rho_Q\in \mathfrak{S} ({\cal H}_{Q})$, 
produce classical outcomes $x\in \mathcal{X}$ with conditional probabilities
\begin{eqnarray}
p_x(\rho_Q) =\mbox{Tr} [ E_x \rho_Q]\;,
\end{eqnarray}
with $\{ E_x\}_{x\in \mathcal{X}}$ being a set of positive operators on $\mathcal{H}(Q)$,
satisfying the normalization condition $\sum_x E_x=\openone$, which defines the 
statistics of the measurement in the POVM (Positive-Operator Valued Measure) 
representation~\cite{BPbook,PETZBOOK,HOLEVObook}.
cq- and qc-channels can both be extended to CPTP maps~(\ref{qchmap}) by 
introducing an ancillary quantum system $Q'$ of dimension equal to the 
cardinality of $\mathcal{X}$, and characterized by an orthonormal set $\{ |e_x\rangle_{Q'}\}_{x\in \mathcal{X}}$~\cite{Holevo04}. 
For instance taking $p_x={_{Q'}\langle} e_x| \rho_{Q'}|e_x \rangle_{Q'}$ with 
$\rho_{Q'}\in \mathfrak{S} ({\cal H}_{Q'})$, the cq-channel defined above induces
the following CPTP mapping from $Q'$ to $Q$,
\begin{eqnarray}
\rho_{Q'} \mapsto \Phi_{\rm cq}(\rho_{Q'}):=  \sum_{x\in \mathcal{X}} {_{Q'}\langle} e_x| \rho_{Q'} |e_x \rangle_{Q'}  \; \rho_Q^{(x)} \;.\label{CQ}
\end{eqnarray}
Analogously the qc-channel induces the following CPTP mapping from $Q$ to $Q'$,
\begin{eqnarray}
\rho_Q \mapsto  \Phi_{\rm qc}(\rho_Q) := \sum_{x\in \mathcal{X}}   |e_x \rangle_{Q'}\langle e_x|   \;\mbox{Tr}[E_x \rho_Q ]\;. \label{QC}
\end{eqnarray}
The concatenation $\Phi_{\rm cq} \circ \Phi\circ  \Phi_{\rm qc}$, with $\Phi$ 
being a generic quantum channel on from $Q$ to $Q$, can be also represented as a CPTP map 
(from  $Q'$  to $Q'$) and describes the typical scenario where a collection of 
classical messages  (represented by elements of the set $\mathcal{X}$) are transferred to 
Bob via a quantum link (represented by the $\Phi$) who ``reads" them through the 
POVM $\{ E_x\}_{x\in \mathcal{X}}$.
In particular when applied to elements of the orthonormal set
$\{ |e_x\rangle_{Q'}\}_{x\in \mathcal{X}}$, $\Phi_{\rm cq} \circ \Phi\circ  \Phi_{\rm qc}$ 
induces a classical stochastic process where $x\in \mathcal{X}$ is mapped into $x'\in \mathcal{X}$ 
with conditional probability
\begin{eqnarray}
p(x'|x) = \mbox{Tr} [ E_{x'} \Phi(|e_x \rangle_{Q'}\langle e_x|)]\;.
\end{eqnarray}

\subsection{Entanglement Breaking and PPT  channels} \label{sec:EBCHANNELS}

The cq- and qc-channels defined in the previous section  are particular instances of a
larger group of CPTP transformations, called Entanglement Breaking (EB).
As the name suggests, a channel $\Phi$ is EB if, when operating on half of a joint input
state $\rho_{QA}$ of $Q$ and of an ancillary system $A$, produces output states
$(\Phi\otimes \id )(\rho_{QA})$  that are {\it separable} (i.e., not entangled)~\cite{HHHH09,GHUNE09} 
-- see Fig. \ref{fig4abc}c.
These maps are closed under convex combination and channel concatenations, that is, 
given $\Phi_1$ and $\Phi_2$ EB, then $p \Phi_1 + (1-p) \Phi_2$
and $\Phi_1 \circ \Phi_2$ are also EB for all $p\in [0,1]$
(more generally concatenating an EB map with a generic CPTP map produces an EB channel).
Necessary and sufficient conditions for being EB can be found in
Refs.~\cite{HSR03,R03} and, for the special case of infinite dimensional systems, in Ref.~\cite{Holevo2008}. 
In particular, $\Phi$ is  EB if and only if its associated CJ state~(\ref{CJiso}) is separable. 
Alternatively $\Phi$ is EB if and only if it is possible to identify a POVM $\{ E_k\}_{k}$ on $Q$ 
and collection of states $\{  \rho_{Q'}^{(k)}\}_{k}$ in the output space $Q'$ such that
  \begin{eqnarray}
  \Phi(\rho_Q)  = \sum_{k}  \rho_{Q'}^{(k)}    \;\mbox{Tr}[E_k \rho_Q ]\;, \label{EBREP}
\end{eqnarray}
for all inputs $\rho_Q$ (this last condition immediately shows that maps~(\ref{CQ}), (\ref{QC}) are indeed EB).

EB channels form a proper subset of PPT channels. The latter are defined as those
channels which produce output states $(\Phi\otimes \id )(\rho_{QA})$ with positive partial transpose (PPT)~\cite{HHH96,P96,R01}.
A necessary and sufficient condition for such a property is that the channel's CJ state~(\ref{CJiso}) is PPT.
Channels which are PPT but not EB are called entanglement binding maps.
Generalizations of EB channels have been presented in~\cite{DPG12} to describe those
CPTP maps that become EB only after a certain number of concatenations, and in~\cite{MZiman2010,FRZ12} 
to describe maps that, when acting jointly on a composite system break entanglement among the subsystems that compose it.

\subsection{Complementary channels and Degradability}\label{DegPPTEB}
\begin{figure}[t]
\centering
\includegraphics[width=0.48\textwidth]{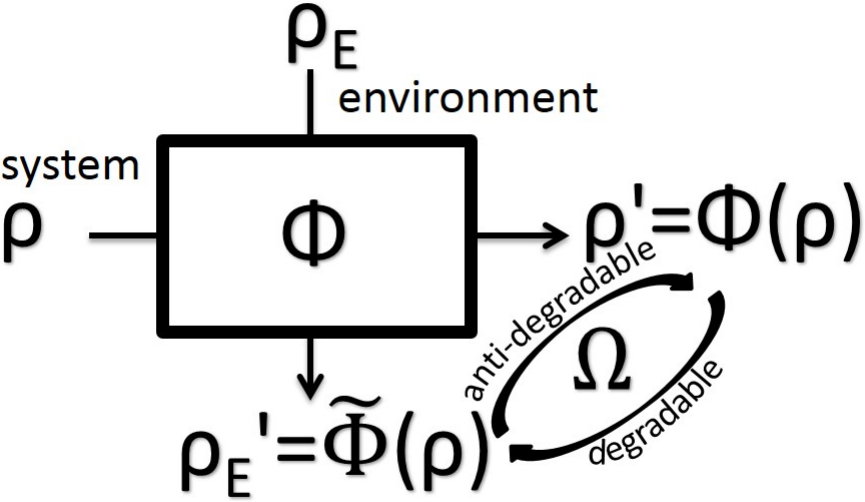}
\caption{Graphical representation of a channel $\Phi$, its complementary one $\tilde{\Phi}$, and the degradability 
properties, in terms of the system and an external environment $E$.}\label{fig2d}
\end{figure}

Associated with the Stinespring representation~(\ref{dia}) is the notion of the
\emph{complementary channel} of $\Phi$ (see Fig. \ref{fig2d}).
The latter is the CPTP map $\tilde{\Phi} \in \mathfrak{P}(Q\mapsto E)$
which  sends the initial states of the system $Q$  into the  states of the environment $E$ through the transformation
\begin{equation}\label{cc}
\rho_Q \mapsto \tilde{\Phi}(\rho_Q) := \Tr_Q \Big[ U_{QE} \big( \rho_Q \otimes|\omega\rangle_E\langle \omega|  \big) U_{QE}^{\dag} \Big] \;,
\end{equation}
where $\Tr_Q$ denotes the partial trace over the system Hilbert space. 
The purity of the environmental state $\omega_E=|\omega\rangle_E\langle \omega|$
ensures the uniqueness of $\tilde{\Phi}$ up to an isometric transformation on $E$.
Channels~(\ref{cc}) defined in terms of unitary dilations~(\ref{dia}) with non pure states $\omega_E$
are called \emph{weak-complementaries} of $\Phi$ and  in general don't enjoy such symmetry~\cite{CG06,CGH06}.

The definition of $\tilde{\Phi}$  allows us to introduce another property of quantum channels, which is called
\textit{degradability} \cite{DS05}. A map $\Phi$ is \textit{degradable} when one can
recover the final environment state $\tilde{\Phi}(\rho_Q)$ just by applying a third CPTP map to
the output system state. More formally, a \textit{degradable} map is such that there exists
a CPTP map ${\Omega} \in \mathfrak{P}(Q\mapsto E)$ satisfying the relation:
\begin{equation}\label{degra}
\tilde{\Phi} = {\Omega} \circ {\Phi}\;.
\end{equation}
Similarly, a channel is called \textit{anti-degradable} when the opposite relation holds, i.e.,
\begin{equation}\label{antidegra}
{\Phi} = {\Omega}\circ \tilde{\Phi} \;,
\end{equation}
for some ${\Omega} \in \mathfrak{P}(E\mapsto Q)$, as shown in Fig. \ref{fig2d}. 
Special examples of anti-degrabable channels
are the {\it symmetric channels} introduced in~\cite{SSW08}: these are CPTP maps for which
$\Phi$ and $\tilde{\Phi}$ coincide (hence they are both degradable and anti-degradable).
Structural properties of degradable and anti-degradable channels have been extensively analyzed
in Ref.~\cite{CRS08}, showing for instance that
EB channels are always anti-degrabable.
Analogous definitions can be obtained for \emph{weak-complementary} channels: in this case one says that
$\Phi$ is \emph{weakly-degradable} if Eq.~(\ref{degra}) holds -- there is no need to define 
a weakly-anti-degradability condition as
the latter can be shown to be equivalent to the  anti-degradabilty condition~\cite{CGH06}.

\subsection{Causal, Localizable, LOCC and Separable channels}\label{sec:semicausal}

Additional structures arise when a quantum channel $\Phi$ acts on a multipartite system,
e.g., a bipartite one, $Q=Q_1Q_2$.
It is useful to imagine that the two subsystems are associated with spatially or temporally
separated laboratories where local CPTP maps can be applied and that can exchange
classical or quantum information.
In particular, bipartite channels can be characterized in terms of:
1) how the output of one subsystem changes if a local transformation is applied
to the input of the other subsystem;
2) which resources (e.g., a pre-shared quantum state, local operations on the subsystems,
classical or quantum communication) are needed to simulate the bipartite channel.

\begin{figure}[t]
\centering
\includegraphics[width=0.48\textwidth]{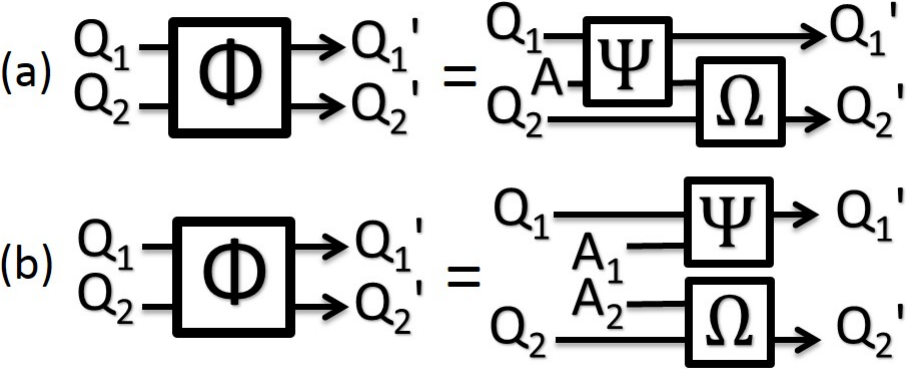}
\caption{Graphical structure of semi-localizable (a) and localizable (b) channels on
bipartite systems $Q_1Q_2$, where the two parties 
act locally with the maps, respectively, $\Psi$ and $\Omega$ on their subsystems and some auxiliary resources.}
\label{fig5ab}
\end{figure}

The notions of {\it causal} and {\it semi-causal} channels
developed in Refs.~\cite{ESW01,PHHH06} provide a means of characterizing how
the output of one subsystem depends on the input of the other. In this context
a quantum channel $\Phi \in \mathfrak{P}(Q\mapsto Q)$ acting on a bipartite system
$Q=Q_1Q_2$ is said to be $Q_1 \nrightarrow Q_2$ \textit{semi-causal}~\cite{BGNP01} if for any
local CPTP map $\Psi \in \mathfrak{P}(Q_1\mapsto Q_1)$  applied to $Q_1$ before the action of $\Phi$, there
is no detectable effect in the subsystem $Q_2$, i.e.,
\begin{equation}\label{12semi-causal}
\Tr_{Q_1} \left[ \Phi \left(\rho_{Q}\right) \right] = \Tr_{Q_1} \left\{ \Phi \left[ \left( {\Psi} \otimes \mathrm{id} \right) \left( \rho_{Q} \right) \right] \right\} \;,
\end{equation}
where $\rho_Q$ is a generic (possibly entangled) input state of the two carriers and
where $\mbox{Tr}_{Q_1}$ denotes the partial trace with respect to $Q_1$.
In other words, for $Q_1 \nrightarrow Q_2$ \textit{semi-causal} maps cross-talking from $Q_1$ to $Q_2$ is prevented.
Similarly, one introduces the notion of $Q_2 \nrightarrow Q_1$ \textit{semi-causal} map.
When both properties are satisfied, the map is called \textit{causal} or \textit{non-signaling}.
Special examples of non-signaling channels are the tensor product channels  $\Phi= \Phi_1 \otimes \Phi_2$ with $\Phi_{1,2}$ being CPTP maps operating locally on $Q_1$ and
$Q_2$ respectively.

Another way of characterizing a bipartite quantum channel is in terms of the physical
resources which are needed to simulate it.
A bipartite channel is said to be \textit{localizable} if
it can be implemented by applying local CPTP maps on the subsystems with the
assistance of a pre-shared bipartite quantum state \cite{BGNP01}.
Notice that the simulation of localizable channels does not require classical nor
quantum communication between the two laboratories.
Formally, this is the case when $\Phi$ can be represented as
\begin{equation}\label{semi-causal}
\Phi(\rho_{Q_1 Q_2}) = \Tr_{A_1 A_2} \left[ (\Psi \otimes \Omega) (\rho_{Q_1 Q_2} \otimes \omega_{A_1 A_2} ) \right]\;,
\end{equation}
where $\omega_{A_1 A_2}$ is a shared bipartite state, and $\Psi$ and
$\Omega$ are quantum channels acting locally on subsystems $Q_1 A_1$ and $Q_2 A_2$,
respectively -- see Fig. \ref{fig5ab}.
Otherwise, $\Phi$ is called $Q_1 \rightarrow Q_2$ \textit{semi-localizable}
if also one-way quantum communication from $Q_1$ to $Q_2$ is required to simulate the channel.
Accordingly in this case Eq.~(\ref{semi-causal}) is replaced by
\begin{equation}\label{semi-localizable}
\Phi(\rho_{Q_1 Q_2}) = \Tr_{A} \left[ (\Psi \circ \Omega) (\rho_{Q_1 Q_2} \otimes \omega_{A} ) \right] \,,
\end{equation}
with $\omega_{A}$ being the state of an ancillary system $A$ which is transmitted
from one laboratory to the other and acts as the mediator between $Q_1$ and $Q_2$,
and $\Psi$ and $\Omega$ are quantum channels acting on the systems $Q_2A$ and $Q_1A$ respectively.

By comparison of (\ref{12semi-causal}) and (\ref{semi-localizable}) it follows that all
$Q_1 \rightarrow Q_2$ semi-localizable maps are $Q_2 \nrightarrow Q_1$ semi-causal, which
in turn implies that all localizable maps are causal.
Moreover, it can be proven that semi-causality implies semilocalizability, hence semi-causal
and semi-localizable maps coincide, although causal and localizable maps do not \cite{BGNP01,ESW01,PHHH06}.

An important class of bipartite quantum channels are finally those that can be simulated
only with local operations and classical communication (LOCC), see \cite{CLMOW2012}
for a recent survey.
These channels are hence termed  LOCC channels.
Classical communication is generally allowed in both directions between the two laboratories, the most general LOCC channel however can always be equivalently obtained by concatenating local CPTP maps and one-way classical communication. A LOCC
transformation can hence be simulated by a finite number of iterations of
the following sequence of operations:
1) on one of the two subsystems, say $Q_1$, a local CPTP map $\Psi_1$ is applied;
2) classical information is sent from $Q_1$ to $Q_2$, possibly conditioned on the
local output of the map $\Psi_1$;
3) conditioned on the received classical information, a local CPTP map $\Psi_2$ is applied on
subsystem $Q_2$;
4) the sequence of operations is repeated with the roles of $Q_1$ and $Q_2$ exchanged.
A closely related class of bipartite channels is that of {\it separable channels},
defined as those channels admitting a Kraus representation in which all the Kraus
operators are in the form of a direct product of operators acting on the local subsystems.
It is easy to see that all the LOCC channels are separable. Interestingly enough,
there exist separable channels which are not LOCC \cite{BennettETAL99}.

\subsection{Examples} \label{sec:examples}

Here some examples of quantum channels are presented.

\subsubsection{Qubit channels}\label{sec:qubit}

Qubit channels are the simplest, yet non trivial, example of quantum channels: 
they are  CPTP transformations $\Phi \in \mathfrak{P}(Q \mapsto Q)$
that map the states of a bi-dimensional quantum system (qubit) into
states of the same system (in this case ${\cal H}_Q = {\mathbb C}^{2}$).
A compact characterization of these channels can be obtained by adopting the Bloch ball representation, according to which
any density operator $\rho$ of the system is uniquely identified with the corresponding (Bloch) vector
$\sf{r} = (r_x, r_y, r_z)\in \mathbb{R}^3$
of length  $|\sf{r}|\leqslant 1$ via the correspondence
 \begin{eqnarray}
\rho = \rho(\sf{r}):= \frac{1}{2} (\openone + \sf{r} \cdot \sf{s}) \;,
\end{eqnarray}
where  ${\sf s} = (\sigma_x, \sigma_y, \sigma_z)^{\top}$ is a column vector formed by the Pauli matrices.
In this framework  any qubit channel $\Phi$
 induces affine transformations  of the form
\begin{eqnarray}
 \sf{r} \mapsto  \sf{r}' = \sf{M} \sf{r} + \sf{t} \;,  \label{affinemapp}
 \end{eqnarray}
with $\sf{M}$ and $\sf{t}$  being respectively  a fixed $3\times3$ real matrix and a fixed three dimensional real vector satisfying certain
 consistency requirements  -- see  Refs.~\cite{KR01,R03,RSW02}.
    In particular qubit unital channels are obtained for $\sf{t}=0$ and $\sf{M}^\top \sf{M}  \leqslant \sf{I}$, the inequality being saturated
    if and only if  $\Phi$ describes  a  unitary transformation
(the latter case corresponds to having ${\sf{M}} \in \mbox{SO}(3, \mathbb{R})$).
Exploiting this fact  and the matrix singular value  decomposition \cite{HJ90}
one can use the unitary equivalence of Eq.~(\ref{UNIEQ})
to identify  a canonical form for the qubit channel $\Phi$ where  the matrix $\sf{M}$ of Eq.~(\ref{affinemapp})  
is written as $\sf{O}' \sf{D} \sf{O}$ with $\sf{D}$ being a real diagonal $3\times 3$ matrix,
and with $\sf{O}'$ and $\sf{O}$ being elements  of  $\mbox{SO}(3, \mathbb{R})$.

\begin{figure}[t]
\centering
\includegraphics[width=0.3\textwidth]{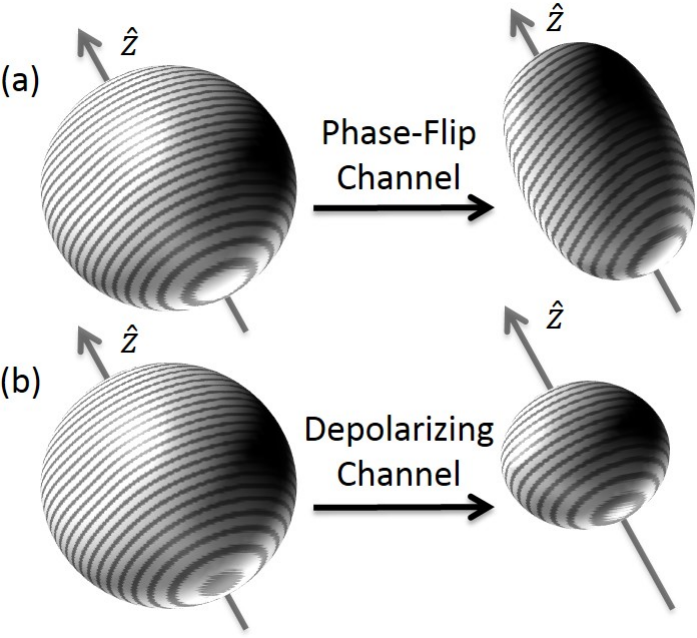}
\caption{Pictorial representation of the shrinking effects on the Bloch sphere via phase-flip (a) and depolarizing (b) channels.}
\label{fig3qbit}
\end{figure}

An important class of qubit channels that have been extensively analyzed in the literature are
those admitting a representation~(\ref{kraus1}) with only two  Kraus  operators $K_0$ and $K_1$.
In the canonical basis formed by the eigenvectors $\{ |0\rangle, |1\rangle\}$ of the Pauli operator $\sigma_z$
they can be parametrized as
\begin{equation}
\label{kraus:qq} K_0=\left(
\begin{array}{cc}
  \cos\theta & 0 \\
  0 & \cos\phi \\
\end{array}
\right),\quad K_1=\left(
\begin{array}{cc}
  0 & \sin\phi \\
  \sin\theta & 0 \\
\end{array}
\right)\;,
\end{equation}
with $\theta, \phi  \in [0,\pi]$, up to unitary rotation.
The corresponding affine mapping~(\ref{affinemapp}) is obtained with
$\sf{M}  =  {\rm diag}\left(\cos(\phi-\theta),\cos(\phi+\theta),(\cos(2\theta)+\cos(2\phi))/2\right)$
and $\sf{t}=\left( 0, 0, (\cos(2\theta)-\cos(2\phi))/2 \right)$.
In the Stinespring representation~(\ref{dia})  these maps describe situations in which the qubit system
interacts with the smallest non-trivial  environment (i.e., another qubit initialized  in a pure state) and can be shown to be
degradable  for $\cos(2\theta)/\cos(2\phi) \geq 0$  and anti-degradable
otherwise~\cite{GF05,WP05,CG07}.
In particular  setting  $\cos(2 \theta)=1$ and $\cos(2 \phi)=
2\eta -1$,  Eq.~(\ref{kraus:qq})  defines the  \textit{amplitude damping} channel
with damping rate~$\eta$.
For $\sin\theta=\pm \sin \phi$ instead one gets unital maps. Specifically
  for $\phi= \theta$, Eq.~(\ref{kraus:qq}) describes
 the \textit{bit-flip} channel that exchanges the states $|0\rangle$ and
$|1\rangle$ with probability $p_x= \sin^2\phi$.
 By applying the unitary matrix (Hadamard transform) $\frac{1}{\sqrt{2}}\left(\begin{array}{cc}
1& 1\\ 1 &-1 \end{array}\right)$ to $K_1,K_2$ in Eq.~(\ref{kraus:qq}) one recovers
a unitarily equivalent channel called the \emph{phase-flip} channel (or \emph{phase damping} channel)
that introduces a $\pi$ shift  between the states $|0\rangle$ and
$|1\rangle$ with probability $p_z= \sin^2\phi$ -- see Fig. \ref{fig3qbit}.
Still Eq.~(\ref{kraus:qq}) for  $\phi = -\theta$ describes the  \textit{bit-phase flip} channel
which with probability $p_y= \sin^2\phi$ exchanges the states $|0\rangle$ and
$|1\rangle$ and also adds a relative $\pi$ shift to them. 

Convex combinations of these three maps plus the identity channel define the class of \emph{Pauli channels}: i.e.,
\begin{equation}\label{PAULI}
\Phi(\rho) = p_0 \rho + p_1 \sigma_x \rho \sigma_x + p_2 \sigma_y \rho \sigma_y +p_3 \sigma_z \rho \sigma_z \, ,
\end{equation}
with non-negative parameters $p_0+p_1+p_2+p_3=1$.
Via the canonical representation detailed in the previous paragraphs any other
unital qubit map can be obtained from~(\ref{PAULI}) through the concatenation~(\ref{UNIEQ}).

\subsubsection{Erasure channels} \label{SEC:ERASURE}

Erasure channels describe those communication scenarios in which errors are somehow 
heralded (i.e., the receiver Bob can determine whether or not something bad has 
happened to Alice's original message): accordingly they provide the simplest 
examples of CPTP maps operating among spaces of different dimensionality. 
Given a system $Q$ described by the Hilbert space ${\cal H}_Q$, an erasure map 
is a stochastic transformation connecting $\mathfrak{S}({\cal H}_{Q})$ with
$\mathfrak{S}({\cal H}_{Q'})$, where ${\cal H}_{Q'} = {\cal H}_Q \oplus |e\rangle$ 
and $\oplus$ denotes the direct sum of the input Hilbert space with an extra 
``erasure" state $|e\rangle$ (the error ``flag") which is orthogonal to each of 
the vectors of $Q$.  
In particular as described in Ref.~\cite{GBP97,BDS97}, the channel will send the 
input $\rho_Q$ to itself with probability $1-p$ and to $|e\rangle$ with probability $p$.

\subsubsection{Weyl covariant channels}\label{sec:weyl}

Given a quantum system of finite dimension $d$ (\emph{qudit})
and the canonical basis $\{|e_{k}\rangle\}_{k=0,\dots,d-1}$,
consider the group $\mathbb{Z}_{d}\times \mathbb{Z}_{d}$
as a discrete phase space and take the unitary representation of such a group in the Hilbert space  $\mathcal{H}$ of the system  as
\begin{equation}
{\sf z}=\left( \begin{array}{c} x \\ y \end{array} \right) \mapsto W_{\sf z}=U^{x}V^{y},
\label{Wdef}
\end{equation}
where $x,y\in $ $\mathbb{Z}_{d}$ and $U$, $V$ are unitary
operators on $\mathcal H$ generalizing the Pauli operators  $\sigma_x$ and $\sigma_z$ in the following way \cite{G99}
\begin{equation}
U|e_{k}\rangle=|e_{k+1(\mathrm{mod}d)}\rangle,\qquad
V|e_{k}\rangle=\exp \left( \frac{2\pi i k}{d}\right)
|e_{k}\rangle.
\label{Pauligendef}
\end{equation}
The operators $W_{\sf z}$ are the \textit{discrete Weyl operators} and satisfy the
canonical commutation relations
\begin{equation}
W_{\sf z}W_{{\sf z}^{\prime }}=e^{(2\pi i /d) {\sf z}^\top {\sf \Sigma} {\sf z}^{\prime}} W_{{\sf z}^{\prime }}W_{\sf z},
\label{ccrdiscrete}
\end{equation}
where the row vector ${\sf z}^\top$ is the transpose of the column vector ${\sf z}$ and where
\begin{equation}
\sf{\Sigma}:=\left(
\begin{array}{cc}
{\sf 0} & -{\sf 1}\\
{\sf 1} & {\sf 0}
\end{array}\right)\;,
\label{Spform}
\end{equation}
is the matrix representation of the symplectic form with ${\sf 0}$ 
(resp. ${\sf 1}$) the null (resp. identity) $d\times d $ matrix.

The algebra $\mathcal{B}(\mathcal{H})$
of operators on $\mathcal{H}$ can be considered
as a Hilbert space supplied by the
Hilbert-Schmidt inner product.
There, the Weyl operators form an orthogonal
basis, ${\rm Tr} \left( W_{\sf z}W_{{\sf z}^{\prime }}^{\dag }\right) = d \delta _{{\sf z}{\sf z}^{\prime }}$,
 hence  for all $O\in
\mathcal{B}(\mathcal{H} )$ one has
\begin{equation}
O=\sum_{\sf z}f_{O}({\sf z})W_{\sf z},
\qquad
f_{O}({\sf z})=\frac{1}{d} {\rm Tr}\left( O W_{\sf z}^{\dag }\right).
\label{Xexp}
\end{equation}

In this scenario a CPTP map $\Phi$ is
said to be Weyl covariant~\cite{FH05} when for all ${\sf z}\in\mathbb{Z}_{d}\times \mathbb{Z}_{d}$  it fulfills the identity
\begin{equation} \label{weylcov}
\Phi\circ {\cal W}_{\sf z} = {\cal W}_{\sf z} \circ \Phi\;,
\end{equation}
 with ${\cal W}_{\sf z}$ representing the quantum channel~(\ref{UNITARYCH}) associated with the unitary $W_{\sf z}$.
 Applying Eq.~(\ref{weylcov}) to the operator $W_{\sf z}$, from (\ref{ccrdiscrete}) it follows that
 $\Phi(W_{\sf z})$ commutes with $W_{\sf z}$.  Hence,
 by means of Eq.~(\ref{Xexp}), one can write
\begin{equation}
\Phi (W_{\sf z})=\phi ({\sf z})W_{\sf z},
\label{Phiaction}
\end{equation}
with $\phi({\sf z})$ a complex-valued function,
termed the `characteristic function of the channel'
(without loss of generality it can be assumed $\phi(0)=1$).

Special examples of Weyl covariant channels are provided by Weyl channels~\cite{A07}
defined as those
CPTP maps which admit  Kraus decomposition in terms of random Weyl operators, i.e.,
$\Phi =\sum_{\sf z}p_{\sf z}{\cal W}_{\sf z}$,
with $p_{\sf z}$ a probability distribution over $\mathbb{Z}_d\times \mathbb{Z}_d$.
These are unital maps and their associated characteristic function is given by
$\phi({\sf z})=\sum_{{\sf z}'}  e^{(2\pi i /d) {\sf z}' {\sf\Sigma}  {\sf z}^T}  p_{{\sf z}'}$.
In particular  any \emph{$d-$depolarizing} channel
\begin{equation}\label{DEPOL}
\Phi(\rho)  =\lambda \rho +(1-\lambda )\frac{1}{d}\openone \;,
\end{equation}
with $\lambda\in [0,1]$
is a Weyl channel having $\phi({\sf z})=\lambda $ for ${\sf z}\neq 0$.
One notices also that since for $d=2$ (Fig. \ref{fig3qbit}) the Weyl operators reduce to the standard Pauli operators including identity,
any unital qubit ($d=2$) channel which is unitarily equivalent to the Pauli channels~(\ref{PAULI})
is a Weyl channel.
It is also possible to define the {\emph{transpose}} $d$-depolarizing transformation,
\begin{equation}\label{DEPOLT}
\Phi(\rho)  =\lambda \rho^\top +(1-\lambda )\frac{1}{d}\openone \;,
\end{equation}
which defines a CPTP map for $\lambda\in\left[1-\frac{d}{d+1},1-\frac{d}{d-1}\right]$ \cite{Fannes04},
(in the above expression $\rho^\top$ is the transpose of $\rho$ with respect to a fix basis).

\subsubsection{Continuous variable quantum channels}\label{CV-qc}

Up to now mainly finite dimensional Hilbert spaces have been considered.
This has been done to avoid technicalities related with the proper definition of the domains of the functionals.
It is true however that the most common implementations of quantum communication lines  are typically realized with continuous-variable
(CV) systems~\cite{BvL05,WPGCRSL11} which at the quantum level are associated with an infinite dimensional space -- consider for instance the
 transferring of classical signals encoded into light pulses propagating along  optical
fibers or in free-space~\cite{CD94}.

CV systems admit a description in terms of a discrete set of (say $n$)  bosonic oscillators,
typically a set of normal modes of the electromagnetic field, defined
by ladder operators $a_1 , a_1^\dag, a_2, a_2^\dag, \cdots, a_n, a_n^\dag$
obeying canonical commutation relations, $[ a_k , a_{k'}^\dag ] = \delta_{kk'}$. Introducing
the generalized ``positions" and ``momenta" coordinates $\{ x_k , y_k \}_{k=1,\dots,n}$,
the density operators $\rho$  of the CV system can be represented in terms of
the (symmetrically ordered) characteristic functions $\chi({\sf z}) = \tr[\rho V({\sf z})]$, where  ${\sf z}:=(x_1, y_1, \dots , x_n, y_n)^{\top}$
defines the vector of phase-space variables and where $V({\sf z}): = \exp{\left[ \frac{1}{\sqrt{2}} \sum_k (x_k+{i} y_k) a_k^\dag - \mbox{h.c.} \right]}$
denote $n$-mode Weyl operators.
The function $\chi({\sf z})$ and the operator   $V({\sf z})$
represent the infinite dimensional counterparts of  $f_{O}({\sf z})$ and  $W_{\sf z}$
introduced in Sec.~\ref{sec:weyl}. 
In particular $V({\sf z})$ fulfills commutation relations analogous to Eq.~(\ref{ccrdiscrete}), i.e.,
\begin{equation}
V({\sf z}) V({\sf z}')=e^{2\pi i \;  {\sf z}^\top {\sf\Sigma}^{(n)} {\sf z}^{\prime}} V({\sf z}')V({\sf z})\;,
\label{ccrcont}
\end{equation}
where now ${\sf \Sigma}^{(n)}$ is the $2n\times 2n$ block matrix defined by ${\sf\Sigma}^{(n)} :=\bigoplus_{k=1}^n \sf{\Sigma}$, with
$\sf{\Sigma}$ the single-mode phase-space canonical symplectic form
deducible from Eq.~(\ref{Spform}) for $d=1$.

Due to their physical relevance and to the relative simplicity of their mathematical
description, a remarkable class of states is the class of Gaussian states~\cite{EP03,FOP05,HOLEVObook}.
They correspond to multi-mode (thermal) Gibbs states of Hamiltonians which are quadratic in the ladder 
operators of the system  and are formally identified by the property of possessing Gaussian characteristic functions, i.e.,
\begin{equation}
\chi({\sf z})=\exp{({i} {\sf m}^{\top} {\sf z}-\frac{1}{2}{\sf z}^{\top} {\sf C}^{(n)}{\sf z})}.
\end{equation}
In this expression $\sf m$ is the vector of first moments
\begin{equation}
 {\sf m}_k ={\rm Tr}\left({\sf x}_k\rho\right),
 \end{equation}
where
\begin{equation}
  \sqrt{2}{\sf x}:=\left((a_1+a_1^\dag),-i(a_1-a_1^\dag),(a_2+a_2^\dag),-i(a_2-a_2^\dag),\ldots\right).
\end{equation}
Furthermore, ${\sf C}^{(n)}$ is the covariance matrix (CM)
\begin{equation}
{\sf C}^{(n)}_{hk} = \frac{1}{2}{\rm Tr}\left({\sf x}_h{\sf x}_k\rho\right)+\frac{1}{2}{\rm Tr}\left({\sf x}_k{\sf x}_h\rho\right)-{\rm Tr}\left({\sf x}_h\rho\right){\rm Tr}\left({\sf x}_k\rho\right),
\end{equation}
obeying  the generalized uncertainty relation \cite{SMD94}
\begin{eqnarray}
 {\sf C}^{(n)}-{i} {\sf\Sigma}^{(n)}/2 \geq 0\;.
 \end{eqnarray}
Concerning quantum channels in CV systems,
attention has been mainly devoted to the study of Gaussian channels, i.e.,
CPTP maps  that map Gaussian input states to Gaussian output states~\cite{HW01,WPGCRSL11}. 
A part from attenuation and thermalization events  arising  from linear interactions with bosonic baths,
they include also  squeezing and linear amplification processes.
When applied to a (not necessarily Gaussian) density operator $\rho$  with characteristic function
$\chi({\sf z})$, a Gaussian channel $\Phi$ will transform it
into an output density operator $\rho'=\Phi(\rho)$ having characteristic function
\begin{equation}
\chi'({\sf z})=\chi({{\sf X}^{(n)}}^{\top}{\sf z}) f({\sf z}) \, ,
\end{equation}
where ${\sf X}^{(n)}$ is a matrix
inducing a linear transformation on the $2n$-dimensional phase-space vector ${\sf z}$, and
the function $f({\sf z})$ is Gaussian, i.e., $f({\sf z})=\exp{({i} {{\sf d}^{(n)}}^{\top}{\sf z}-\frac{1}{2}{\sf z}^{\top} {\sf Y}^{(n)} {\sf z})}$.
The linear term proportional to the vector ${\sf d}^{(n)}$ accounts for a translation
(displacement) of the mean ${\sf m}$, while the quadratic term proportional
to the matrix ${\sf Y}^{(n)}$ adds a term to the CM.
CPTP conditions are ensured if and only if
\begin{eqnarray}
{\sf Y}^{(n)} - i ({\sf\Sigma}^{(n)}   -   {\sf X}^{(n)} {\sf\Sigma}^{(n)} {{\sf X}^{(n)}}^{\top})/2  \geq 0\;. \label{constraint1}
\end{eqnarray}
Gaussian transformations which are also unitary are characterized by the property that
${\sf X}^{(n)}$ is a symplectic matrix (i.e., ${\sf X}^{(n)} {\sf\Sigma}^{(n)} {{\sf X}^{(n)}}^{\top} = {\sf \Sigma}^{(n)}$), and ${\sf Y}^{(n)}=0$.
An $n$-mode Gaussian channel is hence characterized by the triad
$({\sf d}^{(n)},{\sf X}^{(n)},{\sf Y}^{(n)})$ satisfying the constraint~(\ref{constraint1}).
The concatenation of two Gaussian channels with associated triads
$({\sf d}^{(n)}_1,{\sf X}^{(n)}_1,{\sf Y}^{(n)}_1)$
and $({\sf d}^{(n)}_2,{\sf X}^{(n)}_2,{\sf Y}^{(n)}_2)$ is in turn characterized by the triad
$( {\sf X}^{(n)}_2 {\sf d}^{(n)}_1 + {\sf d}^{(n)}_2 , {\sf X}^{(n)}_2 {\sf X}^{(n)}_1 , {\sf X}^{(n)}_2 {\sf Y}^{(n)}_1
{{\sf X}^{(n)}_2}^{\top} + {\sf Y}^{(n)}_2 )$.
It follows that, by applying suitable Gaussian unitaries at the input and output
of the channel, one can always reduce the channel to a canonical form,
in which ${\sf d}^{(n)}=0$, and the matrices ${\sf X}^{(n)}$, ${\sf Y}^{(n)}$ take a particular symmetric form.
For the case of channels acting on one or two modes, the reduction to
canonical forms allows one to classify Gaussian quantum
channels according to invariance under unitary transformations
\cite{Holevo2007,CG06,CarusoEisert2008,SEW05}.

The basic processes of linear attenuation and amplification are modeled by single-mode Gaussian
channels with ${\sf{X}}^{(1)}=\sqrt{\eta}$, ${\sf{Y}}^{(1)} = |1-\eta| /2$.
For $\eta \leq 1$ these channels describe linear losses (with attenuation factor $\eta$), while for
$\eta >1$ they model the process of parametric amplification (with gain $\eta$).
If extra Gaussian noise affects the attenuation or amplification process, one gets the noisy versions of
the lossy and amplifier channel.
In particular, the {\it lossy and noisy} Gaussian channel is defined by
${\sf{X}}^{(1)} = \sqrt{\eta}$ and ${\sf{Y}}^{(1)} = (1-\eta)(N_\mathrm{th}+1/2)$ ($\eta \in [0,1]$
and $N_\mathrm{th} \geq 0$), and the {\it additive noise} Gaussian channel by
${\sf{X}}^{(1)} = 1$ and ${\sf{Y}}^{(1)} = N_\mathrm{add}$ ($N_\mathrm{add} \geq 0$). Notice that the
additive noise can be obtained from the lossy and noisy channel by taking the limit of $\eta \to 1$
and $N_\mathrm{th} \to \infty$ under the condition $(1-\eta)(N_\mathrm{th}+1/2)=N_\mathrm{add}$.

\subsection{Transfer fidelities and channel distances} \label{sec:NEWSEC}

In quantum information distance
 measures  are of fundamental importance:  by determining
 how far apart two states  or two transformations are from each other,
 they are an essential guidance in the optimization of the data-processing.

\subsubsection{Input-output and entanglement fidelity of a quantum channel} \label{sec:fidecha}

A proper way to determine how much a system $Q$ is modified by the action of a channel $\Phi\in \mathfrak{P}(Q\mapsto Q)$,
can be obtained by considering the {\it fidelity} functional $F(\rho_1,\rho_2)$~\cite{U76,J94}
(the definition relevant properties are recalled in Appendix~\ref{sec:dist}). 
Accordingly, for each input $\rho_Q$ one defines the {\it input-ouput} (or {\it transfer}) {\it fidelity} associated with the map  $\Phi$, as
\begin{eqnarray}\label{INPUTOUPUTF}
F(\rho_Q; \Phi) := F(\rho_Q, \Phi(\rho_Q))\;,
\end{eqnarray}
which for a pure state $|\psi\rangle_Q$ is linked to
the error probability $P_e(|\psi\rangle_Q; \Phi)$
of {\it not} getting the right state at the channel output,
via the identity
\begin{eqnarray} \label{errorPROBNEW}
P_e(|\psi\rangle_Q; \Phi)=1 -F(|\psi\rangle_Q; \Phi)  \;.
\end{eqnarray}
An overall  estimate of the disturbance  introduced by the channel can then be obtained
by looking at how different from unity is
 the {\it minimum}  or (alternatively)  the {\it average} of  $F(\rho_Q; \Phi)$ evaluated with respect to all possible pure input states of $Q$, i.e., the quantities
\begin{eqnarray}\label{DEFNUOVADIFMIN}
F_{\min}(\Phi) &:=&  \min_{|\psi\rangle_Q} F(|\psi\rangle_Q; \Phi)  \;, \\
\bar{F}(\Phi) &:=& \int d\mu(\psi) F(|\psi\rangle_Q; \Phi)\;, \label{DEFNUOVADIFAVE}
\end{eqnarray}
the rational being that $F_{\min}(\Phi) =1$, as well as $\bar{F}(\Phi) =1$, can occur  if and only if $\Phi$ coincides with the identity channel $\id$. 
The average in Eq.~(\ref{DEFNUOVADIFAVE}) is performed with respect to the Haar measure $d\mu(\psi)$ of the group whose action on a vector $|\psi\rangle$ is able to generate the entire space of states~\cite{ZYCBOOK});
the minimization in Eq.~(\ref{DEFNUOVADIFMIN}) instead can be generalized to include also mixed states  by exploiting the concavity property of the  fidelity~\cite{NC00,Wbook13}, i.e., $F_{\min}(\Phi) =\min_{\rho_Q} F(\rho_Q; \Phi)$.

To gauge the disturbance of the channel $\Phi$,
one may also consider its {\it entanglement fidelities}~\cite{schumacher96},
defined as the input-output fidelities  of the extended map $\Phi\otimes \id$ when operating on {\it purifications} of the density matrices $\rho_Q$.
Recall that a \emph{purification} of a density matrix $\rho_Q \in \mathfrak{S} ({\cal H}_{Q})$
 is any pure state $|\psi_\rho \rangle_{QR} \in {\cal H}_{Q} \otimes {\cal H}_{R}$ of the enlarged system formed
 by $Q$ and by an ancillary system $R$, which
fulfills  the property  $\rho_Q =\tr_{R}\left(|\psi_\rho \rangle_{QR} \langle \psi_\rho |\right)$~\cite{HHHH09,GHUNE09}.
The entanglement fidelity is then written as
\begin{eqnarray}\label{DEFENTFID}
F_e(\rho_Q; \Phi) &:=& F(|\psi_\rho\rangle_{QR}; \Phi\otimes \id) \;,
\end{eqnarray}
where $\mathrm{id}$ is the identity map on $R$.
It is important to stress that $F_e(\rho_Q; \Phi)$
is independent of the way $|\psi_\rho\rangle_{QR}$ is constructed and of the choice of
the ancillary system:
 as a matter of fact, given $\{ K_j \}_{j}$ a set of Kraus operators of $\Phi$,
it can be expressed as
\begin{equation} \label{defentifd1}
F_e(\rho_Q; \Phi) = \sum_{j}\left| \tr(\rho_Q K_j) \right|^2 \, .
\end{equation}
Operationally the entanglement fidelity functional~(\ref{DEFENTFID}) can be used to detect the detrimental effects on the transmission of half of the entangled  state $|\psi\rangle_{QR}$ through the channel $\Phi$.
This quantity is related to the input-output fidelity~(\ref{INPUTOUPUTF})
via the inequality
\begin{eqnarray}\label{FeFin}
F_e(\rho_Q; \Phi) \leq F(\rho_Q;\Phi) \;,
\end{eqnarray}
implying that values of $F_e(\rho_Q; \Phi)$ close to one force $F(\rho_Q;\Phi)$ to approach unity too.
Slightly weaker versions of the opposite implication can also  be proven -- see e.g.
\cite{HOLEVObook12,KW04}.
In particular given $\epsilon >0$, if  $F(|\psi\rangle_Q;\Phi)\geq 1 -\epsilon$ for all input states $|\psi\rangle_Q$ 
belonging to the support of the density matrix $\rho_Q$, then~\cite{BKN00}
\begin{eqnarray}  \label{FeFin2}
F_e(\rho_Q; \Phi) \geq 1-3\epsilon/2 \;.
\end{eqnarray}
Furthermore,  taking $\rho_Q$ to be  the completely mixed state of $Q$, i.e., the density matrix
$\openone/d$ ($d$ being the dimension of ${\cal H}_Q$) whose purification is a maximally entangled state of $QR$,
from Eq.~(\ref{defentifd1}) it follows that
\begin{equation}\label{EntFidMaxEnt}
F_e(\openone/d,\Phi) = \frac{1}{d^2} \sum_{j} \left| \tr(K_j) \right|^2 \, ,
\end{equation}
which can be put in correspondence with the average fidelity~(\ref{DEFNUOVADIFAVE})
 through the identity~\cite{HHH99,nielsen02}
\begin{eqnarray} \label{correspondenceF}
\bar{F}(\Phi) = \frac{d F_e(\openone/d;\Phi) +1}{d+1}\;.
\end{eqnarray}

\subsubsection{Distance measures for channels} \label{sec:dist2}

Distance measures for
quantum channels (and in general for quantum operations) are typically written as $||| \Phi - \Psi |||$
where $||| \Lambda |||$ denotes a proper norm of the superoperator $\Lambda$.
Suitable choices are
 \begin{eqnarray} \label{fromtoNORM}
|||  \Lambda |||_k &:=&\sup_{ \|O \|_k \leq 1} \|  \Lambda(O) \|_k\;,
\end{eqnarray}
where the index  $k$ identifies a  norm for the operators of the  system.
Specifically for $k=1$, $\| O\|_1 = \tr\sqrt{O^\dag O}$ is the {\it trace norm};
for $k=2$, $\| O \|_2 = \sqrt{\tr{O^\dag O}}$ is the {\it Hilbert-Schmidt norm}; and finally
for $k=\infty$,  $\| O \|_\infty = \sup_{|\psi\rangle} \langle \psi | O|\psi\rangle$ is the standard {\it operator norm}
-- recall that they obey the following ordering $\| O\|_{\infty} \leq \| O\|_2 \leq \|O\|_1$~\cite{HJ90}.
While correctly defined, when  applied to CPTP maps, the norms~(\ref{fromtoNORM}) become unstable under channel extension. 
In particular $|||  \Lambda\otimes \id |||_k$ can explicitly depend
upon the dimensionality of ancillary system for which the $\id$ channel is defined.
In order to amend this, regularizations have been proposed. Of particular relevance
 are the norm of {\it complete boundedness}, or cb-norm \cite{Paulsen2003}, and the
 diamond-norm \cite{Kit97}.
  Given a generic (not necessarily CPTP) map $\Lambda:  \mathfrak{S} ({\mathbb C}^{n})
  \mapsto   \mathfrak{S} ({\mathbb C}^{k})$ 
  they are  defined respectively as
\begin{eqnarray} \label{cb-norm}
||| \Lambda |||_{cb} &:=& \sup_m{||| \Lambda \otimes \mathrm{id}_m |||_{\infty}} \; , \\
||| \Lambda |||_{\diamond} &:=& {||| \Lambda \otimes \mathrm{id}_n |||_{1}} \;,
\label{diamondNORM}
\end{eqnarray}
where $\mathrm{id}_m$ denotes the identity channel on $\mathfrak{S}(\mathbb{C}^m)$. 
While not obvious at least in the case of $||| \cdots |||_{\diamond}$,
both these norms 
 are stable under channel extension. Furthermore they 
 are related through  the identity 
 $|||  \Lambda |||_{cb} = |||  \Lambda^* |||_\diamond$, 
where  $\Lambda^*$ is the dual of $\Lambda$~\cite{JKP09}.

In the context of quantum communication, the properties of  the cb-norm have been extensively reviewed 
in Refs.~\cite{KretschmannTH,KW05,KW04,HW01,Keyl2002,Bela05,JKP09}. 
Here recall that it is well behaved under tensor product composition rule~(\ref{compositionNEW1}), 
since it has the property 
\begin{eqnarray}
||| \Lambda_1 \otimes \Lambda_2 |||_{cb}  = |||  \Lambda_1 |||_{cb} \;  |||  \Lambda_2 |||_{cb}\;.
\end{eqnarray}
Furthermore if  $\Lambda$ is completely positive  then 
 $|||  \Lambda |||_{cb} = ||\Lambda(\openone) ||_{\infty}$. 
Accordingly if $\Lambda$ is CPTP and
$\Lambda^*$ its dual channel~(\ref{equazioneHEIS}), one has  that
$|||  \Lambda |||_{cb}$ can take any value up to $d$ (the dimension of the channel input space)
while $|||  \Lambda^{*} |||_{cb}  =1$ always.

Finally, another useful distance measure for quantum channels is the one introduced by \cite{Grace10} 
as a distance between unitary operations acting on a bipartite quantum system, where only the effect of the operations on one component (the subsystem of interest) is relevant in the measure, while the effect on the other component (environment) can be arbitrary.

\subsection{Channels and entropies}\label{sec:entro}

\begin{figure}
\centering
\includegraphics[width=3.4 in]{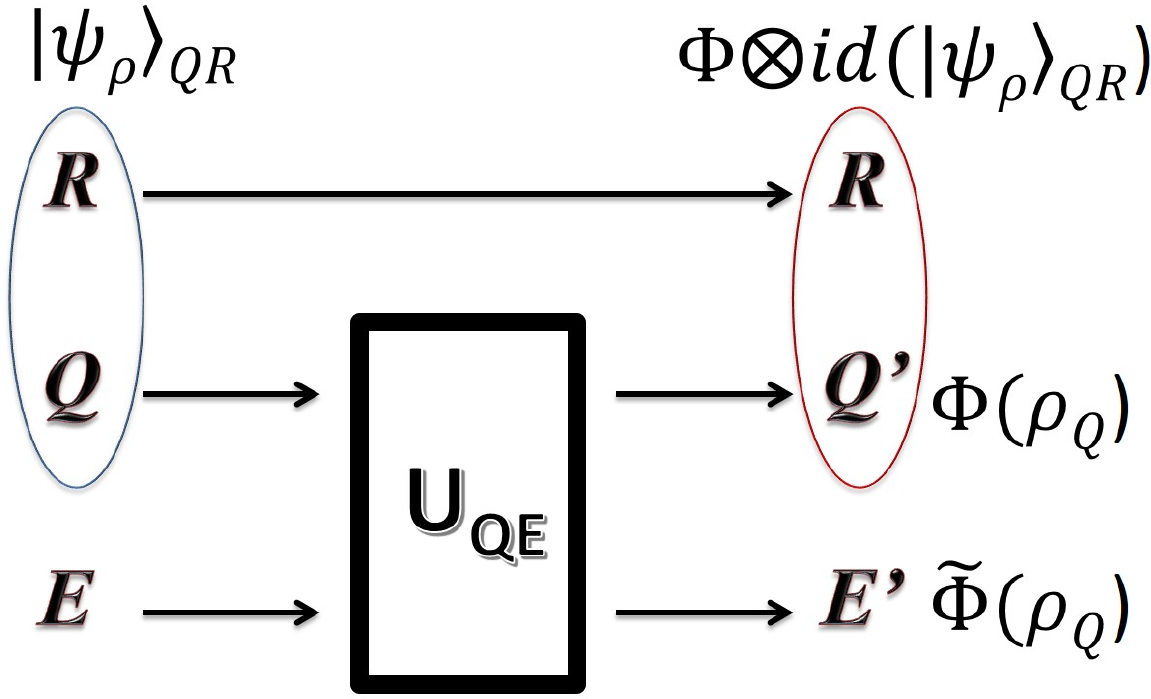}
\caption{Graphical representation of a quantum channel
$\Phi \in \mathfrak{P}(Q\mapsto Q')$
as unitary interaction $U_{QE}$ between the system state $\rho_Q$ and the environmental one $\rho_E$. The action of $\Phi\otimes {\rm id}$ on the purification of $|\psi_\rho\rangle_{QR}$  and the complementary map $\tilde{\Phi}$ are also shown.}
\label{figure04}
\end{figure}

In the study of quantum communication, entropic quantities play a fundamental role in
characterizing quantum channels in terms of their efficiency as communication lines
\cite{BNS98}.
A comprehensive characterization of these functionals can be obtained moving into the so called ``Church of the Larger Hilbert Space",
a construction based on the Stinespring dilation form (\ref{dia}) where also the input state $\rho_Q$ of the system $Q$ is represented
as a reduced density operator of a pure state $|\psi_\rho\rangle_{QR}$ of a larger system $QR$ via a
purification -- Fig.~\ref{figure04}.
Let us denote by
\begin{equation}\label{vNentropy}
S(\rho) := -\Tr\left( \rho \log_2\rho \right) \, ,
\end{equation}
the von Neumann entropy of the density operator~$\rho$~\cite{PETZBOOK,W78,Ohya93}
which generalizes to quantum mechanical systems
the Shannon entropy of a classical random variable $X$ taking values in the alphabet $\mathcal{X}$, defined as 
\begin{eqnarray} \label{SHEN}
H(X):=-\sum_{x \in \mathcal{X}} p(x) \log_2 p(x)\;,
\end{eqnarray}
 with  $p(x)$ being the probability that $X$
 acquires the value $x$~\cite{CTbook,Gbook}.
Then, given a quantum channel $\Phi \in \mathfrak{P}(Q\mapsto Q')$
and an input state $\rho_Q$, there are three important entropic
quantities related to the pair $(\rho_Q,\Phi)$. 
First is  the entropy of the input state $S[Q] := S(\rho_Q)$ ({\em input entropy}), 
which exploiting the fact that the purification $|\psi_\rho\rangle_{QR}$ is pure can also be
expressed as the entropy of the ancillary system $R$,  i.e., $S[Q]=S[R]$.
Second is the entropy of the output
state $\Phi(\rho_Q)$, i.e.,  $S[Q'] := S[\Phi(\rho)]$ ({\em output entropy}). 
Finally there is the
{\em entropy of exchange}~\cite{schumacher96,BNS98} computed as the von Neumann entropy
of the environment $E$ after the interaction with $Q$, i.e., the entropy measured at the output of the complementary
channel $\tilde{\Phi}$ defined in Eq.~(\ref{cc}),
\begin{eqnarray}\label{def-xchange}
S[E'] = S(\rho_Q; \Phi)&:=& S(\tilde{\Phi}(\rho_Q))\\\nonumber
 &=& S\left[ (\Phi\otimes {\rm id})(|\psi_\rho\rangle_{QR}\langle\psi_\rho|) \right]
\end{eqnarray}
(the last identity follows from the fact that the global state of $Q$, $R$, and $E$ is always pure).
A complete  analysis of the relations between these three quantities has been reviewed in~\cite{SN96,HW01}. 
In particular they satisfy the relations
\begin{eqnarray}
S[\Phi(\rho_Q)] + S(\rho_Q;\Phi) &\geq& S(\rho_Q)\;,  \label{superDIS} \\
|S[\Phi(\rho_Q)] - S(\rho_Q;\Phi)| &\leq& S(\rho_Q) \; , \label{triangDIS}
\end{eqnarray}
and the quantum Fano inequality
 \begin{eqnarray} \label{QFANO}
 S(\rho_Q;\Phi)  &\leq& h[F_e(\rho_Q; \Phi)] \\
 &&+ [1-F_e(\rho_Q; \Phi)]\; \log_2(d^2 -1)\;, \nonumber
 \end{eqnarray}
 with $d$ the dimension of the channel input,
 \begin{eqnarray}
\label{binarysh}
h(p) := -p \log_2 p - (1-p)\log_2 (1-p)
 \end{eqnarray}
  the Shannon binary entropy function~\cite{Gbook,CTbook}, and $F_e(\rho_Q; \Phi)$ the entanglement fidelity introduced in Eq.~(\ref{DEFENTFID}).

The input, output and exchange entropies  are the building blocks for constructing several information quantities.
For instance one defines the {\it quantum mutual information} ${I}(\rho_Q ;\Phi)$
between the system $Q$ at the output of the map $\Phi$ and the system $R$ which
enters in the purification $|\psi_\rho\rangle_{QR}$, i.e.,
\begin{eqnarray}\label{q-mutual}
{I}(\rho_Q ; \Phi) & := & S(\rho_Q)+S[\Phi(\rho_Q)]-S(\rho_Q;\Phi) \;.
\end{eqnarray}
This is a non-negative quantity which is known to be concave and sub-additive with respect to density operators on $Q$~\cite{B02,AC97}.

Subtracting $S(\rho_Q)$ from ${I}(\rho_Q ; \Phi)$ one also defines the channel {\it coherent information}~\cite{SN96,BNS98},
\begin{equation}\label{def-coherent}
J(\rho_Q; \Phi) := S[\Phi(\rho)]-S(\rho_Q ;\Phi)  =S[\Phi(\rho)]-S[\tilde{\Phi}(\rho_Q)]\; ,
\end{equation}
where in the last expression it is enlightened that $J(\rho_Q; \Phi)$ can also be expressed as  the difference between the output entropy  of $\Phi$ and of its complementary counterpart $\tilde{\Phi}$.
Differently from $I(\rho_Q; \Phi)$, the function $J(\rho_Q; \Phi)$ is in general neither non-negative, nor convex or sub-additive~\cite{DiSS98,SS07}.
However, both the quantum mutual and the coherent information satisfy data-processing inequalities. 
In particular given $\Phi_1$ and $\Phi_2$ CPTP channels, one has~\cite{HOLEVObook12}
\begin{equation}\label{DATAPROCI}
{I}(\rho_Q ; \Phi_2 \circ \Phi_1) \leq \min\{ {I}_Q(\rho_Q;\Phi_1), {I}(\Phi_1(\rho_Q); \Phi_2) \} \; ,
\end{equation}
while
\begin{equation} \label{DATAPROCJ}
J(\rho_Q ; \Phi_2 \circ \Phi_1) \leq J(\rho_Q;\Phi_1) \; .
\end{equation}

A further entropic quantity useful for characterizing  the channel $\Phi$ is the channel Holevo information.
In constrast to the previous expressions this is a functional of an input ensemble
${\cal E} := \{ p_j ; \rho_{Q}^{(j)}\}_j$ Alice feeds into the channel
(here  $\{ p_j\}_j$ is a probability distribution while $\{\rho_{Q}^{(j)}\}_j$
is a collection of input states).
Accordingly one has
\begin{eqnarray}
\chi({\cal E}; \Phi) &: =& S[\Phi(\rho_Q)] - \sum_j p_j S[\Phi(\rho_Q^{(j)})] \nonumber \\
&=& \sum_j \; p_j \; S(\Phi(\rho_Q)\| \Phi(\rho_Q^{(j)})\;, \label{HOLEVOINFO}
\end{eqnarray}
where $\rho_Q= \sum_j p_j \rho_Q^{(j)}$ is the average state associated with ${\cal E}$ and in the last identity
the quantum relative entropy $S(\rho_1 \| \rho_2) := \Tr \left[ \rho_1 (\log\rho_1-\log\rho_2) \right]$
~\cite{L75,SW00} has been used.
Via the Holevo Bound~\cite{Holevo73a,Holevo73b}  the quantity $\chi({\cal E}; \Phi)$
provides an upper bound on the  information one could retrieve on the
random variable $X$ associated with index~$j$
of the ensemble ${\cal E}$
if allowed to  measure the corresponding states at the output of the channel $\Phi$.
Specifically, indicating with $Y$ the random variables associated with
the estimation of $j$ after a  POVM has been performed on the density matrix
$\Phi(\rho_{Q}^{(j)})$, one has
\begin{eqnarray} \label{HOLEVOBOUND}
I(X:Y) \leqslant  \chi({\cal E}; \Phi)\;,
\end{eqnarray}
with ${I}(X:Y) := H(Y) + H(X)-  H(X,Y)$ being the (Shannon)
mutual information associated with the couple $X$ and $Y$~\cite{CTbook,Gbook}.
The functional $\chi({\cal E}; \Phi)$ obeys the data-processing inequality
\begin{equation}
\chi({\cal E} ; \Phi_2 \circ \Phi_1) \leq \chi(  {\cal E}; \Phi_1)\;,
\end{equation}
for all $\Phi_1$, $\Phi_2$ CPTP maps and for all ensemble ${\cal E}$.
Another quantity related to the Holevo information is the minimum output
entropy of the channel, $S_\mathrm{min}(\Phi) = \min_\rho S[\Phi(\rho)]$, that
quantify the minimum disturbance induced by the channel.
A connection with the coherent information can be established via the identity~\cite{D05,HOLEVObook12}
\begin{eqnarray}\label{identityimpo}
\chi( {\cal E}; \Phi) -\chi({\cal E};\tilde{\Phi})  = J(\rho_Q ; \Phi) - \sum_j p_j J(\rho_Q^{(j)} ; \Phi)\;,
\end{eqnarray}
with $\rho_Q$ being the average density matrix of the ensemble~${\cal E}=\{ p_j ; \rho_Q^{(j)}\}_j$.


\section{From Memoryless to Memory Quantum Channels}\label{sec:fmmqc}

Having in mind the multi-uses communication scenario detailed at the beginning of Sec.~\ref{sec:qcbo},
in which a time-ordered sequence of carriers  $Q:=\{ q_1, q_2, \cdots \}$ propagates from Alice
to Bob along a noisy channel, this section starts by discussing the simplest case where they are
affected by uncorrelated identical maps and then moves on to consider correlations among uses,
i.e., memory effects.

\subsection{Memoryless quantum channels}\label{sec:memorylessquantum}

{\em Memoryless quantum channels} describe those scenarios in which the
noise acts {\em identically} and {\em independently} on each element of the sequence $Q$.
Under this assumption the  multi-use map associated with the communication line
is expressed as a tensor product of a CPTP map $\Phi: \mathfrak{S} ({\cal H}_{q}) \mapsto \mathfrak{S} ({\cal H}_{q})$
that acts on the states of a single carrier~$q$. Therefore, indicating as
${\cal H}_Q^{(n)} :=  {\cal H}_{q_1} \otimes \cdots \otimes {\cal H}_{q_n}$
the Hilbert space of the first $n$ carriers of the system, its input density operators
$\rho_Q^{(n)}\in \mathfrak{S} ({\cal H}_{Q}^{(n)})$ will be mapped into
\begin{equation}\label{mp2}
\Phi^{(n)}(\rho_Q^{(n)}) =\Phi^{\otimes n} (\rho_Q^{(n)})\;,
\end{equation}
with $\Phi^{\otimes n} := \Phi \otimes \cdots \otimes \Phi$.
\begin{figure}
\centering
\includegraphics[width=3. in]{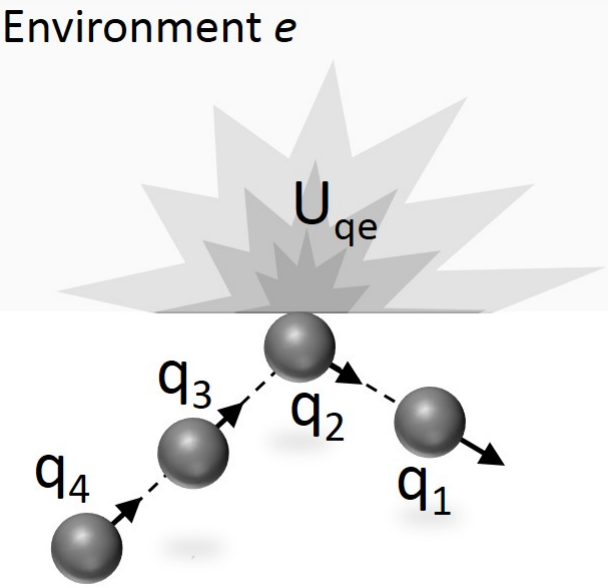}
\caption{Pictorial representation of the model of Ref.~\cite{G05}. The green elements represents
the sequence of carriers that propagates at a rate $\nu$ from Alice to Bob, interacting on the way, with the environment $e$ via the unitary coupling $U_{qe}$. Among two consecutive interaction
the environment tends to relax toward its stable configuration $\omega_e$ via a dissipative process characterized by the relaxing time $\tau$. The memoryless channel configuration is achieved when $\nu \ll 1/\tau$.}
\label{figure0011}
\end{figure}
Equivalently, one can say that the Kraus operators of the memoryless map $\Phi^{(n)}$ can be
expressed as a tensor product $K_{i_1} \otimes \cdots \otimes K_{i_n}$ formed by
independent and identically distributed sequences extracted from  the Kraus set $\{ K_i\}_i$
associated with the single carrier channel $\Phi$.
A simplified, yet informative, model can be found
in Ref.~\cite{G05}. Here the carriers $Q$ are assumed to propagate from Alice to Bob, one by one
and at constant speed, while interacting with an external environmental system via a constant
coupling described by the unitary operator $U_{qe}\in {\cal B}({\cal H}_q \otimes {\cal H}_e)$ whose role is to effectively simulate the interaction between the carriers and the medium which separate the two communicating parties.
In the model the environment $e$ is assumed to undergo a dissipative process which on a time-scale $\tau$
tends to reset it into a stable configuration $\omega_e$  (see Fig.~\ref{figure0011} for a pictorial representation of the scheme).
The memoryless regime is achieved in the limit in which the rate $\nu$ at which the carriers
propagate from Alice to Bob, is much lower than the inverse of the relaxation time $\tau$, i.e.,
$\nu \ll 1/\tau$.
In this limit in fact each carrier couples with identical and independent environmental states.
Defining then $\omega_E^{\otimes n} := \omega_{e_1} \otimes \cdots \otimes \omega_{e_n}$,
this allows one to write
\begin{align}
\rho_Q^{(n)} \, \mapsto \, \mathrm{Tr}_{E} \Big[ & U_{q_n,e_n} \otimes \cdots \otimes U_{q_1,e_1} \nonumber \\
& \big(\rho_Q^{(n)} \otimes \omega_{E}^{\otimes n}  \big) U_{q_1,e_1}^{\dag} \otimes \cdots \otimes U_{q_n,e_n}^{\dag}\Big] \, ,
\label{eqn:memoryless_model}
\end{align}
which reduces to Eq.~(\ref{mp2}) when identifying $\mbox{Tr}_{e} [ U_{qe} (\cdots \otimes \omega_e)U_{qe}^\dag]$
with unitary dilation of the single-use channel~$\Phi$.

\subsubsection{Compound and averaged quantum channels}\label{sec:compound}

Before entering into the subject of memory quantum channels, let us briefly discuss
the situation in which the channel map, though intended as acting like \eqref{mp2}, is not perfectly known to the sender and receiver.
Such a situation can be modeled by considering not a single CPTP map, but rather a set $\{\Phi_i\}_i$ of them.
Here $\Phi_i: \mathfrak{S} ({\cal H}_{q}) \mapsto \mathfrak{S} ({\cal H}_{q})$
and the set $\{\Phi_i\}_i$  can in principle contain a finite or infinite (countable or not) number of CPTP maps.
This leads to the notion of a memoryless \emph{compound quantum channel}, i.e., the family $\{\Phi_i^{\otimes n} : \mathfrak{S} ({\cal H}_{q}^{\otimes n})\to \mathfrak{S} ({\cal H}_{q}^{\otimes n})\}_{n,i}$.
Averaged channels are closely related to compound channels.
The difference is that in the former the sender and receiver know an a priori probability distribution
$\{p_i\}_i$ governing the appearance of the members of compound channel. It means that for any $n\in\mathbb{N}$
one can write the averaged channel map as
\begin{eqnarray}\label{longmemory}
\Phi^{(n)}(\rho_Q^{(n)}) =\sum_i p_i \Phi_{i}^{\otimes n} (\rho_Q^{(n)}).
\end{eqnarray}
Equation~(\ref{longmemory}) describes a scenario in which, with some probability $p_i$ all 
the carriers of the system are operated on by the same identical local transformation $\Phi_i$.
The index $i$ can be interpreted as a ``switch" selecting different memoryless channels, and
(\ref{longmemory}) as the average channel over different
values of the switch.
Classical counterparts of compound and averaged channels were studied since long time ago \cite{BBT59,Wo60,jacobs,ahlswede}.
Compound and averaged quantum channels were introduced only recently \cite{BB09,hayashi08}.
In Sec.~\ref{sec:long} one shall see that these channels are closely related
to a special set of memory channels having long-term memory.

\subsection{Non-anticipatory memory quantum channels}\label{sec:nonant}
Whenever the tensorial decomposition of Eq.~(\ref{mp2}) doesn't apply, one can
speak of {\em memory channels} or {\em correlated noise channels}.
Among the plethora of possibilities, the following will focus only on those
configurations that have physical relevance and have attracted some interest
in the recent literature.
In particular, one shall treat those models in which the noise respects the time-ordering
of the carriers $Q$ so that at a given channel use, the output cannot be influenced by
successive inputs as pictorially shown in the right panel of  Fig.~\ref{fig000}. This property generalizes the notion of {\it semicausality}
discussed in Sec.~\ref{sec:semicausal} to the case of multiple (ordered)  subsystems.
Inspired by the classical theory of communication~\cite{Gbook} one can name
the quantum  communication lines which fulfills such condition, {\em non-anticipatory} quantum channels
(notice however that in the approach of~\onlinecite{KW05} these maps
are called just {\it causal} -- more on this in Sec.~\ref{sec:causal}).

Under the non-anticipatory condition there must exist a family of CPTP maps
${\cal F} := \{ \Phi^{(n)}; n=1,2, \dots \}$ with
$\Phi^{(n)} : \mathfrak{S} ({\cal H}_{Q}^{(n)}) \rightarrow \mathfrak{S} ({\cal H}_{Q}^{(n)})$
which allows one to express the output states of the first $n$ carriers in terms of
the density matrices of their associated inputs, i.e.,
\begin{equation} \label{nonanti}
\rho_Q^{(n)} \mapsto \Phi^{(n)}(\rho_Q^{(n)})\;.
\end{equation}
Clearly the property~(\ref{nonanti}) requires that the family ${\cal F}$ must
fulfill the minimal consistency requirement that for all $m<n$ the element
$\Phi^{(m)}$ should be obtained as a restriction of $\Phi^{(n)}$ over the degrees
of freedom of the first $m$ carriers.
That is, given $\rho_Q^{(n)}\in \mathfrak{S} ({\cal H}_{Q}^{(n)})$ and
$\rho_Q^{(m)} \in  \mathfrak{S}({\cal H}_{Q}^{(m)})$ one must have
\begin{equation} \label{non-ant}
\Phi^{(m)}(\rho_Q^{(m)}) =  \Tr^{(m)} \left[  \Phi^{(n)}(\rho_Q^{(n)}) \right] \;,
\end{equation}
whenever $\rho_Q^{(m)} = \Tr^{(m)}[\rho_Q^{(n)}]$, where $\Tr^{(m)}$ stands for the
partial trace over all the carriers but the first $m$.

\bigskip

\begin{figure}
\centering
\includegraphics[width=3.5 in]{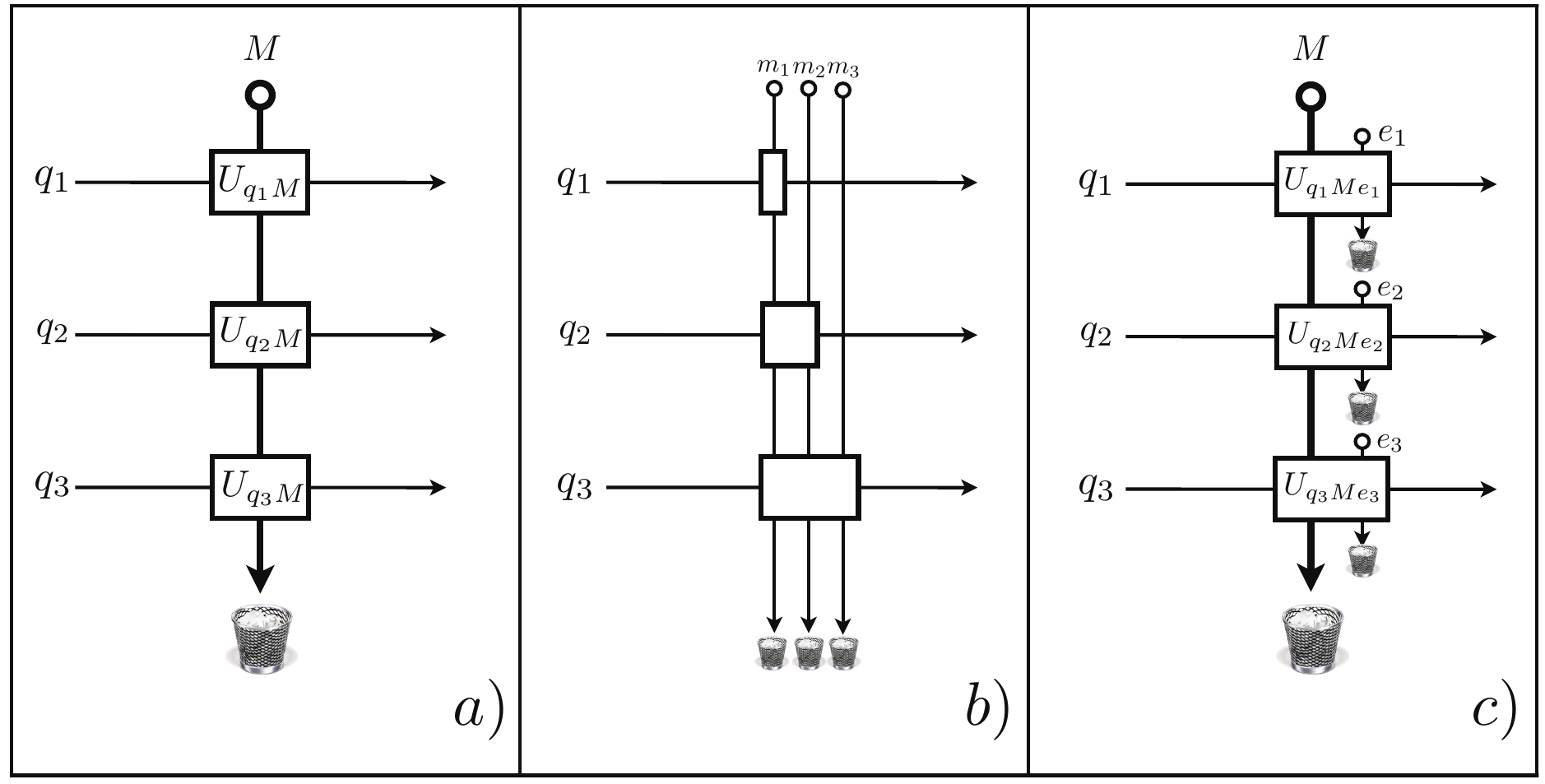}
\caption{Unitary dilations for a non-anticipatory quantum memory channels.
a) graphical sketch of the representations of Eq.~(\ref{eqn:memory_model_10}): here
the noise correlations among the $n$ channel uses can be described via a series of
concatenated unitary interactions with a common reservoir $M$ whose dimension in
general depends (exponentially) upon $n$ ($n=3$ in the example). Notice that while
the carrier $q_1$ might influence the outcome of $q_2q_3$ via their common interaction
with $M$, $q_2 q_3$ cannot influence the output of the first carrier;
b) the environment $M$ can be also represented as a collection of smaller systems
$M_1$, $M_2$, $\dots$  initially prepared into a separable state while, as shown in
Eq.~(\ref{unitaryqjm}), the unitary transformation operating  on the $j$-th channel
use couples it with the first $j$ subsystems  only;
c) unitary dilation~(\ref{eqn:memory_model_1}) where besides $M$ a series of local
environment $e_1$, $e_2$, $\dots$, are also present.
In all the diagrams the unitary operators (represented by the white boxes) are
applied sequentially on the input states of the global system (i.e., the carriers and
the environment) starting for the one on the top of the figure.
The carriers and the environmental states evolve, respectively, from-left-to-right and
from-top-to-bottom while interacting meeting at a white box.
The trash-bin symbol stands for the partial trace operation on the environment.}
\label{figure01}
\end{figure}

As already noticed, in the language introduced in Sec.~\ref{sec:semicausal} non-anticipatory channels
can be classified as {\em semicausal} with respect to the natural ordering of the
channel uses.
The representation of semicausal channels given in Eq.~(\ref{semi-localizable})
can hence be applied, yielding a representation of non-anticipatory quantum channels
in which each carrier couples sequentially with a common \textit{memory system} $M$.
The back-action of $M$ on the message state simulates the memory effects of the transmission.
Accordingly, all the non-anticipatory CPTP maps can be expressed as
\begin{align}
\Phi^{(n)}(\rho_Q^{(n)}) = \Tr_{M} \Big[ & U_{q_nM} \cdots U_{q_1M} \nonumber \\
&\big(\rho_Q^{(n)} \otimes \omega_M \big) U_{q_1M}^{\dag} \cdots U_{q_nM}^{\dag}\Big] \;,
\label{eqn:memory_model_10}
\end{align}
where for all $j=1,2, \dots, n$, $U_{q_jM}$ is a unitary transformation which
describes the coupling of the $j$-th carrier with the memory system $M$, and
where $\omega_M$ is some given state of $M$, see Fig.~\ref{figure01}a.
The unitary transformations $U_{q_jM}$ may in general depend on the carrier label $j$.
Otherwise, if they are independent of $j$ the memory channel has the additional property
of being invariant under translation of the carrier labels.
An explicit proof of Eq.~(\ref{eqn:memory_model_10}) was first given in~\onlinecite{KW05}
in the context of quasilocal algebras (see also Appendix \ref{app:algebras}),
under the assumption of translational invariance
of the noise (more on this will be provided in  Sec.~\ref{sec:causal}).
An alternative proof which doesn't
make use of this hypothesis can be found in Appendix~\ref{sec:appendix}.

In Eq.~(\ref{eqn:memory_model_10}) $M$ is in general a large system whose dimension
$d_M$ is an explicit function of $n$ (in any case it can always be chosen to be less
than or equal to $d^{2n}$ with $d$ being the dimension of a single carrier).
As a matter of fact, as explained in Appendix~\ref{sec:appendix}, one can take $M$ to be a composite system of
components $m_1$, $m_2$, $\dots$, $m_n$ whose dimensions can always be chosen to be not
larger than $d^2$. In this configuration then one can assume $\omega_M$ to be a pure tensor
product state of local terms $|0\rangle_{m_1} \otimes \cdots \otimes |0\rangle_{m_n}$,
and write $U_{q_j M}$ as a transformation which couples the $j$-th carrier \emph{only}
with the first $j$ elements of $M$, i.e.,
\begin{equation}\label{unitaryqjm}
U_{q_j M} = \openone_{m_n} \otimes  \cdots \otimes \openone_{m_{j+1}} \otimes U_{q_j m_j m_{j-1}\cdots m_1}\;,
\end{equation}
with $\openone_{m'}$ being the identity operator on the $m'$ components of the environment, see Fig.~\ref{figure01}b.

An alternative, but fully equivalent, representation for non-anticipatory channels is obtained
by adding to Eq.~(\ref{eqn:memory_model_10}) a collection of local environments which individually
couples with the carriers, i.e.,
\begin{align}
\Phi^{(n)}(\rho_Q^{(n)}) = &\Tr_{ME} \Big[  U_{q_nMe_n} \cdots U_{q_1Me_1} \label{eqn:memory_model_1} \\
& \big(\rho_Q^{(n)} \otimes \omega_M \otimes \omega_E^{\otimes n} \big) U_{q_1Me_1}^{\dag} \cdots U_{q_nMe_n}^{\dag}\Big] \;, \nonumber
\end{align}
where for all $j=1,2, \dots, n$, $U_{q_jMe_j}$ is now the  unitary transformation which
describes the coupling of the $j$-th  carrier with its own local environment $e_j$ and
with the memory system $M$, where $\omega_E^{\otimes n} : = \omega_{e_1} \otimes \cdots \otimes \omega_{e_n}$
as in the memoryless case, and $\omega_M$ is some given state of $M$, see Fig.~\ref{figure01}c.
In principle one can distinguish different setups in which Alice, Bob or Eve (third party) has the control of the
initial/final states of the memory system $M$ \cite{KW05}.
Equation~(\ref{eqn:memory_model_1}) was first introduced by~\onlinecite{BM04} as a model
for representing correlated channels: from Eq.~(\ref{eqn:memory_model_10}) it follows that it
provides a general unitary dilation for {\it every} non-anticipatory quantum maps.
It can also be expressed in terms of an $n$-fold concatenation of a sequence of CPTP maps
acting on a single carrier {\em and} the memory system $M$~\cite{KW05,BM04}.
Such concatenation is shown pictorially in Fig.~\ref{figure01}c and results in the following identity
\begin{equation}\label{nfoldconc}
\Phi^{(n)}(\rho_Q^{(n)}) = \Tr_{M} \Big[ \Phi^{(n)}_{QM} \big(\rho_Q^{(n)} \otimes \omega_M \big) \Big] \;,
\end{equation}
with
\begin{equation}\label{nfoldconc1}
\Phi^{(n)}_{QM} := \Phi_{q_nM} \circ \Phi^{(n-1)}_{QM} = \Phi_{q_nM} \circ ... \circ  \Phi_{q_1M} \;,
\end{equation}
where for $j=1,2, \dots ,n$, $\Phi_{q_jM}: \mathfrak{S} ({\cal H}_{q_j}\otimes {\cal H}_M)
\rightarrow \mathfrak{S} ({\cal H}_{q_j}\otimes {\cal H}_M)$ is a CPTP map
that operates on the $j$-th carrier and on the memory ancilla $M$ and is defined
by the unitary dilation
\begin{equation}\label{eqn:memory_model_2}
\Phi_{q_jM}(\cdots) = \Tr_{e_j} \Big[ U_{q_jMe_j} \big(\cdots  \otimes \omega_{e_j} \big) U_{q_jMe_j}^{\dag}\Big] \;.
\end{equation}
In this representation the evolution of $M$ after the interaction with the carriers is  provided by the transformation
\begin{equation}\label{MEMEV}
\omega_M \mapsto  \Psi^{(n)}(\rho_Q^{(n)}; \omega_M)  := \mathrm{Tr}_{Q} \Big[ \Phi^{(n)}_{QM} \big(\rho_Q^{(n)} \otimes \omega_M \big) \Big]\;,
\end{equation}
which explicitly depends upon the input state of $Q$.

Cases of special interest~\cite{KW05} are those in which, for all $j$, the $\Phi_{q_jM}$
describes the same mapping $\Phi=\Phi_{qM}$ on $\mathfrak{S} ({\cal H}_{q}\otimes {\cal H}_M)$
which, according to Eq.~(\ref{nfoldconc1}) becomes the {\em generator} of the $n$-fold concatenation.
That characterizes memory channels which are non-anticipatory and translation invariant
(i.e., invariant under translation of the information carriers, $q_j \to q_{j+1}$).
Memoryless channels can then be included in this class as a limiting case in which the generator $\Phi$
can be expressed as a tensor product channel that acts independently on the carrier $q$ and on the
memory system $M$.
In terms of the unitary dilation~(\ref{eqn:memory_model_1}) this is equivalent to assuming
that the unitaries $U_{q_jMe_j}$ in Eq.~(\ref{eqn:memoryless_model}) factorize in a tensor
product $U_{q_je_j} \otimes V_M$, where $V_M$ is a unitary operator on the memory system and
$U_{q_je_j}$ acts only on the degree of freedom of the $j$-th carrier and on its local environment $e_j$.

\bigskip

A special subset of non-anticipatory channels is formed by 
\emph{symbol independent} (SI) maps \cite{BDM05}.
They are communication lines where previous input states \emph{do not} affect the action of the channel on
the current input state.
In other words the symbol independent maps are non-anticipatory (or semicausal) with respect to {\em all} possible
ordering of the carriers (in this sense they are hence {\it fully non-anticipatory}).
Accordingly, given a generic subset of the carrier set $Q$, its output state
is uniquely determined by the corresponding input state via a proper CPTP mapping.
Following the terminology introduced in Sec.\ \ref{sec:semicausal}~\cite{PHHH06,BGNP01,ESW01},
they can be said to be {\em non-signaling} (or {\em causal}) channels, meaning that the output states of
any subset of the carriers cannot be influenced by the input state of the remaining carriers.

Channels which are not SI are said to exhibit \emph{intersymbol interference} (ISI)~\cite{BDM05}, 
that is, the input states of previous carriers affect the action of the channel on the current input.
From a physical point of view, in ISI channels there is a non negligible 
back action of the carrier onto the memory during their interaction. 
So the carrier's state (symbol) influences the subsequent actions of the channel. 
On the contrary, in SI channels the carrier does not influence the memory during 
their interaction. Usually this happens because the memory is much larger (in 
terms of degrees of freedom) of the single carrier. A pedagogical example of ISI 
channels is the {\it quantum shift channel}, where each input state is replaced by 
the previous input state, i.e., given the $j$-th carrier $q_j$ whose state is 
$\rho_j$, then $\Phi_{q_j}(\rho_j)=\rho_{j-1}$.

\subsection{Quasi-local algebras approach}\label{sec:causal}

Till now we have followed a constructive approach
in which memory quantum channels  were always thought
of as concatenations of smaller units which, starting from an official ``first carrier" element,
process one quantum signal each.
An alternative view where the communication lines are treated as mappings applied
on infinitely long message strings, is proposed in \onlinecite{KW05,BB08}.
This approach requires some advanced mathematical tools
that are briefly reviewed in Appendix~\ref{app:algebras}.

To set the stage, suppose we have a quantum channel which  transforms input
states of an infinitely extended quantum lattice system (representing the infinite
message string) into output states on the same system.
In~(\onlinecite{KW05}) this map
is formally assigned by working in the Heisenberg picture (see Sec.~\ref{sec:HEIS})  via the introduction of a
completely positive and unital map $\Phi^*
\mathpunct : \B^{\Z} \rightarrow \A^{\Z}$ operating on the quasi-local
algebras $\B^{\Z}$ and $\A^{\Z}$~\cite{BR79} that define the observable quantities on the lattice as described by the receiver Bob and the sender Alice, respectively.
In this context one says that the channel is {\it translational invariant} or (borrowing from~\onlinecite{BB08}) \emph{stationary}
if $\Phi^*$ commutes with the shift operator  on the lattice, i.e.,
 \begin{equation}
            \label{eqn:translinv}
                \Phi^* \circ T_{\cal B}= T_{\cal A} \circ \Phi^*\;,
        \end{equation}
        ($T_{\cal B}$ and $T_{\cal A}$ being the representation of the shift operator on ${\B}^{\Z}$ and ${\A}^{\Z}$ respectively).
Furthermore, $\Phi^*$ is said to be \emph{ergodic} if it is extremal 
in the convex set of stationary channels\footnote{It should 
be noticed that this notion of ergodicity refers to in-parallel 
composition of quantum channels and differs from ergodicity of 
in-series concatenation discussed in \cite{RichWer1996,Ragin2002,BG2007,BCGPY2012} and references therein.}.

Requiring then that future inputs should not affect past measurements, i.e., the non-anticipatory property~(\ref{non-ant}),
Ref.~(\onlinecite{KW05}) introduces the definition of a {\em causal channel}
as a completely positive and unital translational invariant map $\Phi^*$
that fulfills the constraint
          \begin{equation}
            \label{eqn:causal}
                \Phi^*\left(O^{(-\infty,z]} \otimes \1^{[1+z,\infty)}\right) = \Phi^*\left(O^{(-\infty,z]}\right)
                \otimes \1^{[1+z,\infty)}\;,
        \end{equation}
for all $z\in \Z$ and for all $O^{(-\infty,z]} \in \B^{(-\infty,z]}$, where $\B^{(-\infty,z]}$
denotes the set of bounded operators defined on lattice elements up to that associated with the label $z$.
In particular, memoryless configurations are obtained when also the condition
         \begin{equation}
            \label{eqn:memless}
                \Phi^*\left(\1^{(-\infty,z]} \otimes O^{[1+z,\infty)}\right) =\1^{(-\infty,z]}
                \otimes  \Phi^*\left(O^{[1+z,\infty)}\right)\;,
        \end{equation}
applies for all $O^{(-\infty,z]} \in \B^{[1+z,\infty)}$.

Example of causal (non necessarily memoryless)  maps~(\ref{eqn:causal})
 are provided by \emph{concatenated memory channels}~(\onlinecite{KW05}) which
can be easily constructed by adapting the concatenation scheme
of Eqs.~(\ref{nfoldconc})-(\ref{nfoldconc1}) to the quantum lattice formalism.
Within this context Ref.~(\onlinecite{KW05})  proves a  structure theorem
which shows that \emph{any}  map obeying Eq.~(\ref{eqn:causal}) can always be represented as concatenated memory channels produced by an assigned generator (see previous section).

Although cq-channels can be easily included in the above formalism by expressing them as CPTP maps via the embedding~(\ref{CQ}),
it is worth reviewing the approach adopted in~\onlinecite{BB08} to address this special set of maps.  Here
a cq-channel taking values on the classical alphabet $\mathcal{X}$  is  described as a  mapping  which
to each $x\in \mathcal{X}^{\Z}$ (the set of doubly infinite sequences with components from alphabet $\mathcal{X}$)
associates  a complex value linear functional
$W(x,\cdots)$  on $\mathcal{B}^{\Z}$, i.e.,
\begin{eqnarray}
x \mapsto W(x,\cdots) \;. \label{BB08MAP}
\end{eqnarray}
Ultimately, via the Gelfand-Naimark-Segal correspondence~\cite{BR79}, the functional $W(x,\cdots)$ can be  identified with a density operator $\rho_x$ defined on the Hilbert space $\cal H$ carrying a representation $\pi$ of the quasi-local algebra
(see Appendix~\ref{app:algebras}), through the identification $W(x,\cdots) =  {\rm Tr}[\rho_x \pi(\cdots)]$.
In this form the stationary condition~(\ref{eqn:translinv}) of the cq-channel is that
$W(T_{\textrm{in}}x,b)=W(x,T_{\cal B}b)$
for all $x\in \mathcal{X}^{\Z}$ and all $b\in \mathcal{B}^{\Z}$ (here $T_{\textrm{in}}$ and
denote the shift operator on $\mathcal{X}^{\Z}$).
The causality condition~(\ref{eqn:causal}) is instead
\begin{equation} \label{NEWCONDW}
W(x,b)=W(\tilde x ,b) \,,
\end{equation}
for $z\in\Z $, $b\in \mathcal{B}^{(-\infty,z]}$ and all $x,\tilde x\in \mathcal{X}^{\Z} $ $(x_{i}=\tilde x_{i};\forall i\le z)$. 
Similarly, memoryless configurations~(\ref{eqn:memless}) are recovered when Eq.~(\ref{NEWCONDW})
applies also for all
$b\in \mathcal{B}^{[z,\infty)}$
and all $x,\tilde x\in \mathcal{X}^{\Z} $ $(x_{i}=\tilde x_{i};\forall i\ge z)$.

\subsection{Taxonomy of Non-Anticipatory Quantum Memory Channels}\label{sec:tqmc}

Here we review those classes of non-anticipatory quantum channels which have been discussed in the literature.

\subsubsection{Localizable memory quantum channels}

\begin{figure}
\centering
\includegraphics[width=3.0in]{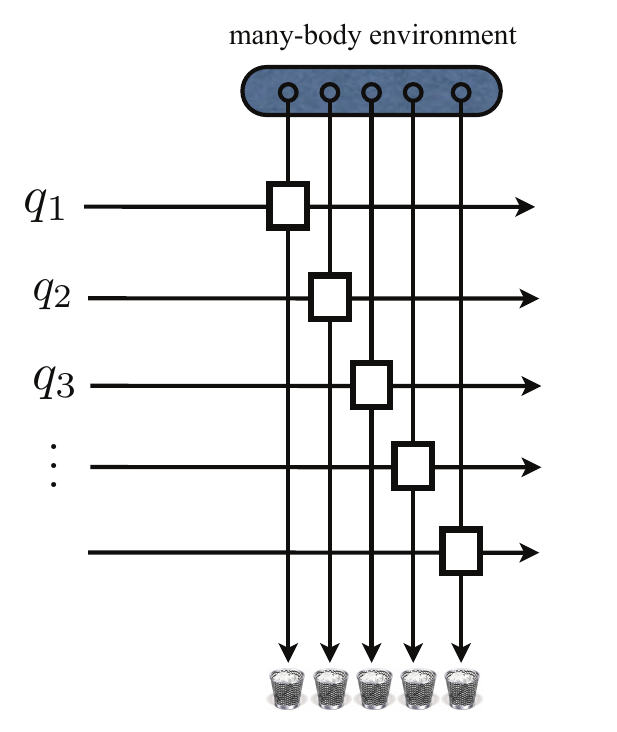}
\caption{Model for a localizable, fully non-anticipatory quantum memory channel.
Here the correlations are introduced by allowing the state of the environment (gray element)
to be initially entangled. As in the previous figures white boxes represents unitary
couplings while the trash-bin indicates partial trace over the corresponding degree of freedom.
These maps are SI and hence non-anticipatory (therefore they also admit unitary dilations
of the form described in Fig.~\ref{figure01}).} \label{fig:memchan2}
\end{figure}

A subset of non-anticipatory quantum channels which
represent the natural multi-partite generalization of the {\em localizable}
maps of Refs.~\cite{PHHH06,BGNP01,ESW01}, reviewed in Sec.\ \ref{sec:semicausal},
has been introduced in Refs.~\cite{GM05,PV07,PV08}.
For such models, the mapping~(\ref{nonanti}) is expressed in terms of (not necessarily identical)
{\em local} unitary couplings with a correlated many-body environmental system
$E :=\{e_1,e_2,\cdots \}$ -- see  Fig.\ \ref{fig:memchan2}. These transformations are clearly SI:
memory effects appear because, differently from the memoryless case~(\ref{eqn:memoryless_model}),
the many-body environment is initialized in a state  $\omega_E^{(n)}$ which does not factorize, i.e.,
\begin{align}
\Phi^{(n)}(\rho_Q^{(n)}) = \Tr_{E} & \left[  U_{q_ne_n} \otimes \cdots \otimes U_{q_1e_1} \left( \rho_Q^{(n)} \otimes \omega_E^{(n)} \right) \right. \nonumber\\
& \left. U_{q_1e_1}^\dag \otimes \cdots \otimes U_{q_ne_n}^\dag \right] \, .
\label{correlated}
\end{align}
It is worth mentioning that a variant of this model~\cite{ROSSINI08} where the local unitary
interaction $U_{q_ne_n} \otimes \cdots \otimes U_{q_1e_1}$ is
replaced by a local Hamiltonian coupling between carriers and environments,
is neither SI nor non-anticipatory.

An alternative representation for the localizable  mappings described by Eq.~(\ref{correlated})  has been also provided in Ref.~\cite{CGP10} by generalizing a
model presented in Ref.~\cite{BB04,BST02} for memoryless channels.
In this approach the channel noise is effectively described as a quantum teleportation
protocol~\cite{BetAL93,BK98,V94} that went wrong because the communicating parties used
non optimal resources (e.g., the state they shared was not maximally entangled).
In the case of (\ref{correlated}) each of the carriers gets teleported independently using the same
procedure, the correlations arising from the fact that the communicating parties use as shared
resource a correlated many-body quantum state.

\subsubsection{Finite-memory channels} \label{sec:finite}

\begin{figure}
\centering
\includegraphics[width=2.5 in]{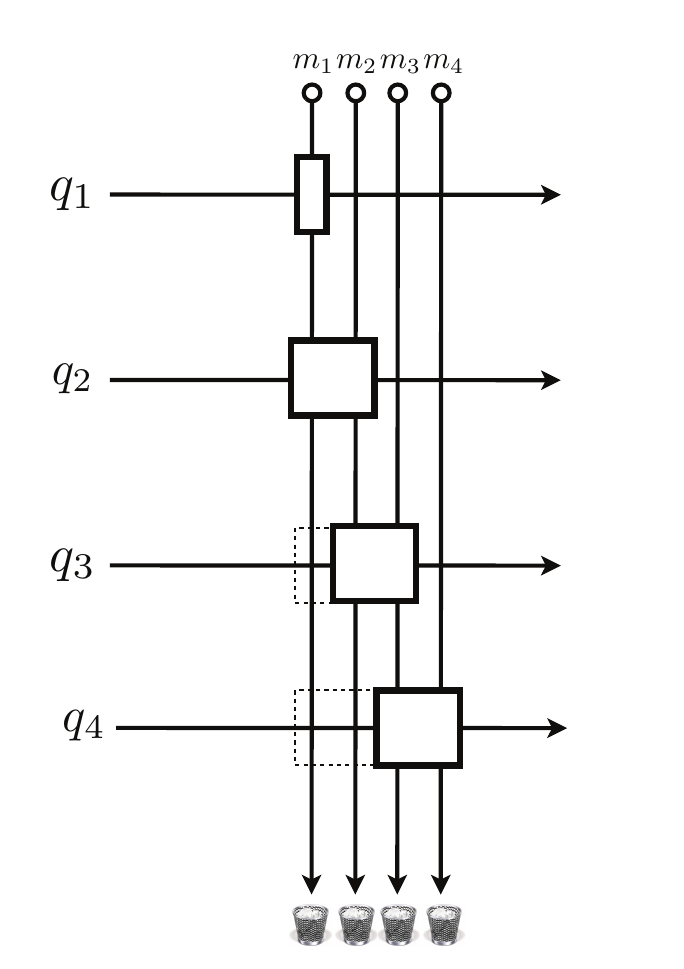}
\caption{Unitary dilation~(\ref{eqn:memory_model_10}) for a finite-memory
non-anticipatory quantum memory channel with correlation length $k=2$.
In the depicted example the total number of channel uses is $n=4$ and each carrier
is supposed to interact with only two  components of the environment.
Consequently the carrier $q_1$ can influence the output carrier $q_3$ only
via $q_2$. For a comparison see the scheme of Fig.~\ref{figure01}b where
instead the first carrier can directly influence $q_3$ via their common
interaction with $m_1$ (symbolized here by dotted lines).}
\label{figure02}
\end{figure}

The expression \textit{finite-memory channels}~\cite{BM04} is
used to indicate those non-anticipatory channels
that admit a representation of  the form~(\ref{eqn:memory_model_1}) with $M$ being finite dimensional.
The dimension of the memory is determined by the number of Kraus operators in the single channel expansion.
Within the representation~(\ref{eqn:memory_model_10}) examples of finite-memory channels
are obtained by assuming that the unitary transformations (\ref{unitaryqjm}) couple
the carriers with no more than a fixed number $k$ of environmental subsystems, the parameter
$k$ playing the role of the correlation length of the channel. More precisely for all $j \geq k$ one has,
\begin{eqnarray}
U_{q_j M} &=& \openone_{m_n} \otimes  \cdots \otimes \openone_{m_{j+1}} \label{unitaryqjm1} \\ \nonumber
&&  \otimes \; U_{q_j m_j m_{j-1}\cdots m_{j-k}}\otimes
 \openone_{m_{j-k-1}} \otimes  \cdots \otimes \openone_{m_{1}}
\end{eqnarray}
(see Fig.~\ref{figure02} for a graphical representation of the case with $k=2$).
Notice that the case of a memoryless channel can be considered as an {\it extreme} example of
finite-memory channels, where $k=1$ and each carrier interacts with a devoted component of the multipartite
environment $M$ (specifically, for each $j$, the carrier  $q_j$ interacts with $m_j$ only).

\subsubsection{Perfect memory channels}\label{sect:perfect}

Memoryless channels have unitary
dilations in which the environment has a dimension which is
at least exponentially growing in $n$ (i.e., $\log[ \mbox{dim}{\cal H}_E^{(n)}] = n \log{d_e}$)
or, equivalently, by possessing a (minimal)
operator sum representations whose Kraus sets contain a number of
elements which is exponentially growing in $n$.
The same property typically holds also for memory channels with the important
exception of the {\em perfect memory} channels~\cite{GBM09,KW05}.
Perfect memory channels are those admitting a representation as in Eq.\ (\ref{eqn:memory_model_1})
where the carriers only interact with the memory system, that is, $U_{q_j M e_j} = U_{q_j M} U_{e_j}$.
The simplest example of such communication lines is obtained
by assuming that the memory system $M$ in Eq.~(\ref{eqn:memory_model_10})
does not scale with $n$ and it  is finite dimensional.
Under this hypothesis the maps $\Phi^{(n)}$ explicitly admit a unitary dilation
with an environment (the system $M$) of constant size.
A comparison with the dimension of the Hilbert space
${\cal H}_Q^{(n)}$ of the information carriers, which grows exponentially
with $n$, shows that information cannot be stuck in the channel environment for
a long time.
As a consequence, in the asymptotic limit of long carrier sequences, no
information is expected to be lost to the environment, yielding optimal communication capacity (see Sec.\ \ref{sec:ctc}).
A typical example is provided by the shift channel (see Sec.~\ref{sec:nonant}) which can be described as
in Eq.~(\ref{eqn:memory_model_10}) by assuming $M$ to have the same dimension of a single carrier and by taking
$U_{q_j M}$ as swapping the states of $q_j$ and $M$. It is worth noticing that in Ref.~\cite{BDM05} it was also conjectured that
the memory channels that, analogously to the shift channel, display only intersymbol interference, can be represented as perfect memory channels.

More generally the class of {perfect} memory channels can be extended to include all the
CPTP maps~(\ref{nonanti}) that admit unitary
dilations~(\ref{eqn:memory_model_10})
in which the dimension $d_M$ of the environmental system $M$   is sub-exponential in $n$, i.e.,
\begin{equation}\label{pro1}
\lim_{n\rightarrow \infty} \frac{1}{n}\log [d_M] = 0 \;.
\end{equation}
As discussed explicitly in Sec.~\ref{sec:pmc}, also in this case the channel is asymptotically noiseless~\cite{GBM09,KW05}.

\subsubsection{Markovian channels}\label{sec:markov}
An important class of non-anticipatory quantum channels
is given by the channels with Markovian correlated noise.
They describe noise models in which the carriers are transformed via the applications of strings of local
CPTP maps
whose elements are randomly generated by a classical Markov process.
Explicitly, Markovian channels admit the following representation:
\begin{eqnarray}
&&\Phi^{(n)}(\rho^{(n)}_Q)  =  \sum_{i_1,\cdots,i_n}
p^{(n)}_{i_n|i_{n-1}}p^{(n-1)}_{i_{n-1}|i_{n-2}}\;\cdots\; p^{(2)}_{i_2|i_1} p^{(1)}_{i_1}  \nonumber \\
&&\qquad \quad  \times  \Phi_{q_n}^{(i_n)} \otimes \Phi_{q_{n-1}}^{(i_{n-1})} \otimes \cdots \otimes \Phi_{q_1}^{(i_1)} ( \rho_Q^{(n)} )\;,
\label{eqn:markov_channel_alternative}
\end{eqnarray}
where $\{ \Phi_{q_j}^{(i)}\}_i$ is a set of CPTP maps operating on the $j$-th carrier,
$p^{(1)}_{i}$ is an initial probability distribution, and, for $j \geq 2$
the $p^{(j)}_{i|i'}$ are conditional probabilities.

The mapping (\ref{eqn:markov_channel_alternative})
is SI (see Sec.\ \ref{sec:nonant}) since modifying the input state of previous (or subsequent)
channel uses does not have any effect on the output states of the carriers that
follow (or preceed).
A unitary dilation of the form (\ref{eqn:memory_model_1})
can be obtained by identifying the initial state $\omega_M$ of the memory $M$ with the vector
$\sum_i \sqrt{p^{(1)}_i}|i\rangle_M$, and by
taking the unitary  $U_{q_j Me_j}$ in such a way that for all vectors $|\psi\rangle_{q_j}$  of the $j$-th carriers one has
\begin{eqnarray}\label{eqn:markovmem1}
U_{q_1Me_1} | \psi,  i' , 0\rangle &=&  K_{q_1}^{(i')}(\ell) | \psi , i' , \phi^{i'}_{\ell}\rangle \;,
\end{eqnarray}
and
\begin{eqnarray}
U_{q_jMe_j} | \psi,  i' , 0\rangle &=& \sum_{i} \sqrt{p^{(j)}_{i|i'}} K_{q_j}^{(i)}(\ell) | \psi , i , \phi^{i'}_{\ell} \rangle
\;,  \label{eqn:markovmem}
\end{eqnarray}
for $j \geq 2$
(in the above expressions  $\{ K_{q_j}^{(i)}(\ell)\}_\ell$ is a set of Kraus operators
for $\Phi_{q_j}^{(i)}$, $|\psi,  i , \phi_{i',\ell}\rangle$ stands for the state
$|\psi\rangle_{q_j}  |i\rangle_{M}  |\phi_{i',\ell } \rangle_{e_j}$, while
$\{ |i\rangle_M\}_k$ and $\{ |\phi_{i,\ell}\rangle_{e_j}\}_{i,\ell}$ are orthonormal
basis for the memory systems $M$ and $e_j$ respectively).

Most of the analysis conducted so far focused on the special  case of homogeneous Markov processes in which
both the $p^{(j)}_{i|i'}$  and the $\Phi_{q}^{(i)}$ do not depend upon the carrier label $j$
(i.e., $p^{(j)}_{i|i'}:=p_{i|i'}$).
Under these conditions one also says that the quantum Markov process is {\it regular}
if the corresponding classical Markov process $p_{i|i'}$ is regular, i.e., if some power of the transition matrix
$\sf\Gamma$ (whose entries are the transition probabilities $p_{i|i'}$) has only strictly positive elements.
In this case, for
$j\rightarrow \infty$ the statistical distribution of the local noise converges to a stationary distribution
$p_i^{(\infty)} := \lim_{n\rightarrow \infty} p_i^{(n)}$, with
\begin{eqnarray}\label{memdef}
p_{i}^{(n)} := \sum_{i'} ({\sf\Gamma}^{n-1})_{i,i'} \; p_{i'}^{(1)}
\end{eqnarray}
being the probability of
getting $\Phi_{q_n}^{(i)}$ on the $n$-th carrier.
The initial probability $p^{(1)}_{i}$ is said to be stationary if it satisfies the eigenvector equation
$\sum_{i^\prime} {\sf\Gamma}_{i,i'} p^{(1)}_{i'} = p_{i}^{(1)}$ (when this happens
$p_i^{(j)}=p_i^{(1)}$ and the local statistical distribution
of $\Phi_{q_j}^{(i)}$ is identical for all the carriers).

The first example of a regular Markov process has been analyzed by \onlinecite{MP02}.
Here the carriers are assumed to be qubits and the CPTP transformations $\Phi_q^{(i)}$ entering in
Eq.~(\ref{eqn:markov_channel_alternative}) are unitary rotations $\Phi_q^{(i)} (\cdots) := \sigma_{i,q} (\cdots)\sigma_{i,q}$
where $\sigma_{0,q}=\openone$ is the identity operator while for $i=x, y, z$, $\sigma_{i,q}$ is the Pauli matrix.
The conditional probability $p_{i|i'}$ which describes the associated classical Markov process was
finally written as
\begin{equation}\label{mem-par}
p_{i|i'} = (1-\mu)p_i^{(1)} + \mu \delta_{ii'} \, ,
\end{equation}
where $\mu\in[0,1]$ is a
correlation parameter (notice that for $\mu=0$ the model describes a memoryless
channel while for $\mu=1$ it describes a long-term memory channel -- see Sec~\ref{sec:long}).
This model of a Markovian correlated Pauli channel shows a remarkable feature when it is used for
the transmission of classical information (see Sec.\ \ref{sec:ctc}). That is, when two successive
uses of the channel are considered, classical information is optimally encoded in either separable
states or maximally entangled states, depending on whether the correlation parameter $\mu$ is below or
above a certain threshold value.
This feature was first conjectured in \cite{MP02}, then proven for certain instances of the model
in \cite{MPV04}, and finally proven for general Markovian correlated Pauli channels in \cite{Daems07}.
Remarkably, this effect is at the root of the superadditivity property of memoryless
quantum channels for transmitting classical information \cite{Has09} (see Sec.\ \ref{SASA}).

An experimental demonstration of the optimality of entangled qubit pairs for encoding classical
information through a correlated Pauli channel was provided by \onlinecite{BDW+04} for mechanically
induced correlated birefringence fluctuations, which in turn induce correlated depolarization \cite{BDB04}.

A generalized model of $d$-dimensional Markovian correlated Pauli channel was considered by \onlinecite{SKBC11}
for the problem of sending classical information using a dense-coding protocol.
An alternative model of two-qubit correlated channel was characterized by \onlinecite{CGMR2008} in terms
of the minimum output entropy.

Going beyond the case of two uses of a qubit channel, Markovian correlated depolarization over an
arbitrary number of channel uses was studied in \cite{KM06b,DKB07}, and the case of Markovian
correlated noise in higher dimensional quantum systems was considered in \cite{KM06,KDC06,KDC06b}.
Generally speaking, the optimality of entangled states for encoding classical information
can be interpreted in terms of a decoherence-free subspace (see Sec.\ \ref{sec:codes})
associated to the correlated noise model: this has been considered for the Hilbert space defined by
multiple uses of a qubit channel in \cite{DKB07} and for the multiphoton Hilbert space associated to
the polarization of light \cite{BB05}.
In a different context, the same phenomenon has been discussed for the problem of quantum communication with
polarized light without a shared reference frame \cite{BRS03}.

Finally, models of Markovian correlated noise in the framework of quantum systems with continuous variables
(see Sec.\ \ref{sec:smcmc}) were first discussed in \cite{CCM+05,CCM+06} for the case of two uses of the
channel, and then extended to the arbitrary number of uses in \cite{SKC09,LMM09,RM07a} (see Sec.\ \ref{sec:smcmc}).

\subsubsection{Fixed--point channels} \label{Sec:fixedpoint}

Within the representation~(\ref{nfoldconc}) a channel is said to be a \textit{fixed--point} memory channel~\cite{BDM05}
if the initial  memory state $\omega_M$ of the representation is left invariant after each interaction with the carriers.
Specifically, recalling the definition~(\ref{MEMEV}) this notion is formalized by the following identity
\begin{equation}\label{partr}
\Psi^{(n)}(\rho_Q^{(n)}; \omega_M) = \omega_M\;, \qquad  \forall \rho_Q^{(n)} \in \mathfrak{S} ({\cal H}_{Q}^{(n)})\;.
\end{equation}
Fixed--point channels can be easily shown to be symbol independent while the opposite is not necessarily true.
Indeed from Eq.~(\ref{nfoldconc1}) one has that the output state of $n$-th carrier
$\rho_{q_n}^\prime:= \mathrm{Tr}_{Q^{(n-1)}}\Big[\Phi^{(n)}(\rho_Q^{(n)}) \Big]$
can be expressed as
\begin{eqnarray}
\rho_{q_n}^\prime &=& \mathrm{Tr}_{Q^{(n-1)}M}\Big[ (\Phi_{q_nM} \circ \Phi^{(n-1)}_{QM})
\big(\rho_Q^{(n)} \otimes \omega_M \big) \Big] \nonumber \\
& = & \mathrm{Tr}_{M}\Big[ \Phi_{q_nM} \big( \Psi^{(n-1)}
\big(\rho_{q_n} ; \omega_M \big)\big) \Big] \nonumber \\
&=&  \mathrm{Tr}_{M}\Big[ \Phi_{q_nM} (\rho_{q_n} \otimes \omega_M)\Big]\;,
\label{fixedpointcond}
\end{eqnarray}
which only depends upon the reduced density operator $\rho_{q_n}$
and not on the previous information carriers (in these expressions $\mathrm{Tr}_{Q^{(n-1)}M}$
indicates the partial trace with respect to $M$ and the first $(n-1)$ carriers).

Markovian memory channels are examples of fixed--points memory channels, in which the memory system can be
represented by the classical variable of the underlying Markov chain. Being classical, the memory
system can be chosen in such a way that it is unaffected by the back-action of the input system.
This representation can be made explicitly by choosing a unitary dilation of the form (\ref{eqn:markovmem1}).
Another example is provided by \cite{PV07,PV08}, in which the input system interacts with the
memory system by a controlled-unitary transformation, where the memory is the control and the
system is the target. In this setting, the resulting memory channel is a fixed--point one if
the initial state of the memory is diagonal in the control basis.

\subsubsection{Indecomposable and forgetful channels}\label{sec:forgetful}

An \emph{indecomposable} channel is one where, for each channel input, the long-term behavior
of the channel is independent of the initial memory state~\cite{BDM05}.
Such independence can be quantified by evaluating the distance
between different trajectories
$\Psi^{(n)}(\rho_Q^{(n)}; \omega_M)$ and $\Psi^{(n)}(\rho_Q^{(n)};\omega'_M)$  associated through Eq.~(\ref{MEMEV}) to two
different initial memory configurations $\omega_M$ and  $\omega'_M$.
Specifically  a finite--memory quantum channel is said to be indecomposable if for any input state $\rho_Q^{(n)}$
and $\epsilon > 0$ there exists an $N(\epsilon)$ such that for $n \geq N(\epsilon)$,
\begin{equation}\label{eqn:indecomp}
D\left[ \Psi^{(n)}(\rho_Q^{(n)}; \omega_M), \Psi^{(n)}(\rho_Q^{(n)};\omega'_M) \right] \leq \epsilon \; ,
\end{equation}
for any pair of initial states of the memory $\omega_M$, $\omega'_M$ (here $D$ is the trace distance -- see Eq.~(\ref{TRACEDISTANCE})).
Equivalently Eq.~(\ref{eqn:indecomp}) can be stated by saying that  for large $n$, $\Psi^{(n)}(\rho_Q^{(n)}; \omega_M)$ converges to
a state of $M$ which {\it depends  on} $\rho_Q^{(n)}$ but {\it not on} $\omega_M$ (compare this with the behavior~(\ref{partr}) of the fixed-point memory channels).
Here one notices that for  finite dimensional system this implies that there exists a family of  CPTP channels $\Theta^{(n)} : \mathfrak{S} ({\cal H}^{(n)}_Q) \rightarrow \mathfrak{S} ({\cal H}_M)$
which fulfills the identity
\begin{eqnarray}\label{indecomp2}
\mbox{Tr}_Q [ \Phi_{QM}^{(n)} ( O^{(n)}_{QM}) ]  \longrightarrow \Theta^{(n)}(O^{(n)}_Q )\;,
\end{eqnarray}
in the limit $n \to \infty$ for all the operators $O^{(n)}_{QM}$ on $Q$ and $M$,
with $O^{(n)}_Q=\mbox{Tr}_M[O^{(n)}_{QM}]$. In the Heisenberg picture -- see Sec.~\ref{sec:HEIS} -- this also can be stated as
\begin{eqnarray}\label{indecomp21}
{\Phi_{QM}^{(n)}}^*(\openone_Q \otimes ...) \longrightarrow  {\Theta^{(n)}}^*(\cdots)  \otimes \openone_{M}\;,
\end{eqnarray}
with ${\Phi_{QM}^{(n)}}^*$ and ${\Theta^{(n)}}^*$ being the dual of $\Phi_{QM}^{(n)}$ and $\Theta^{(n)}$, respectively.

The main features of indecomposable channels have been revisited
through the notion of \emph{forgetful channels}~\cite{KW05}.
The latter has been originally introduced in the quasi-local algebra approach detailed in Sec.~\ref{sec:causal}, where
the quantum memory channels are assumed to be translation invariant and
non-anticipatory.
In the representation~(\ref{nfoldconc}) this definition coincides with the 
limiting condition~(\ref{indecomp2}), which in Ref.~\cite{KW05} is written 
in terms of the cb-norm distance (see Sec.~\ref{sec:dist2}).
In this context a memory channel is said to be {\it strictly forgetful} if there exists a finite integer $m$
such that the rhs and the lhs of Eq.~(\ref{indecomp2}) exactly coincides for all $n\geq m$.
A simple example of forgetful channels can be obtained if $\Phi_{QM}^{(n)}$ 
is defined by the concatenation~(\ref{eqn:memory_model_2}) of a generator map
$\Phi_{qM} := p \;  \id \, + \; (1-p) \; \mbox{SWAP}$, where $p \in [0,1)$,
and $\mbox{SWAP}$ denotes the swap channel which exchanges $q$ and $M$.
In this case the only way for $\Phi_{QM}^{(n)}$  not
to be forgetful is to choose the ideal channel in every step of the concatenation.
However, the probability for this event vanishes
in the limit $n\to\infty$ as $p^n$, implying that Eq.\,(\ref{indecomp2})
holds.

Several criteria for a quantum memory channel to be forgetful have been proposed~\cite{KW05}.
For instance, a sufficient condition is that the cb-norm distance between the rhs and the lhs of Eq.~(\ref{indecomp2})
 falls below $1$ for {\em some} finite $n$.
 From a physical point of view, one could expect a generic quantum
memory channel to be forgetful. Indeed, it can be proven that the
subset of forgetful channels is dense and open (according to the
topology induced by the  cb-norm) \cite{KW05}.

In the case of Markovian channels, the forgetfulness is determined
by the asymptotic properties of the underlying Markov chain: in particular, for a discrete variable
memory system, the channel is forgetful if and only if the underlying Markov
chain converges to a unique stationary state \cite{DD09}.
On the other hand, if the memory system is described by continuous variables,
one could have situations in which the Markov chain has a unique stationary state,
yet the convergence property (\ref{indecomp2}) in cb-norm is not satisfied.
To overcome this limitation, a weaker notion of forgetfulness, named {\it weak forgetfulness},
has been introduced in \cite{LMM09} for Markovian channels.
Although restricted to this setting, its definition coincides with that of
indecomposability [Eq.\ (\ref{eqn:indecomp})], and is equivalent to forgetfulness
for a discrete variable Markov chain.
Beyond this setting, the model of Gaussian memory channels in \cite{LGM10} was proven to be
indecomposable under restricted conditions on the memory initialization, e.g., if the
initial state of the memory is a Gaussian state with finite first and second moments.
Finally, the relation between the forgetfulness of the channel and the chaotic quantum evolution
of the memory system was studied in~\cite{BB10} for a model of a dephasing channel with memory.

\subsubsection{Decaying input memory cq-channels}\label{sec:erg}

Forgetful channels represent configurations in which
the effect of the far past inputs do not strongly affect present and future outputs. Within the quasi-algebra approach
a similar notion has been developed in~\onlinecite{BB08} for the special class of cq-channels (see Sec.~\ref{sec:causal}).
Specifically a cq-channel defined by Eq.~(\ref{BB08MAP}) is said to have \emph{decaying input memory}
if for each $\epsilon>0$ there exists a non-negative integer $m (\epsilon)$ such that
\begin{equation}\label{Wdistance}
|W(x,b)-W(x',b)|\le \epsilon,
\end{equation}
for all $b\in \mathcal{B}^{[n,\infty]}$, $n,\in \Z$, whenever $x_i=x'_{i} $ for $n-m\le i$ and $m\ge m(\epsilon)$.
Notice that $\sum_b|W(x,b)-W(x',b)|$ is a distance between quantum states. 
Then \eqref{Wdistance} says that, starting from the $n$-th use, the outputs of two identical channels $W$ are
almost (i.e., within a distance $\epsilon$) the same provided that the inputs have started to be identical from the
$n-m(\epsilon)$-th use. Hence $m(\epsilon)$ gives an estimation of the memory length.

This provides a `continuity' property of the channel which plays a crucial role in establishing coding theorems,
an idea that will also appear in Section IV.D and goes back to the classic paper by McMillan \cite{McM53}.

\subsubsection{Long-term memory channels}\label{sec:long}

Long-term quantum memory channels describe those communication lines in which the effect of the memory
does not decay with the number of channel uses.
These channels are defined as those memory channels which are not forgetful.

Extreme examples are provided by statistical mixtures of memoryless channels \eqref{longmemory}
of Sec. \ref{sec:compound} as pointed out in \cite{DD07,DSD08}.
The memory correlations of this class of channels can be considered
to be given by a Markov chain which is {\it aperiodic} but {\it not irreducible} \cite{Norris}.
This can be easily seen by noticing that Eq.~(\ref{eqn:markov_channel_alternative}) reduces
to Eq.~(\ref{longmemory})
by setting  $p^{(j)}_{i|i'}=\delta_{ii'}$ for all $j=2, \cdots n$.
Hence, once a particular branch, $i=1, \dots, M$, has been chosen, the successive inputs
are sent through this branch ({\it aperiodicity}) and transition between the different branches (which correspond
to the different states of the Markov chain) is not permitted ({\it reducibility}).

It is worth stressing that the transformation~(\ref{longmemory}) is fully non-anticipatory
(i.e., the input state of any subset of carriers cannot influence the output state of the remaining ones):
as a consequence, fixing an ordering, it can always be represented as in Fig.~\ref{figure01} with a proper
choice of the unitary couplings.

%

\section{Quantum Codes}\label{sec:codes}

Coding theory is the branch of information science studying how to use {\it software}
strategies to counteract the effect of an assigned noise source affecting a communication line or the components  (memory elements) of a database. In a sense it can be described as the
last resort which can be exploited once no further improvements can be obtained at the level of hardware engineering.

Both in the classical and in the quantum setting,
the key idea to prevent the corruption of information is to use {\it redundancy}:  by properly spreading a given message
over many information carriers instead of a single one, one can take advantage of
the structural properties of noise source.
Consider for instance the paradigmatic case in which one wishes to store the information 
contained in (say) $k$ qudits, affected by an assigned error model described by the channel $\Phi^{(k)}$,
into a larger set of $n\geq k$ qudits, affected by the noise $\Phi^{(n)}$.
Then a coding strategy consists in identifying an {\it encoding} CPTP  map
$\Phi_E^{(k \to n)} : \mathfrak{S}({\mathcal H}^{\otimes k})\to\mathfrak{S}({\mathcal H}^{\otimes n})$,
shuffling a state from the smaller space of the $k$ carriers to the
the larger space of the $n$ carries,
and a {\it decoding} CPTP map
$\Phi_D^{(n \to k)} : \mathfrak{S}({\mathcal H}^{\otimes n})\to\mathfrak{S}({\mathcal H}^{\otimes k})$,
moving the information back to the original space,
under the requirement that the resulting channel
$\Phi_D^{(n \to k)}\circ \Phi^{(n)}\circ \Phi_E^{(k \to n)}$
is somehow   ``less" noisy than the original transformation $\Phi^{(k)}$ -- see Fig. \ref{fig_capacities}.
The image $\mathcal{Q}$ of  $\Phi_E^{(k \to n)}$ is called
\emph{quantum error correcting code} (QECC):
it represents the information {\it vault}  where messages are deposited to prevent the noise
from affecting them.  
The ratio $R=(k/n)\times \log d$ represents the communication {\it rate} (in qubits per channel use) of the code, whose
inverse measures how much information spread is involved in the procedure. 
It is worth noticing that if $\Phi_E^{(k \to n)}$ is taken to be an isometry (an option
which often implicitly assumed in QECC) the space $\mathcal{Q}$ becomes a proper vector 
subspace  of ${\mathcal H}^{\otimes n}$ of dimension $d^k$.
When this happens $\Phi^{(n)}\circ \Phi_E^{(k \to n)}$ is just a restriction
of $\Phi^{(n)}$ on $\mathcal{Q}$ and the encoding mapping can be fully specified by simply assigning the latter. 
This also justifies the consideration of a recovery map $\Phi_R^{(n)}$ in place of the decoding map $\Phi_D^{(n\to k)}$ 
(in the following this simplification will be assumed).

\begin{figure}
\centering
\includegraphics[scale=0.39]{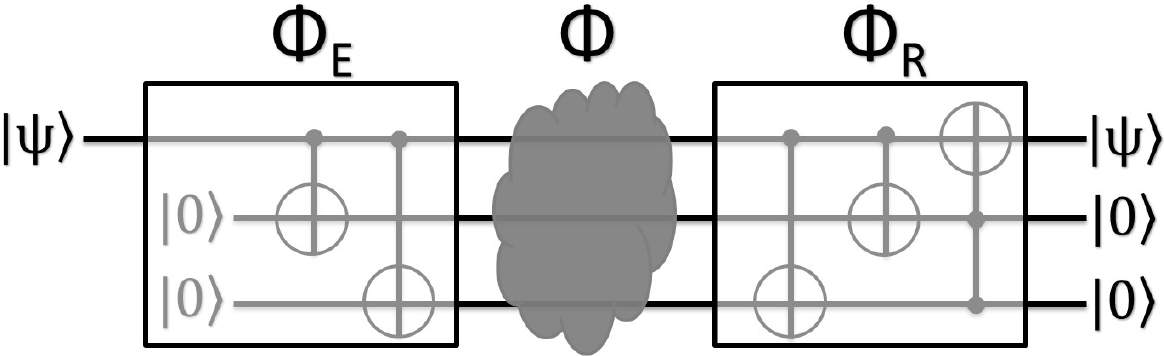}
\caption{The figure shows the encoding and recovery maps for a qubit
repetition code, able to counteract uncorrelated bit-flip errors. The circuital 
implementation of encoding and decoding maps involves controlled NOT operations.
Upon encoding the qubit basis states are spreader over three qubits as $|0\rangle\mapsto |000\rangle$ 
and $|1\rangle\mapsto |111\rangle$. 
The reverse happens upon recovery provided 
that no more than one qubit is affected by bit-flip.} \label{FigCod}
\end{figure}

Different equivalent ways have been devised to evaluate the quality of a given coding procedure
(Cafaro et al. 2011, analyze and compare some of them).
The most commonly used is by means of the input-output fidelities introduced in
Sec.~\ref{sec:fidecha}. In particular good choices are  the minimum or average
fidelity functionals, i.e., $F_{\min}( \Phi_D^{(n \to k)}\circ \Phi^{(n)}\circ \Phi_E^{(k \to n)})$ and
$\bar{F}( \Phi_D^{(n \to k)}\circ \Phi^{(n)}\circ \Phi_E^{(k \to n)})$.
It is important however to distinguish between
two different scenarios: the case where
the messages to be stored/transmitted are purely classical, and the case where instead they are quantum.
In the first scenario the minimization (resp.~average) involved in Eq.~(\ref{DEFNUOVADIFMIN}) 
(resp.~Eq.~(\ref{DEFNUOVADIFAVE})) needs not to be performed over the entire input space  of the $k$ carriers,
but {\it only} with respect to the orthogonal set of states in ${\mathcal H}^{\otimes k}$ encoding the 
classical messages one wishes to protect.
Vice-versa in the second scenario, which implies
the possibility of producing arbitrary superposition of the input signals,  the minimization (resp.~average) is 
performed over the whole input space.
In this last configuration the effectiveness of a correcting code can also be quantified by other distance measures, like the entanglement fidelity $F_{e}( \openone/d^k;\Phi_D^{(n \to k)}\circ \Phi^{(n)}\circ \Phi_E^{(k \to n)})$
defined in Eq.~(\ref{EntFidMaxEnt})
or the cb-norm distance of $\Phi_D^{(n \to k)}\circ \Phi^{(n)}\circ \Phi_E^{(k \to n)}$ from
the identity channel $\id$ on the set of the
$k$ carriers -- see Sec.~\ref{sec:dist2}.
The simplest example of error correcting code is obtained for counteracting bit-flip errors 
by repeating qubit basis states (preferably an odd number of times), see Fig.\ \ref{FigCod}.

When a given code allows for exact protection of the data stored in the $k$ carriers (e.g.,
when  $F_{\min}=1$, implying zero error probability~(\ref{errorPROBNEW})
for all possible channel inputs),  the code is said to be {\it perfect}.
However, in realistic situations codes only allow for arbitrary high fidelity at
a finite rate in the limit of large code length ($n \to \infty$), see Sec.~\ref{sec:ctc}.
For finite code length, one has to find the optimal compromise between rate, fidelity
and the complexity of the coding and decoding operations.
Here, the focus is on the general properties of quantum error correcting codes.
The large body of works in this field  was mainly concerned with the development of strategies  for
independent and identically distributed (i.i.d.) errors, i.e., for noise arising from memoryless quantum channels. 
In this Section, after reviewing the basics of such codes, we analyze their
effectiveness against correlated errors.
Then possible answers to the new challenges posed by memory effects are discussed.
Since the extension of the formalism for QECC (mostly relying on group theory)
from ${\mathcal H}\simeq\mathbb{C}^2$ to higher dimensional $\mathcal H$ is nontrivial,
the presentation will be restricted to qubit systems, i.e., to binary quantum codes.
For non binary codes the reader may refer to \cite{AK01,KKKS06,knill96}.


\subsection{Standard quantum coding theory}\label{newsecQEC}

Consider a memoryless quantum channel $\Phi^{(n)} = \Phi^{\otimes n}$, characterized by a set of
Kraus operators $\{K_{\sf i}\}$, on the Hilbert space ${\mathbb C}^{2\otimes{n}}$ of $n$ qubits
\begin{eqnarray}\label{mappaerrore}
\rho\mapsto{\Phi^{(n)}}(\rho)=\sum_{\sf{i}} {K}_{\sf{i}}\;\rho\; {K}_{\sf{i}}^{\dag},
\end{eqnarray}
where $K_{\sf i}=K_{i_1}\otimes\ldots\otimes K_{i_n}$ describes i.i.d. errors on single qubits.
An error correcting code is identified by a $2^k$-dimensional subspace $\mathcal{Q} \subset {\mathbb C}^{2\otimes{n}}$.
Let us denote as $P_{\mathcal{Q}}$ the projector operator associated with $\mathcal{Q}$.
It is possible to show that the code is able to correct errors belonging to a subset 
$\mathfrak Q \subseteq \{K_{\sf i}\}_{\sf i}$ if and only if there exists a Hermitian matrix $\sf S$ 
such that
\begin{equation}
P_{\mathcal{Q}} K_{\sf l}^{\dagger }K_{\sf m}P_{\mathcal{Q}}={\sf S}_{\sf lm} P_{\mathcal{Q}} \, ,  \label{ecc}
\end{equation}
for any pair of error operators $K_{\sf l}, K_{\sf m}\in{\mathfrak Q}$ \cite{knill02}. 
Due to the unitary freedom of the Kraus representation, condition (\ref{ecc}) holds true 
if and only if a Kraus representation of the map $\Phi^{(n)}$ exists, say $\Phi^{(n)}(\rho)=\sum_{\sf{i}} {K}'_{\sf{i}}\;\rho\; {K'}_{\sf{i}}^{\dag}$,
such that $P_{\mathcal{Q}}{K'}_{\sf l}^{\dagger }{K'}_{\sf m}P_{\mathcal{Q}}=\delta _{\sf lm} s_{\sf m} P_{\mathcal{Q}}$
for some set of non-negative numbers $s_{\sf m}$.
This condition in turn yields that different correctable error operators map the code words in
$\mathcal{Q}$ into mutually orthogonal subspaces. 
This property implies that different errors can be detected by applying a projective measurement
able to distinguish the different orthogonal subspaces. 
Moreover, if the same error applies to two different basis codewords, their scalar product will not change.
Thus, the geometrical interpretation is that each correctable error maps the code space into an orthogonal
subspace without deformations. 

Applying again the unitary freedom of the Kraus representation, one can find a basis 
where any pair of errors acting on a given codeword either produce orthogonal states 
or exactly the same state. 
This phenomenon will occur if the matrix ${\sf S}$ results singular.
In such a case the quantum code $\mathcal{Q}$ is called \emph{degenerate}.

Now, let ${\cal V}^{i_L}$ be the subspace of ${\mathbb C}^{2\otimes{n}} $ spanned by the corrupted images 
$\{{K'}_{\sf l} |i_L\rangle\}_{\sf l}$ of the codewords $|i_L\rangle$ and let $\{|v_r^{i_L}\rangle\}$ be and orthonormal basis of  ${\cal V}^{i_L}$. We may define a subspace ${\cal V}^{i_L}$ for each codeword.
Then, the recovery map $\Phi_R^{(n)}$ will be characterized by Kraus operators $\{R_r\}$ such that 
\begin{align}
R_r=\sum_i |i_L\rangle\langle v_r^{i_L}|.
\end{align} 
In the qubit context we are considering, an $[[n, k]]$ quantum correcting code $\mathcal{Q}$ is given by a 
$2^k$-dimensional subspace of ${\mathbb{C}^2}^{\otimes n}$ encoding $k$ logical qubits 
into $n$ physical qubits ($n\ge k$).

One can assume the error operators to be proportional to the Pauli operators acting on 
the $j$-th qubit and corresponding to no-error, bit-flip error, bit-phase-flip error and 
phase-flip error, i.e., $K_{0_j} \propto \openone_j$, $K_{1_j} \propto \sigma_{x,j}$, 
$K_{2_j} \propto \sigma_{y,j}$, $K_{3_j} \propto \sigma_{z,j}$.
This restriction to Pauli errors represents no loss of generality. Indeed, it can be easily 
shown that if a code corrects a given set of errors it can also correct any linear combination
(by complex coefficients) \cite{NC00}.
It is hence sufficient to restrict to Pauli operators since they are a basis on the space 
of qubit operators.

Given a subset $\mathfrak{Q}\subseteq\{K_{\sf i}\}_{\sf i} $ of errors 
that can be corrected one says that the code $\mathcal Q$ has $\mathfrak Q$-correcting ability.
To each $K_{\sf i}$ one can assign a \emph{weight} $t$, an integer $0\le t\le n$ denoting the number of qubit
where operators $K_{i_j} $ ($j=1,\ldots,n$) act differently from identity.
Then, the correction ability of $\mathcal Q$ can be also expressed by specifying the value of the \emph{distance}
$d=2t+1$ of the code, meaning that $\mathcal Q$ corrects all errors
affecting at most $t$ qubits.

Suppose that  errors are i.i.d.~with probability $p_e$ on each qubit, then
for any of the $\footnotesize\left(
\begin{array}{c}
n\\ t+1
\end{array}
\right)$ ways of choosing $t+1$ locations, the probability that errors occur at every one of those locations
results $p_e^{t+1}$. 
Therefore one has the following upper bound on the probability that at least $t+1$ errors occur in the block of $n$ qubits $\footnotesize \left(
\begin{array}{c}
n\\ t+1
\end{array}
\right) p_e^{t+1}$.
This means that for $p_e$ small the performance of the code $1-F\approx {\mathcal O}(p_e^{t+1})$
is substantially improved over the unprotected data $1-F\approx {\mathcal O}(p_e)$.


An upper bound on the rates achievable by non-degenerate quantum codes is given by
the quantum version of the (classical) Hamming bound \cite{AM96}
\begin{equation}\label{qHb}
2^k\sum_{i=0}^t3^i {n \choose i} \leq 2^n \, ,
\end{equation}
which for large $n$ and $\frac{d}{n}$ fixed yields the approximate bound
$R\le1-\frac{d}{2n}\log_23-h(\frac{d}{2n})$ with $h$ the binary entropy~(\ref{binarysh}).

There are also upper bounds that apply to all quantum codes, not just non-degenerate ones,
like the quantum Singleton bound \cite{KL97}
\begin{equation}\label{qSb}
n \geq 4t+k \, .
\end{equation}
On the other hand a lower bound on the rates, confirming that good codes indeed exist \cite{C97},
comes from the quantum version of the Gilbert-Varshamov theorem, stating that a $[[n, k]]$ quantum 
code of distance $d = 2t+1$ exists with
\begin{equation}
k \geq \max \left\{ k' \, \, | \, \, 2^{k'} \sum_{i=0}^{2t}3^i {n \choose i} \leq 2^n \right\} \, .
\end{equation}
For large $n$ and $d/n$ fixed  one gets the approximate bound 
$R \ge 1-\frac{d}{n}\log_23-h(\frac{d}{n})$.

Unfortunately the explicit construction of quantum codes is not an easy task.
Historically the first quantum code that appeared was a $[[9,1]]$ code with $d=3$ \cite{Shor1995}
whose basis codewords read
\begin{equation}
\left[\frac{1}{\sqrt{2}}\left(|000\rangle + |111\rangle\right)\right]^{\otimes 3} \, , \quad  \left[\frac{1}{\sqrt{2}}\left(|000\rangle - |111\rangle\right)\right]^{\otimes 3}.
\end{equation}
Its construction relies on simple argument.
A three qubit code would suffice to protect against a single bit flip (see Fig.\ \ref{FigCod}).
The reason the three qubit clusters are repeated three times is to protect against phase errors as well.

Then attempts were made following classical \emph{linear} codes \cite{H85}.
In the classical setting, the state of $n$ bit system is represented by a binary
string of length $n$. An error affecting this string can be also
represented by a binary string of the same length, where the $1$'s indicate 
the locations of the bits that have been flipped.
The action of an error string on a code words string is hence represented by a 
summation modulo two. The space of binary strings of length $n$, endowed with the summation modulo
two, defines the linear space $\mathbb{F}_2^n$.
A classical $[n,k]$ linear code $\mathcal{C}$ is defined as a subspace $\mathcal{C}\subseteq \mathbb{F}_2^n$.
The subspace can be characterized by a set of $k$ generators 
or equivalently by a \emph{parity check} $(n-k)\times n$ matrix $\sf H$ such that ${\sf H}{\sf v}^\top=0$, $\forall {\sf v}\in\mathcal{C}$.
Errors ${\sf e}_i$ taking ${\sf v}\in {\mathcal C}$ into ${\sf v}+{\sf e}_i$
can be detected by applying the parity check ${\sf H}({\sf v}+{\sf e}_i)^\top={\sf H}{\sf e}_i^\top=synd({\sf e}_i)$ (\emph{error syndrome}) and corrected iff they give rise to distinct syndromes, i.e., ${\sf H}({\sf e}_i+{\sf e}_j)^\top\neq 0$ for $i\neq j$.
If $\mathcal{C}$ is with distance $d=2t+1$ it means that it is able to correct up to $t$ errors, i.e., bit-flip errors in at most $t$ bits. The set of errors correctable by  $\mathcal{C}$ will be denoted by $\mathfrak{C}$.

A great advantage of linear codes over general error correcting codes is their compact specification.
Using them, quasi classical [or CSS (Calderbank, Shor, Steane)] codes were constructed in the following way \cite{CS96,S96}.
Consider two classical linear codes $\mathcal{C}_1$ and $\mathcal{C}_2$ such that $\mathcal{C}_2^{\perp}\subseteq \mathcal{C}_1$, where $\mathcal{C}_2^{\perp}$ is the dual code to $\mathcal{C}_2$, i.e., 
consisting of those bit strings that are orthogonal to the codewords of $\mathcal{C}_2$.
If $\mathcal{C}_1$ is a $[n, k_1]$ code with distance $d_1$ and $\mathcal{C}_2$ is a $[n, k_2]$ code with distance $d_2$, then the corresponding CSS 
quantum code is a $[[n,k_1+k_2-n]]$ code with distance $\min\{d_1,d_2\}$. 
Its basis codewords are 
\begin{equation}
\label{113}
\frac{1}{\sqrt{|{\cal C}_2^\perp|}}\sum_{{\sf w}\in {\cal C}_2^\perp} |{\sf u}+{\sf w}\rangle, \qquad {\sf u}\in {\cal C}_1.
\end{equation}
Performing the Hadamard transform on each qubit of the code one can switch from ${\cal C}_1$ to ${\cal C}_2$
to account for $\sigma_z$ errors beside $\sigma_x$ ones.
As matter of fact it takes \eqref{113} to
\begin{equation*}
\frac{1}{\sqrt{2^n|{\cal C}_2^\perp|}}\sum_{{\sf x}\in {\mathbb F}_2^n}
\sum_{{\sf w}\in {\cal C}_2^\perp} (-1)^{{\sf x}\cdot ({\sf u}+ {\sf w})} |{\sf x}\rangle.
\end{equation*}
Since $\sum_{{\sf w}\in {\cal C}_2^\perp} (-1)^{{\sf x}\cdot {\sf w}} \neq 0$
if ${\sf x}\in {\cal C}_1^\perp$ and zero otherwise, one is left with a state
$\propto
\sum_{{\sf x}\in {\cal C}_1^\perp}  |{\sf x}\rangle$. This latter,
being ${\cal C}_1^\perp\subseteq {\cal C}_2$, can be considered as an instance of 
basis code words
\begin{equation}
\label{114}
\frac{1}{\sqrt{|{\cal C}_1^\perp|}}\sum_{{\sf w}\in {\cal C}_1^\perp} |{\sf u}+{\sf w}\rangle, \qquad {\sf u}\in {\cal C}_2.
\end{equation} 
Therefore, to correct errors it is enough to implement the parity check of
${\cal C}_1$ by measuring in the $\sigma_z$ basis (for bit flip errors) and that of ${\cal C}_2$ by measuring in the $\sigma_x$ basis (for phase flip errors). An example along this line is provided by the
$[[7; 1]]$ code with $d=3$ (Steane, 1996).

Another, more general, way to construct quantum codes is to exploit the group structure of the set of
errors as it is done for \emph{stabilizer codes} \cite{G97}.
The method can be summarized as follows.
First notice that the set of Pauli errors on $n$ qubit
can be written as\footnote{Actually $\sigma_y=i\sigma_x\sigma_z$, however the imaginary unit, as well as any 
global phase factor, is irrelevant in quantum error correction.}
\begin{eqnarray}
\mathcal{P}_n:=\left\{ \bigotimes_{j=1}^n \sigma_{x,j}^{X(j)} \bigotimes_{j=1}^n \sigma_{z,j}^{Z(j)} \, \, \Big|
\, \, X(j),Z(j) \in \mathbb{F}_2\right\},
\end{eqnarray}
and forms a multiplicative group known as Pauli group.
One can represent the elements of $\mathcal{P}_n$ as $2n$-dimensional binary vectors
\begin{equation}
e\in \mathcal{P}_n \leftrightarrow
({\sf e}_X |{\sf e}_Z)\equiv {\sf e}
\in \mathbb{F}_2^n\times\mathbb{F}_2^n,
\label{everelation}
\end{equation}
where ${\sf e}_X$ (resp. ${\sf e}_Z$) is the $n$-bits vector of components $X^{(j)}$ (resp. $Z^{(j)}$) specifying on which qubits the $\sigma_x$ (resp. $\sigma_z$) error occurs and $({\sf e}_X |{\sf e}_Z)$ is the joint  ${\sf e}_X,{\sf e}_Z$ vector.
Then, one considers an Abelian subgroup $\mathcal{G}\subseteq \mathcal{P}_n$
\begin{equation}
\mathcal{G} = \operatorname{span}\left\{g_i\in{\cal P}_n \, \, | \, \, 1\le i\le n-k\right\},
\label{stabgen}
\end{equation}
where $g_1,g_2,\ldots,g_{n-k}$ are independent of each other.
Notice that the operators in ${\cal P}_n$ have eigenvalues $\pm 1$.
An $n$-qubit vector $|x\rangle$ is said to be stabilized by the group $\mathcal{G}$
if it is a common eigenvector with eigenvalue $+1$.
The set of vectors stabilized by $\mathcal{G}$ forms a $2^k$-dimensional subspace
\begin{equation}
{\mathcal Q}=\left\{
|x\rangle\in{\mathbb C}^{2\otimes n} \, \, \big| \, \, g |x\rangle=|x\rangle, \forall g \in\mathcal{G}\right\}.
\end{equation}

The subspace ${\mathcal Q}$ is a $[[n,k]]$ code. 
Notice that all the errors belonging to the group $\mathcal{G}$
will leave the code word unaffected. 
The other errors will in general change the $n$-qubit states.
To detect which error has occurred one measures the set of $n-k$ commuting
observables $g_i$. 
The results of these measurements are either $+1$ or $-1$. 
The corresponding set of measurement results plays the role of the error
syndrome. Errors with non trivial error syndromes are detectable, while
errors with different error syndromes are correctable.


One can describe stabilizer codes using the same formalism of 
classical linear codes.
By using the vectors in $\mathbb{F}_2^n\times\mathbb{F}_2^n$
corresponding to the generators $g_1,g_2,\ldots,g_{n-k}$ it is possible to write down the
following $(n-k)\times 2n$ parity check matrix
\begin{equation}
{\sf H}:=\left(
\begin{array}{c}
{\sf g}_{1,X}|{\sf g}_{1,Z}\\
\vdots\\
{\sf g}_{n-k, X}|{\sf g}_{n-k, Z}
\end{array}
\right).
\label{Hstab}
\end{equation}
for a $[2n,k]$ classical linear code $\mathcal C$ 
(its $j$-th row is given by the vector $({\sf g}_{j,X}|{\sf g}_{j,Z})$)
corresponding to ${\mathcal Q}$.
Then the analysis of correcting $\mathfrak Q\subseteq \mathcal{P}_n$ errors by 
${\mathcal Q}$ can be traced back to that of
correcting ${\mathfrak C}\subseteq \mathbb{F}_2^n\times\mathbb{F}_2^n$ errors by
${\mathcal C}$.

In this way it results \cite{gaitan}
that ${\mathcal Q}$ has $\mathfrak Q$-correcting ability iff
for every $({\sf e}_{1,X}|{\sf e}_{1,Z})$, $({\sf e}_{2,X}|{\sf e}_{2,Z})$ $\in{\mathfrak C}$ it is
\begin{equation}
{\sf H}({\sf e}_{1,X}+{\sf e}_{2,X} | {\sf e}_{1,Z}+{\sf e}_{2,Z})^{\top}\neq 0.
\label{neqsyn}
\end{equation}
The above condition states that
$\mathfrak Q$ is correctable by $\mathcal Q$ iff
$synd({\sf e}_{i}) \neq synd({\sf e}_{j})$ for all ${\sf e}_i,{\sf e}_j\in{\mathfrak{C}}$ (which corresponds to \eqref{ecc}).

A stabilizer code of distance $d$ has the property that each element of $\mathcal{P}_n$ of weight 
$t$ less than $d$ either lies in the stabilizer or anti-commutes with some element of the stabilizer.
An example is provided by the $[[5,1]]$ code with $d=3$ introduced in \cite{L96} and saturating
the quantum Hamming bound \eqref{qHb}.
It is worth remarking that a systematic method to find stabilizer generators exists based on the connection with vectors over Galois field $GF(4)$ \cite{CRSS98}.

If the subgroup $\mathcal{S}\subseteq \mathcal{P}_n$ is not abelian,
it can be used as well to construct a QECC provided that entanglement
between encoder and decoder is available \cite{BDH06}.
The trick consists in extending the generators of ${\mathcal S}$ (by attaching extra Pauli
operators at their end) in order to generate a new group ${\mathcal S}^\prime$ that is
Abelian and for which the above theory can be applied.
These are entangled-assisted QECC and the notation $[[n,k;m]]$, with $m$ denoting the number 
of entangled ancilla qubits ($n-m-k$ giving the number of unentangled ancilla qubits), is used.
Entanglement-assisted codes may lead to rates higher than their non-entangled counterparts.
The reason is that entanglement allows to increase the dimension of the decoding Hilbert space to $2^{n+m}$
compared to $2^{n}$ for unentangled ancillas.
This also leads to a revision of the Hamming bound \eqref{qHb} with $2^{n+m}$ on the r.h.s. \cite{B02}.

Finally, notice that in constructing codes, besides pursuing the highest
possible rate, one should also take into account the complexity of the encoding/decoding procedures.
This can be evaluated by means of the number of elementary steps, i.e., number of
elementary gate operations, needed. Efficient encoding/decoding requires polynomial (actually near linear) 
scaling of complexity vs block code length $n$. 
Luckily, stabilizer codes are efficiently encodable/decodable \cite{G97},
but usually do not achieve the channel capacity (see Sec.\ \ref{sec:ctc}).

\subsection{Codes concatenation}

Unfortunately, using the previous approach it is quite hard to construct good codes with large distances.
Families of codes that offer good performance by increasing the distance are the toric codes \cite{Kit03}
and the quantum version of Reed-Muller codes \cite{Steane02}.
Aside from them, a particularly simple way to construct codes that can correct multiple errors
is to concatenate several single-error correcting codes, i.e., codes with $d=3$.
For the sake of simplicity, we illustrate the case of two layers of concatenation and
consider single qubit encoding.
Assume the inner code (first layer) is a $[[ n_{1}, k_{1}]]$ stabilizer code $\mathcal{Q}_{1}$
with distance $d_{1}$, and the outer code (second layer) is a $[[ n_{2}, 1]]$ stabilizer code $\mathcal{Q}_{2}$
with distance $d_{2}$.
The concatenated code $\mathcal{Q}=\mathcal{Q}_{1}\circ \mathcal{Q}_{2}$ maps $k_{1}$ qubits into
$n=n_{1}n_{2}$ qubits, with code construction
parsing the
$n$ qubits into $n_{2}$ blocks $B\left( b\right) $ ($b=1$,..., $n_{2}$) each
containing $n_{1}$ qubits.
Explicitly, the concatenated code $\mathcal{Q}$ is constructed as follows.
For any codeword $\left\vert c_{\text{out}}\right\rangle$ of
the outer code $\mathcal{Q}_{2}$,
\begin{equation}
\left\vert c_{\text{out}}\right\rangle =\sum_{i_{1} \dots
i_{n_{2}}}\alpha _{i_{1} \dots i_{n_{2}}}\left\vert i_{1} \dots
i_{n_{2}}\right\rangle \, ,
\end{equation}
with $\left\vert i_{1}\text{...}i_{n_{2}}\right\rangle =\left\vert
i_{1}\right\rangle \otimes $...$\otimes \left\vert i_{n_{2}}\right\rangle $,
replace each basis vector $\left\vert i_{j}\right\rangle $
by a basis vector $\left\vert \phi_{i_{j}}\right\rangle $ of the inner code
$\mathcal{Q}_{1}$, so that
\begin{equation}
\left\vert c_{\text{conc}}\right\rangle :=
\sum_{i_{1}\dots i_{n_{2}}}\alpha _{i_{1} \dots
i_{n_{2}}}\left\vert \phi _{i_{1}}\right\rangle \otimes \dots \otimes
\left\vert \phi_{i_{n_{2}}}\right\rangle \, .
\end{equation}
Notice that the above mentioned construction produces a $[[ n_{1}n_{2}
\text{, }k_{1}]] $ code with distance $d \geq d_{1} d_{2}$.

If there are $L$ levels of concatenation of the same single qubit code, and $p_e$ is the error probability on single qubit,
it is possible to show that the code failure probability is bounded by \cite{gaitan}
\begin{equation}
p_e^{(L)}\le p_0\left(\frac{p_e}{p_0}\right)^{2^L}\,,
\end{equation}
where $p_0$ is an estimate of the threshold error probability that can be tolerated and depends 
on the chosen single qubit code. Hence, provided that $p_e<p_0$, one can make the code failure 
probability as small as one wishes by adding enough levels to the code.

Finally, it is worth remarking that minimum distance is not everything. It helps in constructing good codes (codes with arbitrarily small error probability).
However, very good codes (codes with arbitrarily small error probability and achieving maximum rate)
can be constructed even with bad (small) minimum distance. The reason can be understood 
by means of a metaphor due to Berlekamp, see e.g. \cite{MKbook03}, and applicable in both classical and quantum frameworks. A blind bat lives in a cave and flies about the centre of the cave which corresponds to one codeword with its typical distance from the centre controlled by the error rate. The boundaries of the cave are made up of stalactites that point in towards the centre of the cave. The longest stalactites determine the minimum distance. If there is only a tiny number of such long stalactites, they are relatively unlikely to cause errors when the bat flies beyond the safe distance. It will collide most frequently with more distant (shortest) stalactites, owing to their greater number. So the take home message is that a given code must only be able to correct ``typical" errors.


\subsection{Decoherence free subspaces}\label{DFS}

Since the idea of i.i.d. errors was underlying the standard theory
of quantum error correcting codes, it is natural to expect lowered performance when 
employed on memory channels.

A case study is provided by a regular Markovian channel (see Sec.\ \ref{sec:markov})
where the CPTP maps $\Phi_{q_j}^{(i_j)}$ in Eq.\ \eqref{eqn:markov_channel_alternative}
are of the form $\Phi_{q_j}^{(i_j)}(\ldots) = \Phi^{(i)}(\ldots) = {K}_{i}(\ldots) {K}_{i}^{\dag}$,
with unitary Kraus operators ${K}_{i}$ and
$p_{i_{j}{{\vert}}i_{j-1}}^{(j)} = p_{i|i'} = \left( 1-\mu \right) p_i + \mu \delta_{i,i'}$.
A sequence of $n$ uses of the memory channel are hence represented by the map
\begin{align}
\Phi_{\mu} \left( \rho_Q^{(n)} \right) = &
\sum_{i_{1}, \dots, i_{n}}
p_{i_{n}{{\vert}}i_{n-1}} p_{i_{n-1}{{\vert}}i_{n-2}} \cdots
p_{i_{2}{{\vert}}i_{1}} p_{i_{1}} \nonumber\\
& \times \left( {K}_{i_{n}}\otimes \cdots \otimes {K}_{i_{1}}\right) \rho_Q^{(n)}
\left( {K}_{i_{n}}\otimes \cdots \otimes {K}_{i_{1}}\right) ^{\dagger } \, .
\label{AMarkov}
\end{align}
The correlation parameter $\mu$ roughly quantifies the degree
of memory of the considered channel. For $\mu=0$ one obtains the case of
i.i.d.\ (memoryless) noise, while the limit $\mu=1$ describes completely correlated errors.

In Refs.\ \cite{CM10a,CM10b} it has been shown that the performance of stabilizer codes
(evaluated by means of entanglement fidelity~(\ref{EntFidMaxEnt})
 is lowered by increasing $\mu$.
The same relation between the fidelity and the correlation parameter has been observed
for a model of {\it long term memory channel} (see Sec.\ \ref{sec:long})
obtained by convex combination of uncorrelated and completely correlated quantum channels
\begin{equation}
\Phi \left( \rho \right) =
\left(1- \mu \right) \Phi _{\mu=0}\left( \rho \right)
+\mu\, \Phi_{\mu=1}\left( \rho \right)
\text{,}  \label{Amu01}
\end{equation}
where $\Phi _{\mu=0}$ and $\Phi _{\mu=1}$ are given by (\ref{AMarkov}) in the limiting cases of $\mu =0$ and $\mu=1$
(corresponding to uncorrelated and completely correlated errors respectively).
Actually, the effect of the memory is to take the code probability of error
back to a linear dependence on the single error probability (see also  \cite{klesse05}).\footnote{
It should be noted however that the condition of independent errors
is not equivalent to the condition of memoryless quantum channel.
Indeed, one can say that independent errors are those for which the probability
of $k$ errors is of the order of $\epsilon^k$, given that the single error has a
(small enough) probability $\epsilon$.
On the other hand, in the setting of memoryless channels each qubit
independently interacts with its own environment, and different environments
do not interact among themselves.
While memoryless channels give rise to independent errors, the converse is not
necessarily true. As a matter of fact, there are situations where
although qubits do not interact independently with their environments, the generated errors
still satisfy the independence condition --- provided that qubits do not directly interact with
each other \cite{hwang01,ABF09}.
Thus, in such cases standard QECCs work well enough.}


On the other hand, a suitable strategy to deal with completely correlated errors is represented
by Noiseless Codes, also known as Decoherence Free Subspaces (DFS).
This is a `passive' quantum error correction method where
the key idea is that of avoiding decoherence by encoding quantum information
into special subspaces that are protected from the interaction with the
environment by virtue of some specific dynamical symmetry \cite{PSE96,DG97,ZR97,LCW98,lidar03}.

It turns out that a subspace $\mathcal{Q}$ is a DFS if and only if all Kraus operators, when restricted to $\mathcal{Q}$,
are equal, up to a multiplicative constant, to a given unitary transformation $U_{Q}$.
In the case of imperfect initialization, i.e., state not initialized inside a DFS,
in a suitable basis the Kraus operators are described by a matrix of the form
\begin{equation}\label{3}
{\sf K}_{\sf k}=\left(
\begin{array}{cc}
s_{\sf k}{\sf U}_{Q} & 0 \\
0 & {\sf M}_{\sf k}
\end{array}
\right) \, ,
\end{equation}
where ${\sf M}_{\sf k}$ is an arbitrary matrix that acts on the orthogonal complement $\mathcal{Q}^{\perp}$ 
and may cause decoherence there \cite{SL05}.
Equation (\ref{3}) implies
\begin{equation}\label{2}
{\sf K}_{\sf k}^{\dagger }{\sf K}_{\sf l}=\left(
\begin{array}{cc}
{\sf S} _{\sf kl}{\sf I}_{\mathcal{Q}} & 0 \\
0 & {\sf M}_{\sf k}^{\dagger }{\sf M}_{\sf l}
\end{array}
\right) \, ,
\end{equation}
where ${\sf S} _{\sf kl} = \bar{s}_{\sf k} s_{\sf l}$.
Applying condition \eqref{ecc} to the present setting, it follows that DFS can be viewed as
a special class of QECCs, where upon restriction to the code space $\mathcal{Q}$,
all Kraus operators are proportional to the unitary $U_{Q}$.
It is worth noticing that in the DFSs case the matrix ${\sf S}$ has rank $1$.
Hence, a DFS is an example of maximally degenerate quantum error correcting code.
In the case of perfect DFS encoding,
the necessary and sufficiency conditions are less restrictive than Eq.\ (\ref{3}) -- see \cite{SL05}.

An example of the effective application of DFS encoding can be obtained for the case of
the `completely correlated' channel (see Sec.\ \ref{sec:long}).
By putting $n=2$, ${K}_{0}=\openone$, ${K}_{1}=\sigma_z $, and $p_0=1-p$, $p_1=p$ in Eq.\ \eqref{AMarkov}
one obtains a quantum channel with Kraus operators
\begin{eqnarray}
K_{00} = \sqrt{p} \openone \otimes \openone \, , && K_{01} = 0 \, , \nonumber \\
K_{10} = 0 \, ,                    && K_{11} = \sqrt{(1-p)} \; \sigma_z \otimes \sigma_z \, . \label{1a}
\end{eqnarray}
A DFS is given by $\operatorname{span}\{|01\rangle,|10\rangle\}$ where one can safely encode a qubit
\begin{equation}\label{pac}
\left\vert 0_{L}\right\rangle = \left\vert 01\right\rangle \, , \quad\left\vert 1_{L}\right\rangle =\left\vert 10\right\rangle \, .
\end{equation}
By extending this argument, one can say that in the case of completely correlated errors
it is possible to exploit the invariance of a subspace to encode information reliably.

In Ref.\ \cite{PV10}, it has been provided a generalized quantum
Hamming bound for nondegenerate codes, which depends
on the rank of the CJ state (see Sec.\ \ref{sec:qcbo}) associated to the noise process
and holds for any kind of (possibly correlated) channel model.
The original Hamming bound \eqref{qHb}, which was formulated for the
case of independent noise on the encoding systems is then recovered as a
particular case.
On the other hand, for completely correlated noise it has been shown how
to exploit degeneracy to violate the generalized quantum
Hamming bound and achieve perfect quantum error correction
with fewer resources than those needed for non-degenerate codes.
As an example consider the following channel
\begin{eqnarray}
\Phi(\rho^{(n)}_Q)=p\; \rho+\sum_{i=1,\ldots,n,j>i} \nonumber
\left(p_{X,ij}\; \sigma_{x,i}\sigma_{x,j}\; \rho^{(n)}_Q \;\sigma_{x,i}\sigma_{x,j}  \right. \\
\quad \left. +\; p_{Y,ij} \; \sigma_{y,i}\sigma_{y,j}\; \rho \;\sigma_{y,i}\sigma_{y,j}
+p_{Z,ij} \; \sigma_{z,i}\sigma_{z,j}\; \rho \;\sigma_{z,i}\sigma_{z,j}
\right),\nonumber\\
\label{PhiPauli}
\end{eqnarray}
where the input state is left unchanged with probability $p = 1-\sum_{i=1,\ldots,n,j>i}(p_{X,ij}
+p_{Y,ij} +p_{Z,ij})$, while it undergoes Pauli errors $\sigma_x$,
$\sigma_y$ and $\sigma_z$ on qubits $i$ and $j$ with probabilities
$p_{X,ij} $, $p_{X,ij} $ and $p_{Z,ij} $ respectively.
By evaluating the rank of the CJ state associated with the map \eqref{PhiPauli} one obtains
the generalized quantum Hamming bound of Ref.\ \cite{PV10}:
\begin{equation}
2^k \left[1 + 3 {n \choose 2} \right] \leq 2^n \, .
\label{qhbsimple}
\end{equation}
Then, by considering for instance $k = 1$ one gets $n=7$ as smallest integer
satisfying the bound. However,
one can also construct codes with lower
values for $n$.
For instance, this is the case of the code
\begin{equation}
|0_L\rangle = |000\rangle,\quad  |1_L\rangle = |111\rangle .
\label{repcod}
\end{equation}
Now notice that these basis codewords are not affected by the action of
$\sigma_z$ on any pair of qubits. Consequently
the action of $\sigma_x$ on a pair of qubits is identical to the action of
$\sigma_y$ on the same pair of qubits. In other words
the code is degenerate.
Therefore, one has only to
correct errors due to $\sigma_x$ operators. This can be realized through
a projective measurement onto the subspaces
${\cal S}_{00} = span\{|000\rangle , |111\rangle\}$,
${\cal S}_{01} = span\{|100\rangle , |011\rangle\}$,
${\cal S}_{10} = span\{|010\rangle , |101\rangle\}$ and
${\cal S}_{11} = span\{|001\rangle , |110\rangle\}$.
If the measurement outcome is "$00$", no errors have affected the qubits;
on the contrary, if the measurement outcome is "$01$", errors have affected qubits 2 and 3
and can be corrected by applying there $\sigma_x$.
Similarly, all the other possible errors can be detected and corrected.
It is hence clear that this code violates the quantum Hamming bound \eqref{qhbsimple}
thanks to the invariance of the coding subspace under the action
of pair of $\sigma_z$ which allows for perfect error correction.

Having seen that DFSs are suitable to encode information in the presence of completely correlated errors,
it is natural to expect that their performance decreases by reducing the degree of errors' correlation \cite{DKB07}.
Actually Refs.\ \cite{CM10a} and \ \cite{CM10c} confirm this fact for
the Markovian model of \eqref{AMarkov} and for the model of \eqref{Amu01} respectively.
Then, one may argue that for the memory channel models \eqref{AMarkov}, \eqref{Amu01},
where the memory effects are described by a single parameter $\mu$,
there must be a threshold value $\mu^\star$ that allows one to select the best code between the
standard and the noiseless ones \cite{CM10a,CM10c,arrigo08}.

\subsection{Designing quantum codes for correlated errors}

The specific features of error models can be used to design new quantum codes that better cope with correlated errors.

The results of the previous Subsection suggest that it might be convenient to concatenate
decoherence-free subspaces with standard quantum error correcting codes in
order to achieve higher entanglement fidelity values in both low and high
correlations regimes.
This kind of concatenation was first introduced in Ref.\ \cite{lidar99}, and
it was investigated in the context of memory channels in Ref.\ \cite{clemens04} and
subsequently in Ref.\ \cite{CM10c}.

As an illustrative example,
consider to encode one logical qubit into a decoherence free subspace ($\mathcal{Q}_{DFS}=\mathcal{Q}_{\text{outer}}$)
spanned by the basis codewords
\begin{equation}
\left\vert 0_{L}\right\rangle = |+-\rangle,\quad\left\vert 1_{L}\right\rangle =|-+\rangle,
\end{equation}
and then encode each qubit of this basis into a
three-qubit bit repetition code (\ref{repcod}) ($\mathcal{Q}_{\text{%
bit}}=\mathcal{Q}_{\text{inner}}$).
One obtains that the basis codewords of the
concatenated code $\mathcal{Q}=\mathcal{Q}_{\text{bit}} \circ \mathcal{Q}_{DFS}$
are given by,
\begin{eqnarray}
\left\vert 0_{L}\right\rangle &=&\frac{1}{2}\left( \left\vert
000000\right\rangle -\left\vert 000111\right\rangle +\left\vert
111000\right\rangle -\left\vert 111111\right\rangle \right) \text{,}  \notag
\\
&&  \notag \\
\left\vert 1_{L}\right\rangle &=&\frac{1}{2}\left( \left\vert
000000\right\rangle +\left\vert 000111\right\rangle -\left\vert
111000\right\rangle -\left\vert 111111\right\rangle \right) \text{.}\notag
\label{CW}
\end{eqnarray}
\begin{figure}
\centering
\includegraphics[scale=0.39]{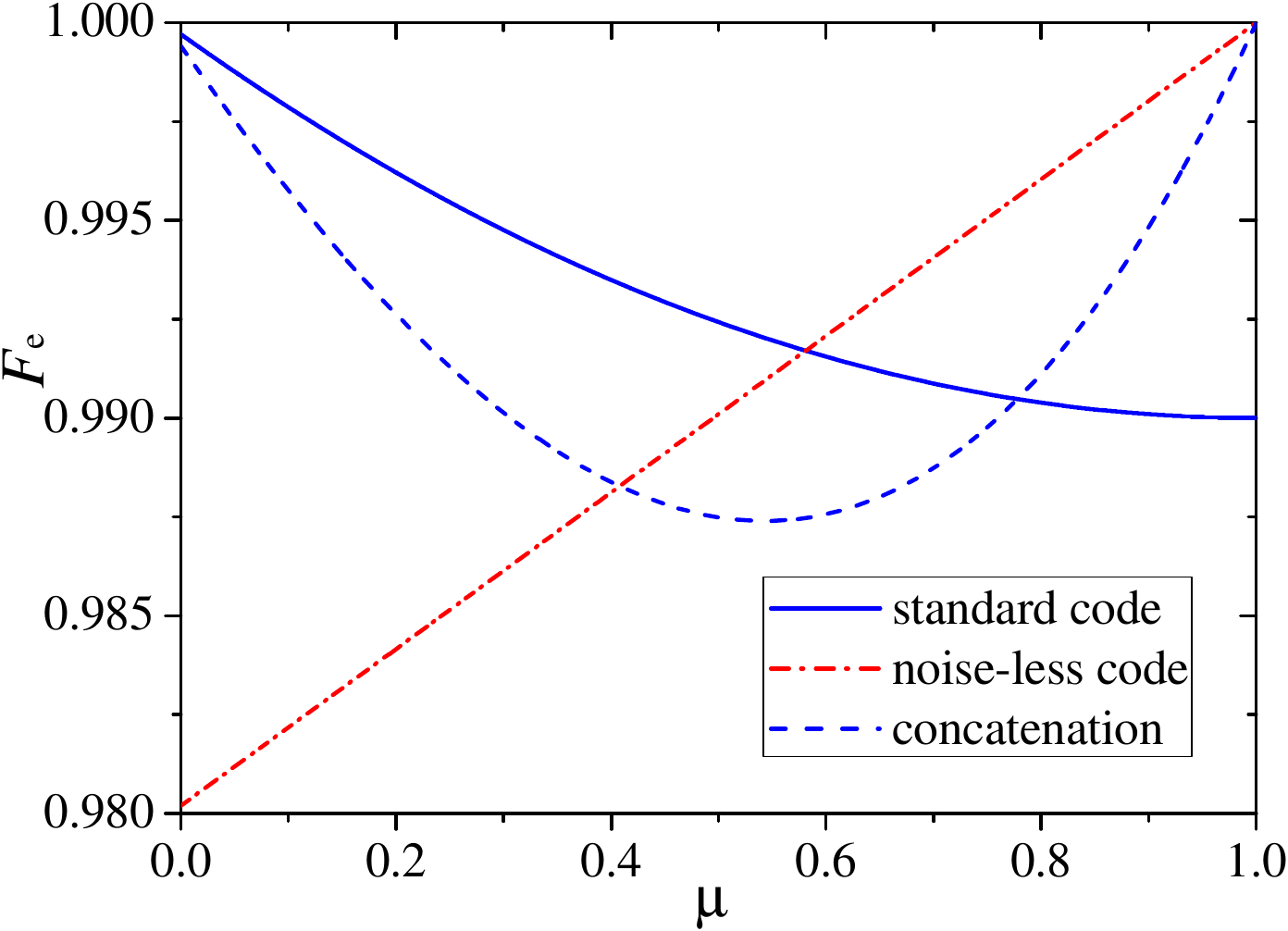}\quad\includegraphics[scale=0.39]{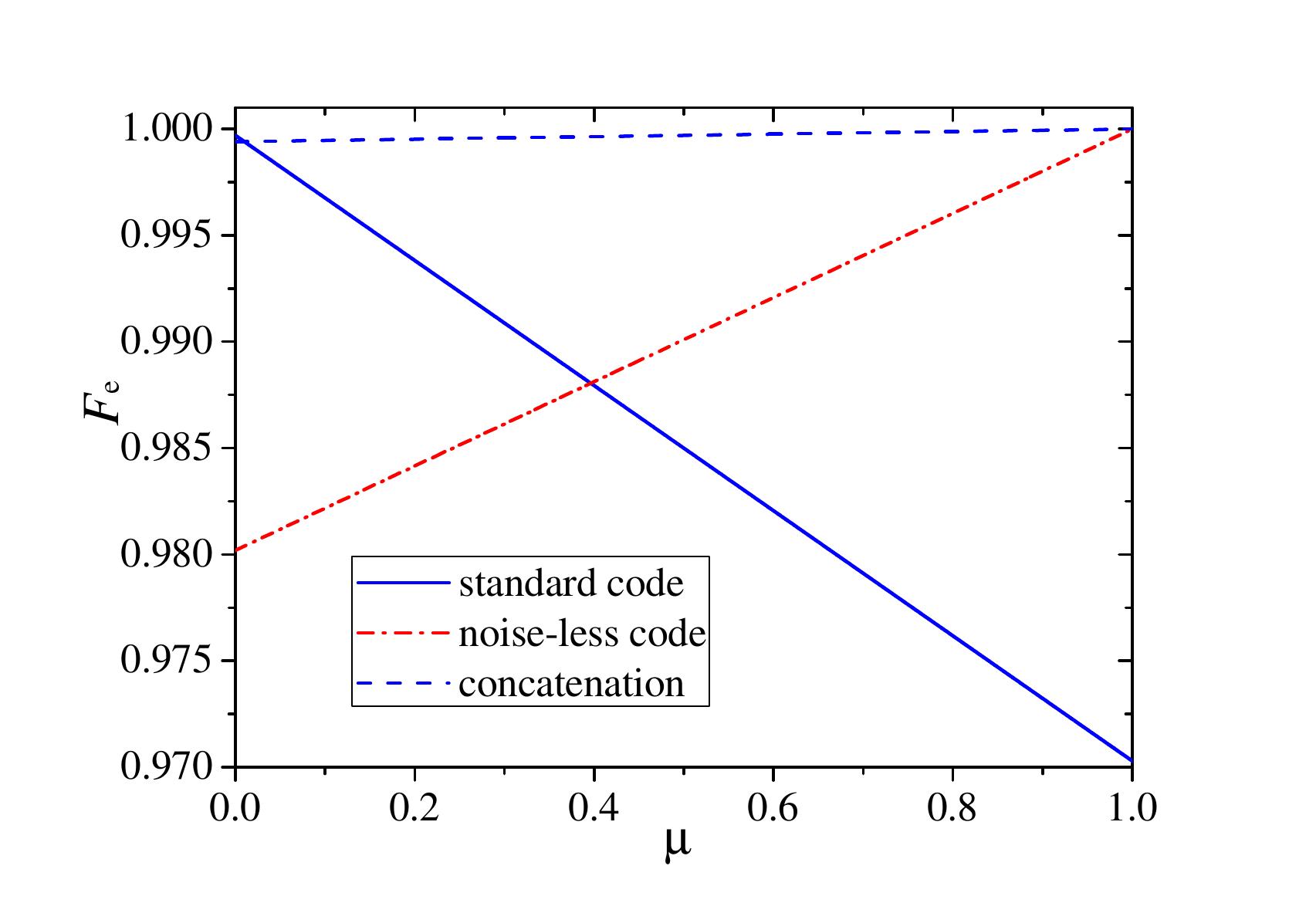}
\caption{Entanglement fidelity for the model \eqref{AMarkov} top figure
and \eqref{Amu01} bottom figure with $p=10^{-2}$. 
Solid lines correspond to standard code. 
Dot-dashed lines refer to noiseless code. Dashed lines represent their concatenation.
} \label{figqecc}
\end{figure}
The entanglement fidelity for the concatenation of a repetition code and a noiseless code for the
models of \eqref{AMarkov} and \eqref{Amu01} with ${K}_{0}=\openone$, ${K}_{1}=\sigma_x$, $p_0=1-p$ and $p_1=p$ is reported in Fig.\ \ref{figqecc}.
It turns out that in the first case the concatenated code does not work well for partially correlated errors.
It is always better to use either the outer or the inner code alone depending on whether one
is below or above the threshold value $\mu^\star(p)$. On the contrary, in the second case
the concatenated code works optimally almost everywhere.
Hence, one may argue that for the model of \eqref{Amu01} the concatenation procedure is
particularly advantageous in the presence of partially correlated errors.

Another error model often employed is that of burst errors.
Such errors can be considered as affecting a sequence of qubits
as opposed to random single qubit errors.
They are well studied in the classical
framework where corresponding error correcting codes have been developed \cite{PW72}.
In \cite{VRA97}, quantum analog of burst-error correcting codes have been considered.
Hamming and Gilbert-Varshamov type bounds have been derived showing that
these codes are more efficient than codes protecting against random errors.
In fact, to protect against burst errors of width $b$
(that is, errors occurring on a number $b$ of consecutive qubits with $b$ a fixed constant), it is enough to map
$n-\log_2 n-O(b)$ qubits to $n$ qubits, while in the case of $t$ random errors at least $n - t \log n$
qubits should be mapped to $n$ qubits.

A linear code $\mathcal{C}$ has
burst-correcting ability $b$ iff, for every burst ${\sf w}_1$ and
${\sf w}_2$ of width $\leq b$ it is ${\sf H}({\sf w}_1+{\sf w}_2)^\top\neq 0$, 
with ${\sf H}$ the parity check matrix of $\mathcal{C}$.
In \cite{VRA97}, an explicit construction of quantum codes for correcting
burst errors has been presented starting from classical binary cyclic codes.
A classical binary cyclic code $\mathcal{C}$ is such that
if $(c^{(1)},c^{(2)},\ldots ,c^{(n)})$ is in $\mathcal{C}$, then so is
$(c^{(n)},c^{(1)},\ldots , c^{(n-1)})$ \cite{H85}.

The definition of quantum burst-correcting codes straightforwardly follows from \eqref{everelation}. Consider
the set $\mathfrak{Q}$ of quantum errors (hence the corresponding set $\mathfrak{C}$ of classical error)
such that both
\begin{eqnarray}
{{\mathfrak C}}_X&=&
\left\{ {\sf e}_X  \in {\mathbb F}_2^n \;|\; \exists {\sf e}_Z  \in{\mathbb F}_2^n\Rightarrow ({\sf e}_X | {\sf e}_Z) \in {{\mathfrak C}} \right\},\\
{{\mathfrak C}}_Z&=&
\left\{ {\sf e}_Z  \in {\mathbb F}_2^n \;|\; \exists {\sf e}_X  \in{\mathbb F}_2^n\Rightarrow ({\sf e}_X | {\sf e}_Z) \in {{\mathfrak C}} \right\},
\end{eqnarray}
are bursts of width $\leq b$. Then any quantum code $\mathcal{Q}$
having $\mathfrak{Q}$-correcting ability is called a $b$-burst quantum correcting code.

Now, suppose to have a $(3b+1)$-burst-correcting binary $[n.k]$ cyclic code $\mathcal{C}$
 then, to construct the $b$-burst-correcting $[[n,k]]$ quantum code one can proceed as follows.

Let the $(n-k)\times n$ matrix $\tilde{\sf H}$ be a parity check matrix
for the classical code $\mathcal{C}$. Let $\tilde{\sf H}_{\rightarrow m}$ denote the matrix that is obtained from
$\tilde{\sf H}$ by cyclically shifting the columns $m$ times to the right.
Since $\mathcal{C}$ is cyclic, also $\tilde{\sf H}_{\rightarrow m}$ is a parity check matrix of $\mathcal{C}$.
Consider the stabilizer quantum code $[[n,k]]$ defined by the parity check matrix (see \ref{Hstab})
\begin{equation}
{\sf H}= \left( \tilde{\sf H}+\tilde{\sf H}_{\rightarrow b}\, |\, \tilde{\sf H}+\tilde{\sf H}_{\rightarrow 2b+1}\right ).
\end{equation}

Let ${\sf e}=({\sf e}_X |{\sf e}_Z)$ and ${\sf e}'=({\sf e}'_X| {\sf e}'_Z)$ be bursts of width $\leq
b$, with ${\sf e}\neq {\sf e}'$ and take
\begin{equation}
{\sf w}={\sf e}_X+{\sf e}'_X+({\sf e}_X+{\sf e}'_X)_{\rightarrow b}+{\sf e}_Z+{\sf e}'_Z+
({\sf e}_Z+{\sf e}'_Z)_{\rightarrow 2b+1},
\end{equation}
where ${\sf e}_{\rightarrow b}$ denotes the
vector obtained by cyclically shifting ${\sf e}$ to the right $b$ times.
Then, it is easy to check that ${\sf w}\neq 0$ and ${\sf w}$ is the sum of two
bursts of width $\leq 3b+1$. Hence ${\sf w}\not\in{\mathcal C}$ and
\begin{equation}
{\sf H} ({\sf e}+{\sf e}')^{\top}=\tilde{\sf H} {\sf w}^{\top}\neq 0,
\end{equation}
which guarantees the ability of the quantum code $[[n,k]]$ to correct $b$-burst errors.

The existence of classical $(3b+1)$-burst-correcting binary cyclic codes 
with length $n=2^m-1$ and dimension $k=n-m-(3b+1)$ is known
with $m$ only depending on $b$ \cite{PW72}.
Hence these classical codes lead to almost optimal quantum codes
(compared with the bound $n-\log_2 n -O(b)$ given above).

\bigskip

By increasing the length of the bursts, one should increase the length of the burst code as well.
Alternatively it might be possible
to resort to the \emph{interleaving} technique.
By using this method,
the codewords can be distributed amongst the qubit stream so that consecutive words are never next
to each other. On de-interleaving they are returned to their original positions so that any errors that have
occurred become widespread. This ensures that any burst (long) errors now appear as random (short) errors.

Classically the interleaving of $m$ codewords $({\sf c}_{1}, {\sf c}_{2}, \cdots, {\sf c}_{m})$ of an $[n,k]$ code
is achieved  by permuting the positions of bits in codewords as follows
\begin{widetext}
\begin{eqnarray}
\left( {\sf c}_{1}, {\sf c}_{2}, \cdots, {\sf c}_{m} \right)
&=&
\left(
\left( {\sf c}_1^{(1)}, {\sf c}_1^{(2)}, \cdots, {\sf c}_1^{(n)} \right),
\left( {\sf c}_2^{(1)}, {\sf c}_2^{(2)}, \cdots, {\sf c}_2^{(n)}\right),
\cdots,\left( {\sf c}_m^{(1)}, {\sf c}_m^{(2)}, \cdots, {\sf c}_m^{(n)}\right)
\right)
\nonumber\\
&\to&
\left(\left({\sf c}_1^{(1)}, {\sf c}_2^{(1)}, \cdots, {\sf c}_m^{(1)}\right),
\left({\sf c}_1^{(2)}, {\sf c}_2^{(2)}, \cdots, {\sf c}_m^{(2)}\right),
\cdots,\left({\sf c}_1^{(n)}, {\sf c}_2^{(n)}, \cdots, {\sf c}_m^{(n)} \right)
\right).
\label{interl}
\end{eqnarray}
\end{widetext}
The procedure is equivalent to constructing the code as an $m\times n$ array where
every row is a codeword of the original $[n,k]$ code $({\sf c}_{1}, {\sf c}_{2}, \cdots, {\sf c}_{m})$.
Now, a burst of length $\le b m$ can have at most $b$ symbols in any row of this array.
Since each row can correct a burst of length $\le b$, the code can correct all bursts of length $\le bm$
(the parameter $m$ is the interleaving degree).

Therefore,
given an $[n,k]$ classical code correcting bursts of length $\le b$, then interleaving this code to the degree $m$ produces an $[nm,km]$ classical code correcting bursts of length $\le bm$ \cite{PW72}.

Moving to the quantum framework, in order to interleave quantum codes, one needs to exchange the qubits one by one, following \eqref{interl}.
Therefore, the basic step of the quantum interleaving simply is a swapping operation between two qubits.
Then, the classical result can be extended as follows \cite{K00}:
interleaving an $[[n,k]]$ quantum code correcting bursts of length $\le b$ to the degree $m$ produces an $[[nm,km]]$ quantum code correcting bursts of length $\le bm$.


\subsection{Convolutional Codes}

It is possible to extend the notion of stabilizer codes introduced in Sec. \ref{newsecQEC}
to codes which allow for an overlap between the
individual steps of the encoding operation \cite{C98,C99,OT03,OT04}.
From classical coding theory, codes with these properties are called convolutional codes.
Although not specifically designed for memory channels, they are intimately related to them.

A quantum convolutional stabilizer code is defined by the generators of its stabilizer group just like a block
stabilizer code (see Eq.\eqref{stabgen}).
Consider a convolutional code encoding $k$ logical qubits per $n$ physical qubits, such that  every block has an output of $n + m$ qubits. 
$n$ of those are output qubits, while $m$ are passed on to the next step.
Then, an $[[n, k]]$ $m$-convolutional stabilizer code is given by the Abelian stabilizer group
\begin{equation}
\mathcal{G} = span \{g_{j,i} = \openone^{\otimes j n}\otimes
g_{0,i}| 1\le i \le n-k, 0 \le j\},
\end{equation}
where $g_{0,i} \in\mathcal{P}_{n+m}$ and all $g_{i,j}$ are independent of each other.

Note that the total number of physical qubits (length of the code) is left unspecified. Actually it is useful to set it to infinity by considering a Pauli group ${\cal P}_{\infty}$ with elements defined on a semi-infinite chain of qubits, but acting nontrivially only on a bounded number of them (as it happens with quasilocal algebra, see Appendix B). Then the generators are considered to be padded from the right with identities $\openone$.

The structure of the stabilizer group
generators can be summarized, following \eqref{Hstab}, by a semi-infinite matrix
\begin{equation}
{\sf H} = \left(
\begin{picture}(170,65)
\put(0,15){\framebox(70,47){$\begin{array}{c} {\sf g}_{0,1} \\ \vdots \\ {\sf g}_{0,n-k} \end{array}$}}
\put(50,-35){\framebox(70,47){$\begin{array}{c} {\sf g}_{1,1}  \\ \vdots \\ \quad {\sf g}_{1,n-k} \end{array}$}}
\put(130,-55){\makebox(0,0)[lb]{$\ddots$}}
\multiput(70,-35)(0,4){12}{\line(0,1){2}}
\put(45,-20){\vector(0,1){30}}
\put(45,00){\vector(0,-1){30}}
\put(43,-26){\rotatebox{90.0}{\makebox(0,0)[lb]{$n-k$}}}
\put(55,-40){\vector(1,0){15}}
\put(65,-40){\vector(-1,0){15}}
\put(57,-48){\makebox(0,0)[lb]{$m$}}
\put(75,-40){\vector(1,0){45}}
\put(115,-40){\vector(-1,0){45}}
\put(93,-48){\makebox(0,0)[lb]{$n$}}
\end{picture}
\right) \label{eq:general_structure}
\end{equation}
Each line of the matrix represents one of the $g_{j,i}$ and each
column a different qubit. Thus, any given entry of ${\sf H}$ is a Pauli
matrix for the corresponding qubit and generator.  The rectangles
represent which qubits are potentially affected by the
action of the generators. Clearly, when $m = 0$ one has each block separately and
obtain a block code.

Actually, it is possible to use the invariance by $n$ qubit translation of the generators to find a
shorter description.
One defines the shift (delay) operator $D$ acting on any element $A\in\mathcal{P}_{\infty}$ by
\begin{equation}
D[A] = \openone^{\otimes n} \otimes A.
\end{equation}
Then, the generators of the code can be written as:
\begin{equation}
g_{j,i} = D^j[g_{0,i}], \ 0 \leq j, \ 1 \leq i \leq n-k.
\end{equation}
Using this, one only needs to consider the first $n-k$ generators. All the others are obtainable by repeated applications of $D$.

In addition to applying $D^j$ to an element of the Pauli
group $A$ (with bounded support), it is also possible to consider
a polynomial $P(D) = \sum_j \alpha_j D^j$ and apply it as
$P(D)[A] = \prod_j \alpha_j D^j[A]$.
That critically relies on the fact that all copies of $A$ shifted by $D^j$ commute (see e.g., \cite{OT04}).

\begin{figure}
\centering\includegraphics[width=2.5in]{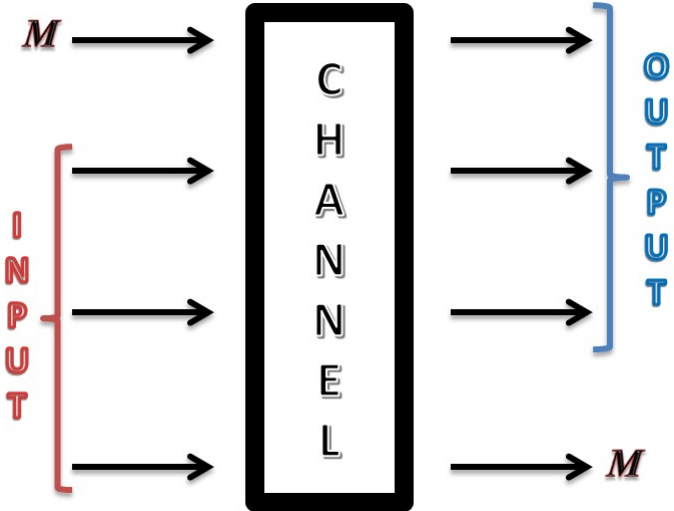}
\caption{Diagram of the channel encoding the $[[3,1]]$ $1$-convolutional stabilizer code.
The data qubit is the third on the input side. The first input qubit is the memory input, the last output qubit is the memory output. }\label{conv-fig}
 \end{figure}

It is worth noticing that the encoding operations for convolutional codes can be described as quantum
memory channels, due to the fact that some of the output qubits of the $n$th encoding step will be
used as inputs in the $(n + 1)$th step of the encoding, thus the blocks overlap. These qubits correspond to the
memory system of the memory channel describing the encoding map. The encoding operation is the same
in every step (neglecting initialization and finalization), thus every step is described by
the same channel (see Fig. \ref{conv-fig}).
This can be stated more precisely saying that
for every $[[n, k]]$ $m$-convolutional stabilizer code one can find an encoding operation
which is described by a the concatenation of a Weyl covariant memory channel with uni-modular characteristic function (see Section \ref{sec:weyl}).
The channel has $n$ input and output qubits and uses $m$ qubits of memory. One use of the channel corresponds to one
block in the encoding \cite{G10}.

Unfortunately convolutional codes also carry disadvantages. Because information is transmitted
from one block to the next, errors can spread as well. Depending on the encoding algorithm errors that spread without
bound on the output side can occur. These are called \textit{catastrophic} errors and have to
be avoided by employing non-catastrophic convolutional codes \cite{GR06}.
Necessary and sufficient conditions for an encoder to be non-catastrophic are provided by \cite{PTO09}
and the minimal amount of resources to satisfy them have been determined in \cite{HHW11}.
Furthermore, an attempt to relate such conditions to the property of strict forgetfulness of the memory channel representing 
the encoder has been made in \cite{G10}.
Lacking the boundaries between code blocks, convolutional codes exhibit the same ``continuous structure" 
as channels with memories. As such, they could result particularly suited to protect from correlated noise.



\section{Capacities of Quantum Channels}\label{sec:ctc}

A natural question that arises after having examined
correcting codes is what are the maximum communication rates achievable
in quantum channels. For classical channels the highest rate (number of bits per channel uses) of
reliable information transmission attainable via the
application of encoding and decoding error correcting procedures defines
the capacity \cite{CTbook,Gbook}.
In this context reliability refers to the requirement that the transferred messages have to be received
without possibility of misunderstanding, i.e., the  communication errors have to be removed by the selected
coding strategy. One speaks of zero-error capacity  when imposing this constraint for codes of finite length (i.e., codes which operate on a finite number of information carries or channel uses)~\cite{Sh56,KO98}.
  However in many cases of physical and technological interest, it is more reasonable and mathematically more convenient  to enforce
  such condition only in the asymptotic  limit of infinitely long messages. Under this paradigm in fact
  explicit expressions for the channel capacity are available as a function of the noise model which is tampering the communication line.
For instance in the case of a memoryless classical channel
characterized by  the conditional probability
$p(y|x)$ of producing
the output symbol $y$ when fed with input $x$, the
associated capacity can be expressed as~\cite{Sh48}
\begin{eqnarray} \label{shannoncap}
 C_{SH} = \max_{p(x)} I(X:Y)\;,
\end{eqnarray}
where  the maximization is performed over all probability distributions on $x$, and where ${I}(X:Y)$ is the corresponding  Shannon mutual information -- see Sec.~\ref{sec:entro}.
The proof  leading to (\ref{shannoncap}) relies on the notion of typical sequences~\cite{CTbook,Gbook} and it
does not provide an explicit recipe for determining the optimal coding and decoding strategies
(this is why error correcting codes and capacities are often treated as distinct
subjects with no exception for the quantum realm --- except few notable exceptions, e.g.\ {\it polar codes} \cite{Arik09}).
Yet Shannon's result establishes a fundamental  benchmark that is useful
to test the effectiveness of any coding procedure  -- an informal and clear
introduction to these topics can be found in~\cite{PRESKILL} or~\cite{GALINDO}.
Strong versions of the {\it converse}
Shannon theorem have been proved~\cite{WoBOOK,ARIMOTO}, which establish that if the rate 
of communication of a (memoryless) classical channel exceeds $C_{SH}$, then the error probability of any coding scheme converges 
to one in the limit of many channel uses.

The notion of capacity based on asymptotic reliability has also an important operational meaning
stated by the  {\it Reverse Shannon Theorem},
a {result which, strangely enough, was only formulated  and proved only recently
within the context of quantum communication~\cite{B99,B02,Cuff2008}.
According to it, for any classical noisy channel of capacity $C_{SH}$, if the sender and receiver share
an unlimited supply of random bits, an expected $n C_{SH} + o(n)$  uses of a noiseless binary channel are sufficient to exactly  simulate $n$ uses of the original channel.

As anticipated in Sec.~\ref{sec:intro}, the generalization of the above ideas to the quantum setting leads to the introduction
of a plethora of channel capacities, depending on whether classical or quantum information
has to be transmitted, and whether additional resources, as pre-shared entanglement, are exploited.
An unification of these quantities under a common formalism based on resource inequalities has been presented in~\cite{DEVEUNI04,DEVEUNI08,ABE09,ABE03,HSIEH10a,HSIEH10b}. 
The following sections will not report on this approach: instead
they shall focus on clarifying
the operational definitions of these quantities in a
framework which doesn't make explicit reference to the structure of the communication line. 
Then coding theorems will be reviewed, which allow one to express the capacities  in terms of suitable entropic quantities, starting
from the case of memoryless channels (the best understood and characterized so far) and then moving  to the more complex scenario of memory channels.

\begin{figure}[t]
\centering
\includegraphics[width=0.49\textwidth]{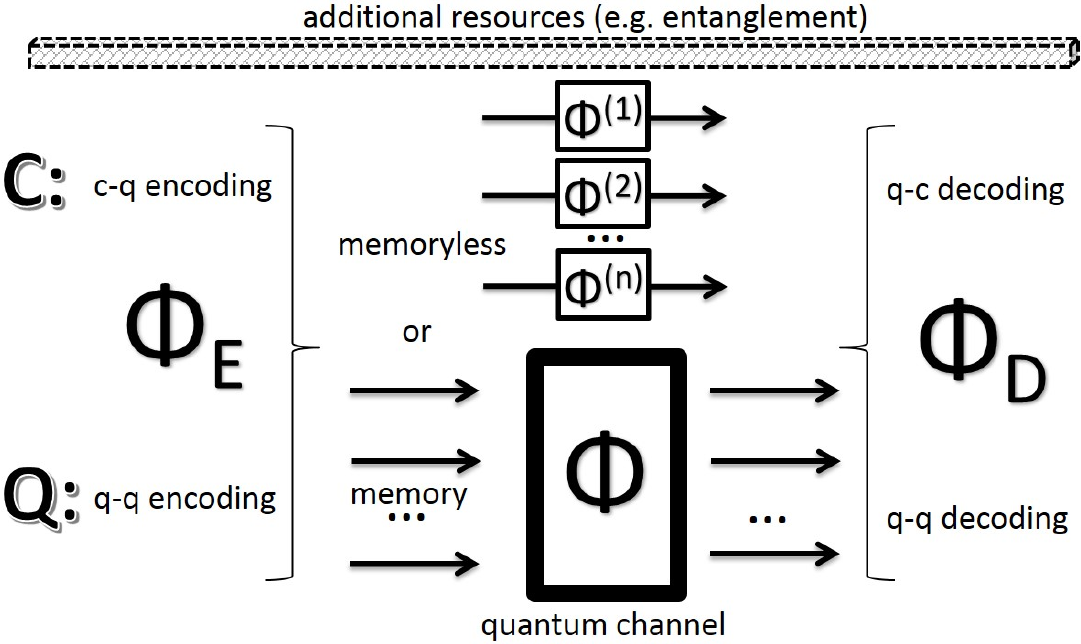}
\caption{Classical ($C$) and quantum ($Q$) capacities of a (memoryless or memory) quantum channel $\Phi$
in terms of all possible classical/quantum encoding and decoding schemes, with the potential use of additional resources as shared entanglement.}
\label{fig_capacities}
\end{figure}

\subsection{Operational definitions}\label{ssec:oper}

\subsubsection{Sending bits or qubits on a quantum channel}\label{sec:sendingbits}

The classical (resp. quantum) capacity $C$ (resp. $Q$)  of a quantum  channel defined
by the CPTP maps~$\Phi^{(n)}$ of Eq.~(\ref{nonanti})
 is the maximum  rate $R$ at which
classical (resp. quantum) information, encoded on a set of quantum carriers,
can be sent reliably
from the sender Alice  to the receiver Bob~\cite{Shor1995} -- see Fig. \ref{fig_capacities}.
 As in the classical setting~\cite{Sh48}
 the rate is measured as  the ratio $R=k/n$ among the number $k$ of bits (resp. qubits)
transmitted and the number $n$ of carriers employed  (the  ``redundancy''
of the code according to Sec.~\ref{sec:codes}).
Similarly the  reliability condition  is introduced by requiring
 that in the asymptotic
limit of $k\rightarrow \infty$ the error probability of the procedure can be made arbitrarily small
(or, equivalently, the fidelity of the transmission will approach unity),  while
keeping $R$ constant. In view of these operational definitions, $C$ and $Q$  can be
expressed   as the following limit~\cite{BS98,B02}
\begin{align}
 &\lim_{\epsilon \to 0} \limsup_{k \to \infty} \left\{ \frac{k}{n} : \;
 \exists\;  {\Phi_E^{(k \to n)},  \exists\;  \Phi_D^{(n \to k)}}\,,
 \right. \label{clascap}
\\  \nonumber
& \!  \left.
\min_{m\in {\cal M} } F \left(|m\rangle ;
\Phi_D^{(n \to k)} \circ {\Phi}^{(n)} \circ \Phi_E^{(k \to n)}\right)
 > 1-\epsilon \right\},  \nonumber
\end{align}
where, analogously to the notation introduced at the beginning of Sec.~\ref{sec:codes}
and as  sketched  in Fig.~\ref{fig_capacities},
$\Phi_E^{(k \to n)}$ and $\Phi_D^{(n \to k)}$ are respectively encoding and decoding channels
 mapping elements $|m\rangle$ from  a reference input space ${\cal M}$ (the {\it messages} Alice wishes to send to Bob) to states $\Phi_E^{(k \to n)} (m) := \Phi_E^{(k \to n)} (|m \rangle \langle m |)$  of $n$ carriers (the {\it codewords} of the
 procedure), and $F$ is  the input-output fidelity function introduced in Sec.~\ref{sec:fidecha}.
  In particular to fix the units properly,  the expression for $C$ is obtained by taking
 ${\cal M}$  to be a collection of $2^k$ orthogonal vectors which, without
  loss of generality  can be  identified  with the elements
$\{|0\rangle,|1\rangle\}^{\otimes k}$ of the computational basis of $k$ qubits.
 On the other hand,
for the quantum capacity $Q$ the set ${\cal M}$ coincides with  the whole ${\mathbb C}^{2 \otimes k}$ (as the latter includes
the elements of the canonical basis, it trivially follows that for a given communication line one has $Q \leq C$).
The limits in Eq.~(\ref{clascap}) is finally computed by first taking a supremum limit in $k\rightarrow \infty$ (which always exists) and
then sending the error parameter $\epsilon$ to zero to enforce the transmission
fidelity to  approach unity for all input messages. In the above expression
this is explicitly enforced by requiring $1-\epsilon$ to
lower bound the  minimum value achieved on ${\cal M}$ by the transmission fidelity.
Such rather strong requirement however can
be relaxed by replacing it with a similar constraint that only applies on the {\emph{average}} transmission fidelity:
this doesn't affect the limit in Eq.~(\ref{clascap}) and hence neither the definitions of
$C$ and~$Q$~\cite{KW04,Keyl2002}.
Similarly   the same value of   $Q$ one gets from~(\ref{clascap}) can also be obtained
by substituting  the minimum in $F$ with the entanglement   fidelity introduced in
Eq.~\eqref{FeFin}~\cite{BKN00,KW04}\footnote{
All the definitions introduced so far
assume a notion of capacity in which the error probability
is required to nullify  only in the asymptotic limit  of large enough $k$.
As in classical communication theory~\cite{Sh56,KO98}  however, a more stringent reliability requirement can be 
enforced, i.e., imposing that the min-fidelity on ${\cal M}$ should  equal $1$ for a finite number $n$ of channel  uses.
Under this condition one is led to the definition of {\it zero-error} classical and quantum capacities \cite{Medeiros05}.
This corresponds to the maximal communication rate achievable by {\it perfect codes}, as introduced in 
 Sec.\ref{sec:codes}.}.

Finally, one observes that the definition~(\ref{clascap})
yields a natural data-processing inequality for the capacities. For instance, given $C$ the classical capacity of a
quantum channel whose CPTP mapping $\Phi^{(n)} = {\Phi^{'}}^{(n)}\circ {\Phi^{''}}^{(n)}$
is obtained by concatenating other two CPTP maps, one has
\begin{eqnarray} \label{CAPCONC}
C \leq \min\{ C', C''\}\;,
\end{eqnarray}
where $C'$ and $C''$ are respectively the classical capacities of channels described by
${\Phi^{'}}^{(n)}$ and ${\Phi^{''}}^{(n)}$ 
(the same relation applying also for  the quantum capacities  $Q$, $Q'$ and $Q''$
as well as for  all the other capacities reviewed in the following sections with the notable exception of those discussed in
Sec.~\ref{concap} where the constraints may introduce spurious effects in the optimization, see e.g., Ref.~\cite{GM05}).
 The proof of this rather intuitive fact follows by observing that when passing from $C$ to
 $C'$ one can interpret ${\Phi^{''}}^{(n)}$ as part of the decoding procedure of map ${\Phi^{'}}^{(n)}$.
 Since the $C'$ is obtained by optimizing the transmission rate with
respect to {\it all} possible decodings, including those which do not use ${\Phi^{''}}^{(n)}$
as a preliminary stage, it follows that $C'$ is certainly not smaller than $C$.
Similarly
when passing from $C$ to $C''$, one can interpret ${\Phi^{'}}^{(n)}$ as part of
the encoding stage for ${\Phi^{''}}^{(n)}$: again since $C''$ is the optimal rate with
respect to all encoding maps one has that it is certainly not smaller than $C$.
An application of the above analysis to two channels $\Phi_1^{(n)}$ and $\Phi_2^{(n)}$
which are unitarily equivalent~(\ref{UNIEQ}) shows that they must  possess the same capacities: in this case indeed the CPTP concatenation which links the two maps can always be reversed, producing
both the inequality $C(\Phi_1^{(n)}) \leq C(\Phi_2^{(n)})$ and its counterpart $C(\Phi_2^{(n)}) \leq C(\Phi_1^{(n)})$.

\subsubsection{Capacities assisted by ancillary resources}  \label{sec:capass}

Entanglement is a fundamental resource in quantum information theory. In the context of quantum communication
this fact is testified by the teleportation~\cite{BetAL93} and super-dense coding~\cite{BW92} protocols.
The former  provides a nontrivial way of transmitting arbitrary quantum states when quantum carriers are not available 
but only bits can be exchanged  through a classical communication  line.
The latter instead, in the presence of noiseless quantum communication line, allows to send  2 bits of classical information
per transferred physical qubit.
The necessary additional resource for both procedures is a shared entangled state between the sender Alice and receiver Bob.
An application of these ideas to noisy communication lines introduces the notion of {\it entanglement-assisted} classical capacity
$C_{ea}$ (resp. quantum  capacity $Q_{ea}$) of a quantum channel $\Phi$~\cite{B02,B99}.
Operationally they are defined as the maximum rate of reliable transmission of classical 
(resp.~quantum) information when the sender and the receiver have at their disposal an 
unbounded number of pre-shared maximally entangled states as ancillary side resources. 
Formal expressions are hence obtained through the same limit given in Eq.~(\ref{clascap})
with the difference that now  the transformations $\Phi_E^{(k \to n)}$ and $\Phi_D^{(n \to k)}$
map elements of the reference input space ${\cal M}$ to and from the joint space associated with 
the $n$ carriers of the channel plus the local quantum memories where Alice and Bob are storing their
 prior entangled states (specifically $\Phi_E^{(k \to n)}$
 acts on Alice's memories, and  $\Phi_D^{(n \to k)}$
 on Bob's memories).
 While by definition $C_{ea}$ and $Q_{ea}$
 provide natural upper bounds for the unassisted
 counterparts $C$ and $Q$ respectively,
 a direct application of the teleportation and super-dense
  coding protocol shows that for any given quantum channel
   they are related by the identity
  $C_{ea} =2 Q_{ea}$~\cite{B99}.

Unlimited classical communication between Alice and Bob
is another example of an ancillary resource which is known
to  increase the ability of transferring arbitrary
quantum states over a quantum channel via the application of entanglement distillation protocols
-- admitting   only forward classical communication from Alice to Bob is instead of no use in this respect~\cite{BetAL96}.
This yields the notion of (two-way) classical assisted quantum capacity $Q_2$~\cite{B02,B99} 
which again can be formally  expressed as in Eq.~(\ref{clascap}) by replacing the concatenation 
$\Phi_D^{(n \to k)} \circ {\Phi}^{(n)} \circ \Phi_E^{(k \to n)}$ with an arbitrary LOCC process 
intermediated by the action of the channel (see Sec.\ \ref{sec:semicausal}).
Quantum capacity assisted by providing access to non trivial zero quantum capacity side channels
has been analyzed in~\cite{SSW08}.

Both the classical and quantum capacities of a channel can finally also be improved by allowing  a feedback communication line (either quantum or classical) which permits the receiver   Bob of the messages   to  signal at the sender Alice. 
Interestingly enough  it has been shown that using feed-back in the presence of prior shared entanglement is of no use, i.e.,
$C_{ea+FB}= C_{ea}$ and $Q_{ea+FB} = Q_{ea}$~\cite{Bowen04a,Bowen04b}.

A  partial ordering among  some of the  quantities introduced in this section has been presented in \cite{BetAL06}.

\subsubsection{Private classical capacity of a quantum channel} \label{sec:private1}

The private classical capacity $C_p$ of a quantum channel
is defined as the maximum rate at which
classical information can be transmitted  privately from the sender to the receiver.  Formally this is enforced by requiring that
a third party (Eve)  who has access to the channel environment and who is trying to recover Alice messages to Bob,
will get it with an error probability that is approaching unity  in asymptotic limit of infinitely long messages \cite{D05,Cai06}.
Again $C_p$ can be expressed as the limit in Eq.~(\ref{clascap}) by further constraining the coding and decoding procedures to satisfy an entropic inequality that implement
the privacy requirement. Specifically this is obtained by upper bounding  with  $\epsilon$  the Holevo information~(\ref{HOLEVOINFO})  of the complementary channel (\ref{cc})  of $\Phi^{(n)}$ and associated with the uniform ensemble
${\cal E} = \{ p_m=2^{-k} ,  \Phi_E^{(k \to n)}(m)\}_{m\in {\cal M}}$ generated by the
encoding mapping selected by Alice. As the complementary map is the transformation
that links  the channel inputs to the images they produce on the environment (see Sec.~\ref{DegPPTEB}),
this choice, via the Holevo bound~(\ref{HOLEVOBOUND}),
 ensures that the information Eve can recover on Alice's messages
vanishes when taking the limit $\epsilon \rightarrow 0$.
By construction $C_p$ is
always smaller or equal to the corresponding  $C$ and greater or equal to
$Q$, i.e.,
\begin{eqnarray}
C \geqslant C_p  \geqslant Q\;,
\end{eqnarray}
(the last inequality being associated with the fact that the
ability of sending  all vectors of ${\mathbb C}^{2 \otimes k}$ with unit fidelity ensures that no information on the transferred states is passing from Alice to the environment).

\subsubsection{Constrained capacities}\label{concap}

The operational definitions of  capacities can be
modified in order to account for possible constraints on the input (or output)
states on the channel.
For instance, three weaker versions of the classical capacity of a quantum channel have been identified~\cite{BS98}.
Specifically one defines the {\it product-state} (or {\it classical-quantum}, or {\it Holevo})
classical capacity $C_{cq}$ (which following a rather universal conventional hereafter will be indicated with
the symbol $C_1$) by requiring that the employed coding maps entering in Eq.~(\ref{clascap}) produce only separable 
codewords, that is, $\Phi_E^{(k \to n)} (m)$ is a separable state of the $n$ carriers for all messages $|m\rangle$ in  $\mathcal{M}$
[one notes incidentally that the analogous of $C_1$ for the classical private capacity $C_p$, i.e.,
the product-state private classical capacity $C_{p,1}$, has been defined in~\cite{D05} -- see Sec.~\ref{sec:private-theo}].
Similarly one defines a quantum-classical capacity $C_{qc}$ by leaving the encoding channel unconstrained but imposing  the decoding channels
$\Phi_D^{(n \to k)}$ to be LOCC with respect to the outputs of different uses of the channel.
Finally assuming LOCC operations for $\Phi_D^{(n \to k)}$   and  separability for the codewords  $\Phi_E^{(k \to n)} (m)$ one defines the
classical-classical capacity  $C_{cc}$.
The unconstrained capacity  $C$ (often identified
in this context also as the quantum-quantum or
$C_{qq}$ capacity)  is a natural
(and in general strict~\cite{Has09}) upper bounds for the others. Similarly, for any assigned quantum channel, $C_{cc}$ is a natural lower bound
for $C_{cq}$ and $C_{qc}$, the ordering between
the last two being at present unknown.

Of special interest are also a class of physically motivated constrained
capacities obtained by introducing a family of observables $\{ A^{(n)} \}_{n=1,\dots ,\infty}$ and
by imposing that for any $n$ the mean value of $A^{(n)}$ is bounded on the ensemble of states at the input
of $n$ uses of the quantum channel.
The capacity of a quantum channel under such a constraint can be defined as in
Eq.\ (\ref{clascap}) under the additional requirement that for any $k$ and $n$
\begin{equation}
\mathrm{Tr} \left( A^{(n)} \rho^{(n)} \right) \leq a \, ,
\end{equation}
with
$\rho^{(n)} = \Phi_E^{k \to n}(\openone/2^{k})$,
 $\openone/2^{k}$ being the average state over the set ${\cal M}$.
In particular, a relevant role is played by additive observables, for which
one can put $A^{(n)} = n^{-1} \sum_{k=1}^n A_k$, where $A_k \equiv A$ is the observable
for a single input quantum system at the input of the $k$-th use of the channel.

The notion of constrained capacity naturally applies in the context of CV channels (see Sec.\ \ref{CV-qc}).
Indeed, due to the fact that the carrier Hilbert space is infinite-dimensional
it turns out that the capacity of a CV channel can be infinite \cite{HW01}.
As a matter of fact, an infinite value for the capacity corresponds to
the encoding of information into larger and larger sectors of the Hilbert space.
Clearly, that is in contradiction with the finiteness of the resources
employed in physical realizations, e.g., the finiteness of the mean energy.
It is hence meaningful to introduce a notion of capacity under a physically
motivated constraint.
Most natural choices are to impose a constraint on the mean value of
the energy or the number of bosonic excitation per mode.
In the latter case one has $A_n = n^{-1} \sum_{k=1}^n a_k^\dag a_k$, where
$\{ a_k, a_k^\dag \}$ are the canonical ladder operators at the channel input.
From a technical point of view, these choices, besides being physically sound,
guarantee that the set of states satisfying the constraint form a compact set.
This is a crucial feature to ensure that the coding theorems for Gaussian channels
under constrained mean excitation number (or energy) yield expressions formally analogous to the unconstrained case,
with the optimization being performed over input ensembles satisfying the constraint \cite{Holevo03,Holevo05,Holevo97, Holevo06}. Since the excitation number and the energy are quadratic in the canonical variables, their mean values can be expressed in terms of the first and second moments of the characteristic function of the
input states, see Sec.\ \ref{CV-qc}.
In particular, the condition of having no more that $N$ mean excitations per mode
is expressed in terms of the first and second moments
\begin{equation}\label{eq:enconstrcov}
\frac{\mathrm{Tr}(\sf{C}^{(n)})+|\sf{m}|^2}{2n} \leq N+\frac{1}{2} \, .
\end{equation}

\subsubsection{A superoperator norm approach to quantum capacities}

The limit which defines $C$ and $Q$ in  Eq.~(\ref{clascap}) indicates
that for sufficiently large $k$
there exists $\Phi_D^{(n \to k)}$ and $ \Phi_E^{(k \to n)}$
which makes
$\Phi_D^{(n \to k)} \circ {\Phi}^{(n)} \circ \Phi_E^{(k \to n)}$
close to the identity transformation $\text{id}_{\cal M}$ on ${\cal M}$.
Specifically for  the quantum capacity $Q$,
$\text{id}_{\cal M}$ is the identity super-operator on
 ${\mathbb C}^{2 \otimes k}$, while  for the classical capacity $C$, the map
 $\text{id}_{\cal M}$  is the fully dephasing channel on  ${\mathbb C}^{2 \otimes k}$ 
 which leaves the elements of its computational basis $\{|0\rangle,|1\rangle\}^{\otimes k}$  unchanged.

Based on this observation a definition of
capacities which is fully equivalent
to the approach of  Sec.~\ref{sec:sendingbits}
can be  given in terms of the cb-norm superoperator distance defined in Sec.~\ref{sec:dist2}.
In this approach~\cite{Keyl2002,KretschmannTH}, a positive quantity $R$ is said to be an  achievable rate for the channel $\Phi$
if for all sequence $\{k_i,n_i\}_{i \in  \mathbb{N}}$ with $\lim_{i\rightarrow\infty} k_i = \infty$ and $\limsup_{i\rightarrow\infty} k_i/n_i < R$ one has
\begin{eqnarray}
&&\lim_{i\rightarrow\infty} \inf_{\Phi_D,\Phi_E} |||
\Phi_E^{*(k_i \to n_i)} \circ \Phi^{*(n_i)} \circ{\Phi_D}^{*(n_i \to k_i)}\nonumber\\
&& \qquad \qquad\qquad \qquad \quad  -\text{id}_{\cal M}^*
|||_{cb}=0 \;,
\label{cdNORMcap}
\end{eqnarray}
where $\Phi_E^{*(k_i \to n_i)}$,  $\Phi^{*(n_i)}$ and ${\Phi_D}^{*(n_i \to k_i)}$ are the
duals~(\ref{equazioneHEIS}) of the the maps $\Phi_E^{(k_i \to n_i)}$,  $\Phi^{(n_i)}$ and ${\Phi_D}^{(n_i \to k_i)}$
defined in~(\ref{clascap}), while $\text{id}_{\cal M}^*$ is the dual of the identity map on ${\cal M}$.
With this prescription the values of $Q$ and $C$,
are then identified as the supremum of the corresponding
achievable rates. A similar construction was presented also in Refs.~\cite{KW04,HW01}:
here however  the cb-norm distance was used directly in the Schr\"{o}dinger channel representation.

\subsection{Coding theorems for memoryless channels}\label{ssec:codingth}

Coding theorems provide expressions for the communication capacities of memoryless quantum channels~(\ref{mp2}) in terms
of suitable entropic functions of the input and output states of the channel. While referring the reader to~\cite{Wbook13,
HOLEVObook12,winterthe} for a detailed review of the subject,
here are reported the main results concerning the entanglement-assisted classical capacity, the classical capacity
and its product-state version, the private classical capacity, and the quantum capacity.
Limitations and applicability of these expressions for the case of memory channels will be discussed in the following sections.

\subsubsection{The Holevo-Schumacher-Westermoreland coding theorem}

Preliminary attempts to compute the classical capacity of quantum channels were
presented in Ref.~\cite{Haus95,Haus96}.
A closed expression for the product-state classical capacity introduced in Sec.~\ref{concap}
was finally  provided by the Holevo-Schumacher-Westermoreland (HSW) coding theorem \cite{Holevo98,SW97}. 
It mimics the Shannon formula~(\ref{shannoncap}) by establishing that
for a memoryless channel~$\Phi$, $C_1(\Phi)$ can be
expressed in terms of a maximization of the associated
output Holevo information~(\ref{HOLEVOINFO}) over the set of input state ensembles ${\cal E} = \{ p_j, \rho_j\}$
(possibly satisfying some additional input constraints), i.e.,
\begin{equation}\label{oneshotunassist}
C_{1}(\Phi) = \max_{\cal E} \chi({\cal E}; \Phi)
\end{equation}
(owing to the concavity of von Neumann entropy~\cite{PETZBOOK,W78}, the maximization can in fact be
always restricted to ensemble of pure states).
Besides the original derivations~\cite{Holevo98,SW97}, several independent proofs of Eq.~(\ref{oneshotunassist})
are known~\cite{Holevo04,winter,HN03,lloyd11,GiovannettiPRA85,hayashi09,hayashi07,Oga99,Oga02,DD07,Sen2011}.
As is typical with many coding theorems the general argument
beyond the HSW result consists of two parts: {\em i)} an inequality which
establishes that the rhs of Eq.~(\ref{oneshotunassist})
is an upper bound for the channel capacity
({\emph{converse}} part of the theorem);
and {\em ii)}   a {\emph{direct}} part
which proves the existence of a coding
procedure that saturates such bound
asymptotically in the length of code.
Part {\em i)} can be established via the {\it classical} Fano inequality~\cite{CTbook} (relating the average information lost in a classical noisy channel to the error transmission probability) and  the  Holevo
bound inequality~\cite{Holevo73a,Holevo73b} on the achievable information of a quantum source  -- see Eq.~(\ref{HOLEVOBOUND}). 
A sketch of this proof can be found in Appendix~\ref{a:DERIVO}.
This approach is sufficient to show that any rate exceeding the capacity will necessarily produce an error
probability which is non zero even in the limit
of infinitely many channel uses.
It worth noticing however that, in contrast with classical information theory, establishing a 
{\it strong version} of the converse part of the theorem (i.e., proving that the transmission error probability 
will necessarily reach 1  as soon the rate exceeds the capacity threshold)
is particularly demanding in the quantum setting. As a matter of fact, strong converse coding theorems
have been derived only for limited classes of finite dimensional channels~\cite{KONIGWEHNER09,winter,Oga99,WWYnew13}  while  explicit
counter-examples have been provided which show
that in general they do not apply when considering continuous variable systems ~\cite{winterwilde}.

The {\it direct} part of the coding theorem which yields to
Eq.~(\ref{oneshotunassist})  is based instead on the notion of typical subspace for quantum sources~\cite{Ohya93,schumacher95}.
From this it  follows that given an
ensemble  ${\cal E} = \{ p_j, \rho_j\}$ and an  integer number $N$
 fulfilling the condition
$N\leqslant 2^{n \chi({\cal E}; \Phi)}$,
one can
identify  $N$ codewords $\rho_1^{(n)}$, $\cdots$,
$\rho_N^{(n)}$  of the form
$\rho_j^{(n)} = \rho_{j_1} \otimes \rho_{j_2} \otimes \dots \rho_{j_n}$ and an associated POVM that allows Bob to discriminate among the output counterparts
of the $\rho_j^{(n)}$ (i.e., the density matrices
$\Phi(\rho_{j_1}) \otimes \Phi(\rho_{j_2}) \otimes \dots \Phi(\rho_{j_n})$) with an error probability that can be bounded below any assigned threshold by sending $n$ to infinity.
Accordingly one is thus lead to  coding-decoding
schemes which guarantee  faithful transfer of classical messages
at rates $R =\frac{\log_2 N}{n} \leqslant \chi({\cal E}; \Phi)$
(the upper limit being attainable for large enough $n$), which
approach~(\ref{oneshotunassist}) when taking the supremum over~${\cal E}$.

It is important to notice that the POVM  which comes with the proof of the HSW theorem is 
explicitly a joint one: this is the reason why the rhs of Eq.~(\ref{oneshotunassist}) coincides with
$C_1(\Phi)$ and not with  the $C_{cc}(\Phi)$ capacity of Sec.~\ref{concap}  (the latter indeed is the maximum rate
attainable when allowing only LOCC operations among the various channels outputs). 
As a matter of fact, closed-form expressions for $C_{cc}(\Phi)$ and  for $C_{qc}(\Phi)$  are at present still missing. 
An explicit formula  for the unconstrained capacity $C(\Phi)$ defined in Sec.~\ref{ssec:oper}  
instead can be obtained as a simple generalization of  Eq.~(\ref{oneshotunassist}).
This is done by adopting a block-coding strategy
which, for all $n$, allows one to represent the density matrices produced
by the (possibly non separable)  encoding maps $\Phi_E^{(k \to n)}$  of Eq.~(\ref{clascap}) as  tensor
states over blocks of channels uses on which $\Phi^{\otimes n}$ acts as a single carrier map with an associated
product-state capacity $C_1(\Phi^{\otimes n})$.
The resulting  capacity of $\Phi$ can then be obtained
by taking the limit over $n$ of the associated rates, i.e.,
 \begin{equation}\label{C-from-C1}
C(\Phi) = \lim_{n\to\infty}\ \frac{1}{n} C_1(\Phi^{\otimes n}).
\end{equation}

\subsubsection{The private classical capacity theorem}\label{sec:private-theo}

The capacity formula for the private classical capacity $C_p(\Phi)$
introduced in Sec.~\ref{sec:private1} was given  in Ref.~\cite{D05,Cai06}.
As for the HSW theorem discussed in the previous section
it is derived by first providing a closed expression for
its product-state version $C_{p,1}(\Phi)$(i.e., the private
classical capacity attainable by using only separable codewords), and then using block coding to compute $C_p(\Phi)$  via the identity
 \begin{equation}\label{Cp-from-Cp1}
C_p(\Phi) = \lim_{n\to\infty}\ \frac{1}{n} C_{p,1}(\Phi^{\otimes n})\;.
\end{equation}
Analogously to Eq.~(\ref{oneshotunassist}),
 $C_{p,1}(\Phi)$ is a functional of the
 Holevo information~(\ref{HOLEVOINFO}). In this case however one has
 \begin{equation} \label{capsec}
C_{p,1}(\Phi) =
\max_{{\cal E}} \left[ \chi({\cal E} ; \Phi) -
\chi( {\cal E}; \tilde{\Phi}
)] \right]\;,
\end{equation}
where
$\tilde{\Phi}$ is the complementary channel of $\Phi$ as defined in Sec. \ref{DegPPTEB}.

 The converse part of the proof which leads to Eq.~(\ref{capsec}) is obtained by joining   the
 classical Fano inequality and  the Holevo Bound at the output of the map $\Phi$ with
  the privacy
requirement imposed on the Holevo information of the complementary channel~$\tilde{\Phi}$
of $\Phi$.
Vice-versa the direct part of the coding theorem is based
on the fact that, using a typical subspace argument, for each $\epsilon>0$ and for each ensemble ${\cal E}=\{ p_j , \rho_j\}$ satisfying
$\chi({\cal E} ; \Phi) >  \chi( {\cal E}; \tilde{\Phi})$,
 one can  identify up to $N\simeq  {2^{n \chi({\cal E} ; \Phi)
}}$ states of the form $\rho_{j_1} \otimes \rho_{j_2} \otimes \cdots \otimes \rho_{j_N}$  which,
when organized in groups of $M\simeq  {2^{n \chi({\cal E} ; \tilde{\Phi})}}$ elements each,
{\it i)} can be faithfully discriminated on  Bob side,  
{\it ii)} fully overlap  on  Eve side.
Accordingly by assigning to each of such group the {\em same} classical message, Alice can
now transfer up to $N/M \simeq 2^{n [ \chi({\cal E} ; {\Phi}) - \chi({\cal E} ; \tilde{\Phi})]}$ 
distinct messages to Bob without Eve being able to read them.

\subsubsection{The quantum capacity theorem}

The expression for the quantum capacity $Q(\Phi)$ of a memoryless channel $\Phi$ is 
\cite{lloyd97,D05,Shorlecture2002}:
\begin{equation}\label{qcap}
Q({\Phi}) = \lim_{n\to\infty} \frac{1}{n} Q_1( \Phi^{\otimes n} ) \, ,
\end{equation}
where
\begin{equation} \label{DEFQ1}
Q_1( \Psi) = \max_{\rho} J(\rho; \Psi) \, ,
\end{equation}
with $J( \rho; \Psi)$ is the associated coherent information~(\ref{def-coherent})
and the maximization is over all input states -- see also~\cite{HAYDEN08}.

The converse part of the coding theorem can be obtained  as an application of  the quantum
Fano inequality~(\ref{QFANO}) which, when imposing a lower limit to the entanglement fidelity,
forces the dimensionality of ${\cal M}$ to be bounded in terms of the channel coherent information~\cite{BNS98,BKN00} (in  Appendix~\ref{BOUNDSQA} a detailed derivation of this relation is presented
for the general case of non-necessarily memoryless channels).
Following~\cite{D05}, a relatively simple proof of the direct part of the theorem instead can be
obtained as a modification of the coding theorem for the private classical capacity.
The idea here is to extract from the codes which lead to privacy of the classical messages,
those which allow also the preservation of the coherent superpositions among the various
codewords. It turns out that this can be enforced by restricting the maximum in (\ref{capsec})
to only those ensemble ${\cal E}$ formed by pure states elements: a condition which,
thanks to~(\ref{identityimpo}), allows one to identify as achievable rates for quantum
communication those obtained from Eq.~(\ref{capsec})  in which $\chi({\cal E} ; \Phi) -
\chi( {\cal E}; \tilde{\Phi})$  gets replaced by $J(\rho;\Phi)$.

\subsubsection{The Bennett-Shor-Smolin-Thapliyal
theorem and the Quantum Reverse Shannon theorem}

The entanglement-assisted classical capacity of a memoryless channel is given
in terms of the quantum mutual information defined in Eq.\ (\ref{q-mutual}): i.e.,
\begin{eqnarray}\label{ct_Cea}
C_{ea}(\Phi) = \max_{\rho} I(\rho; \Phi) \; ,
\end{eqnarray}
where the maximization is over any input $\rho$ (the corresponding
quantum capacity version $Q_{ea}(\Phi)$ being half of $C_{ea}(\Phi)$  as already anticipated in Sec.~\ref{sec:capass}).
This result was proven in~\cite{B99,B02} by generalizing the dense-coding protocol~\cite{BW92} to the case of
noisy memoryless channel.  In dense-coding, the sender and the receiver share a maximally entangled
state in a Hilbert space of finite dimension, say $d^2$.
The sender encodes classical information by applying $d^2$ generalized $d$-dimensional Pauli unitaries
to one half of the maximally entangled states, which is then sent through the channel.
These transformations maps the given states into $d^2$ orthogonal states, which the receiver can
reliably distinguish.
More generally Alice, the sender, may encode classical information by applying generic CPTP maps on her 
half of the maximally
entangled state and then sending it through the channel.
However \onlinecite{B02} proved that the use of encoding by generalized Pauli unitaries is optimal.
This is obtained using the results of \cite{Holevo98,SW97} about encoding classical information
into quantum states, which apply also in this settings (see also \cite{HSIEH2008}).
It is worth noticing that the encoding by generalized Pauli unitaries constrains $\rho$ to be the
maximally mixed state, $\rho = \openone/d$. 
This limitation is circumvented by considering the typical
input states in the asymptotic limit of many uses of the channel, in which
the average input state is always the maximally mixed one on the typical subspace.
Remarkably, due to the subadditivity of the quantum mutual information, the expression for $C_{ea}(\Phi)$
of a memoryless channel does not require the regularization over the channel uses.

An important property of $C_{ea}(\Phi)$ is provided by the Quantum Reverse Shannon
theorem (Bennett et al., 2009; Berta, Christandl, and Renner 2011, Berta et al. 2013, and Berta, Renes, and Wilde 2013) which generalizes the
 Reverse Shannon theorem discussed in the introductory paragraphs of the
Sec.~\ref{sec:ctc}. It establishes that the input-output mapping of $n$ uses of a memoryless quantum
channel $\Phi$ can be simulated using $n C_{ea} (\Phi) + o(n)$
uses of a noiseless (qubit or bit) channel by providing the sender and the receiver an unlimited
supply of prior shared entanglement.

\begin{figure}[t]
\centering
\includegraphics[width=0.49\textwidth]{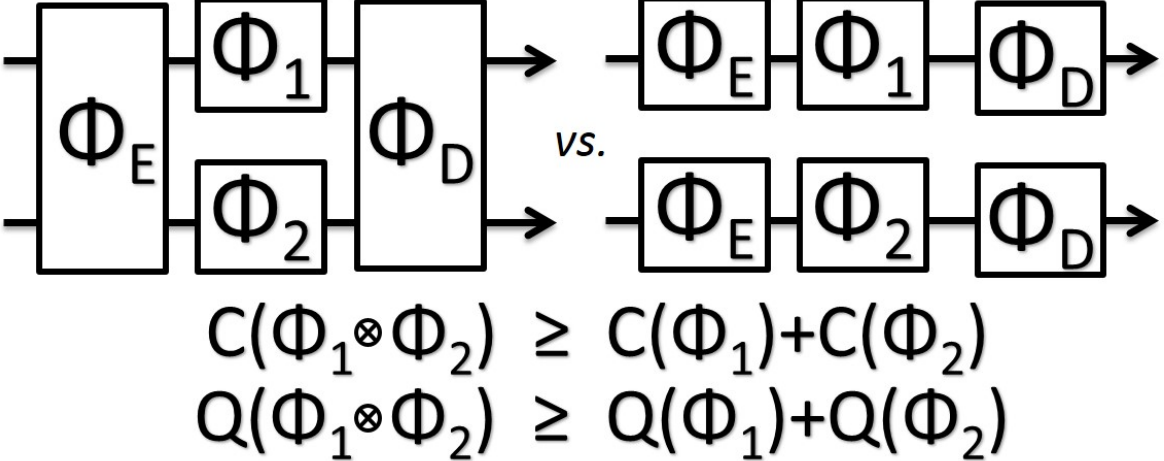}
\caption{According to the superadditivity property of channel capacities,
the capability of transmitting classical or quantum information over a tensor product of two
maps $\Phi_1$ and $\Phi_2$ is, in general, larger than the sum of the individual capacities.
This can be intuitively understood by the fact that global encoding/decoding schemes
allow one to explore a larger part of the Hilbert space, with respect to the local ones.}
\label{fig7sa}
\end{figure}

\subsubsection{Superadditivity and superactivation}\label{SASA}

The expressions in (\ref{C-from-C1}), (\ref{Cp-from-Cp1}) and (\ref{qcap}) for the classical, private and quantum capacity
of a memoryless quantum channel require the computation of the regularized limit over the
number of uses of the channel, $n\to\infty$.
Then one has the inequalities $C({\Phi}) \geq C_1({\Phi})$, 
$C_p({\Phi}) \geq C_{p,1}({\Phi})$, and $Q({\Phi}) \geq Q_1({\Phi})$.
For a given memoryless channel $\Phi$, if the first inequality is strict, that is, $C({\Phi}) > C_{1}({\Phi})$,
one says that the Holevo information of $\Phi$ is {\it superadditive}, otherwise it is said to be {\it additive}.
Similarly, one says that the coherent information is superadditive whenever $Q({\Phi}) > Q_1({\Phi})$,
and additive otherwise -- see~e.g.~\cite{SMITHREV}.

At a higher level of complexity,  when two different channels, $\Phi_1$ and $\Phi_2$, are used in parallel the
inequalities $C(\Phi_1 \otimes \Phi_2) \geq C(\Phi_1) + C(\Phi_2)$,
$C_p(\Phi_1 \otimes \Phi_2) \geq C_{p}(\Phi_1) + C_{p}(\Phi_2)$,
and $Q(\Phi_1\otimes\Phi_2) \geq Q(\Phi_1) + Q(\Phi_2)$ follow from simple
coding arguments (for instance the rate  $C(\Phi_1) + C(\Phi_2)$ can  always be
attained by feeding the inputs of $\Phi_1$ and $\Phi_2$ independently
with their corresponding  optimal codes) -- see Fig. \ref{fig7sa}.
If it happens that one of these inequalities is strict, then one says that 
the classical, resp. private, resp. quantum capacity is superadditive under 
tensor product of the channels $\Phi_1$ and $\Phi_2$ [notice that the 
additivity of (say) the Holevo capacity of $\Phi_1$ and $\Phi_2$ doesn't
necessarily guarantee the additivity of $C$ under the tensor 
product $\Phi_1\otimes \Phi_2$].

These (super)additivity issues are instances of a general (super)additivity problem
in quantum information theory \cite{Holevo2007a}.
While it was know early on that the coherent information can be superadditive \cite{DiSS98,SS1996},
hence the regularization over $n$ in Eq.~(\ref{qcap}) is in general necessary, the problem of determining whether the Holevo
information is additive or superadditive under tensor product of quantum channels has been for a long time an open problem.
The additivity of the Holevo information was shown to be equivalent to the
additivity of other quantities in quantum information theory \cite{Shor2004},
most notably the entanglement of formation \cite{BetAL96} and the minimum output entropy \cite{KR01},
and was put in connection with the behaviour of a family of operator norms under composition
of quantum channels \cite{AHW00,HW08}.
Only recently it has been established that the Holevo information can indeed be superadditive
for certain quantum channels \cite{Has09}, also implying the superadditivity of the
minimum output entropy and the entanglement of formation. Extensions of the results presented in~\cite{Has09} 
can be found in Refs.~\onlinecite{FKM10,BaH10,ASKnew10}.

However, notwithstanding the fact that the Holevo information is {\it generally} superadditive,
it has been proven to be additive for certain classes of quantum channels \cite{Hir06,AM09}, among which are the qubit unital channels \cite{King02} and
the entanglement breaking channels \cite{Shor2002a}.
In particular it is worth observing that in the case of qc-channels~(\ref{QC}) the classical capacity formula~(\ref{C-from-C1})
reduces to the Shannon capacity (\ref{shannoncap}) for a classical channel with conditional probability
$p(y|x) = \langle e_y| E_x |e_y\rangle$.
Moreover, as observed in \cite{Holevo07} if a quantum channel is (super)additive for the
Holevo information, so is its complementary channel.
The coherent information  has been proven to be additive for degradable channels
(see in Sec.\ \ref{sec:qcbo}), for anti-degradable channels (for which the quantum capacity
is always zero \cite{BDS97}), and for entanglement-breaking channels \cite{CRS08} and PPT maps
(see Sec.\ \ref{DegPPTEB}) \cite{HHH96,P96}.
Regarding the private classical capacity
 $C_p(\Phi)$ it is known that it coincides with its product-state version $C_{p,1}(\Phi)$
 for degradable and anti-degradable maps~\cite{D05}: in particular in both cases one has
 $C_p(\Phi) = C_{p,1}(\Phi) = Q(\Phi) = Q_1(\Phi)$, which for anti-degradable maps
 implies $C_p(\Phi)=0$.

A remarkable example of superadditivity for the quantum capacity,
called superactivation,
has been provided by \onlinecite{SY10} by building up from previous
results on the  quantum assisted capacities~\cite{SSW08}.  In particular, it has been shown
that it is possible to find channels $\Phi_1$ and $\Phi_2$ with zero quantum capacity, i.e.,
$Q(\Phi_1)=Q(\Phi_2)=0$, for which, by parallel use of the two communication
lines $\Phi_1$ and $\Phi_2$ in $\Phi_1 \otimes \Phi_2$, it becomes possible to transmit quantum information, i.e.,
$Q(\Phi_1 \otimes \Phi_2)>0$.
Specifically, $\Phi_1$ and $\Phi_2$ are given by an anti-degradable channel and a PPT channel which possesses a non-zero private classical capacity~\cite{HHHOP05,HPHH08}.
Notice that the two channels have zero quantum capacity for different reasons: the first as a consequence of the
no-cloning theorem and the second due to the fact that entanglement cannot be distilled from PPT state.
However, while it cannot be used to distill entanglement, there exist PPT channels that can still be
used to establish a secret key between the sender and the receiver.
See \cite{BOS11} for other examples in terms of \textit{depolarizing} maps and for a more general construction.

It is finally worth noticing that in the context of zero-error classical capacity (see footnote in Sec.~\ref{sec:sendingbits}) superactivation effects  have been observed in Refs.~\cite{DS08,DUAN}.

\subsection{Coding theorems  for  memory channels}\label{sec:eb}

The operational definitions of channel capacities, introduced in Sec.\ \ref{ssec:oper},
apply to both memoryless and memory quantum channels.
Indeed, they express the optimal classical and quantum information
transmission rates between two parties, no matter how complex is the internal structure of the
communication line.
However, the memory setting is often more complicated than the memoryless one.
One notices in particular that when dealing with non-anticipatory channels, introduced in Sec.~\ref{sec:nonant}, different notions of coding procedures and capacities can be
defined depending on {\it whom}, among Alice, Bob or a third party (Eve), controls (or uses for the encoding and decoding procedures)
the initial and final states of the memory system $M$. For instance
for the same  communication line  one can
 introduce the classical capacities $C_{AB}$, $C_{AE}$,
$C_{EB,\mu}$, $C_{EE,\mu}$, with the first (second) index representing the party controlling the
initial (final) memory state and $\mu$ being Eve's choice for the initial state of the memory $M$ (when considered) -- same classification holds for the other forms of capacity, i.e., quantum,
private classical, etc..
The differences between these various choices have been analyzed in Ref.~\cite{KW05}:
below, for the sake of simplicity, only the situation in which the third party (Eve) has full control of the memory $M$ will be considered.

\subsubsection{Entropic bounds}

The presence of correlations introduced by noise   makes more remote the possibility of formalizing capacities in terms of
entropic quantities when residing in the memory setting. A useful strategy is
to derive bounds on the capacities, with particular attention to upper bounds,
and then show whenever possible their achievability (thus providing coding
theorems).
The first attempts in this direction were presented in Ref.\ \cite{BM04},
where for the case of finite-memory channels, i.e., maps with memory of finite dimension (described in Sec.\ \ref{sec:finite}),
 the bounds  of Eqs.~(\ref{cbounds}) and (\ref{qbounds}) discussed  below (as well as an analogous inequality for the
 entanglement assisted capacity)  were derived and shown to be achievable
  for a class of Markovian channels.

Simple geometric considerations allow one to conclude that both $C$ and $Q$, independently of
the noise model can never be larger than $\log_2 d$, where~$d$ is the dimension of the Hilbert space of an individual
carrier (the rational being that in the space of $n$ carriers one cannot fit  more than $d^n$ orthogonal
states). This threshold however  is  not particularly informative as  it does not depend upon the CPTP mapping which describes the action of the channel (its value being  achieved only by noiseless channels, i.e., by  identity or unitary maps). 
Tighter upper limits on $C$ and $Q$ can be derived from the Holevo Bound~(\ref{HOLEVOBOUND}) and the quantum Fano
inequality~(\ref{QFANO}) respectively.
Specifically as explicitly shown in Appendix~\ref{BOUNDSCA} for the classical capacity  one gets
\begin{eqnarray}\label{cbounds}
            C \leq \lim_{n\to\infty} \frac{1}{n} \max_{\cal E}
            \chi({\cal E}; \Phi^{(n)}) \, ,
\end{eqnarray}
where $\chi(\dots)$ is the Holevo information defined as in Eq.~(\ref{HOLEVOINFO}) and the maximization
is performed over the ensembles of the first $n$ input carriers. Similarly for the  quantum capacity
one has
\begin{eqnarray}\label{qbounds}
Q \leq \lim_{n\to\infty} \frac{1}{n}  \max_{\rho} J(\rho,{\Phi}^{(n)}) \,,
\end{eqnarray}
where the maximization is now performed over the set of density matrices of the first $n$ carriers,
and where $J(\dots)$ is the coherent information defined in Eq.~(\ref{def-coherent}) --
see Appendix~\ref{BOUNDSQA}.

By direct comparison with Eqs.~(\ref{C-from-C1}) and
(\ref{qcap}),
one notices that for memoryless channels (i.e., when $\Phi^{(n)} = \Phi^{\otimes n}$) the
bounds given above coincide with the exact values of the corresponding capacities.
If the channel has memory correlations however, this feature is typically  lost apart from
some special configurations that will be analyzed in the following.

It is worth stressing that the inequalities~(\ref{cbounds}) and (\ref{qbounds})  refer to the limit
of infinite number of channel uses, i.e., they are asymptotic. Besides them, one
could also consider bounds referring to a finite number of channel uses (so called one-shot setting).
Notice that any situation in which a channel is used a finite number of times with
arbitrarily correlated noise can be equivalently described as a single use of a larger channel.
Bounds on the one shot classical capacity have been found in~\cite{WR12}
by using relative entropy type measure defined via hypothesis testing, while
bounds on the one shot quantum capacity have been derived in~\cite{BD10}
in terms of a generalization of relative Renyi entropy of zero order.

\subsubsection{Perfect memory channels} \label{sec:pmc}

Perfect memory channels admit a Kraus representation with a number of Kraus operators
growing sub-exponentially with the number of channel uses $n$ -- see Sec.\ \ref{sect:perfect}.
In other terms, the size of the environment is not large enough to `contain' the information sent
from Alice to Bob, which is exponentially increasing in $n$, and so asymptotically the loss of information
into the environment is negligible.
This intuitively explains that perfect memory channels are asymptotically noiseless and have maximal capacities, i.e., $C =Q= \log_2 d$.

Specifically, one can verify that \cite{GBM09,KW05} given a perfect memory channel $\Phi^{(n)}$ [see Eq.~(\ref{eqn:memory_model_10})],
for sufficiently large $n$ there exists a coding procedure  which allows zero-error classical communication for reference set ${\cal M}$ of  size
\begin{equation}
|{\cal M}| \geq \frac{d^n}{d_M^2} \, ,
\end{equation}
$d_M$   satisfying Eq.~(\ref{pro1})
and $d^n$ being the size of the $n$ carriers.
The corresponding  rate is hence $R \geq \log_2 d - \frac{2}{n}\log_2 d_M$, that for $n\rightarrow \infty$ converges to the
optimal value $\log_2 d$, implying hence $C= \log_2 d$.
Analogously, for sufficiently large $n$ there exists a
zero-error quantum communication with a reference set ${\cal M}$ of dimension
\begin{equation}
|{\cal M}| \geq \frac{ d^n }{d_M^4+d_M^2}
\end{equation}
with rate $R \geq \log_2 d -
\frac{1}{n}\log_2[d_M^4+d_M^2]$ which, again, for
$n\rightarrow \infty$ converges to the optimal value $\log_2 d$.

\subsubsection{Forgetful channels}\label{coding:forgetful}

Forgetful channels are characterized by the property that the effects of the
initial memory state become negligible with time, i.e., memory effects die away exponentially fast,
as discussed in Sec.\ \ref{sec:forgetful}.
This feature allows one to prove that the upper bounds of Eqs.~(\ref{cbounds}) and (\ref{qbounds})
 can be actually
asymptotically achieved \cite{KW05}.
Such important result can be demonstrated by invoking a {\it double-blocking} encoding procedure
which effectively maps forgetful channels into memoryless ones.

Consider a $n$ fold concatenation of a memory channel $\Phi^{(n)}$. If the channel
is {\it strictly forgetful} (see Sec.\ \ref{sec:forgetful}) there exists a finite integer $m$ such that for
all $n \geq m$ the final state of the memory system does not depend on its initial state.
In such a case, it is possible to group the channels $\Phi^{(n)}$ into blocks of length $m+l$,
encoding the input in the $l$ channels and ignoring the intermediate $m$ ones.
Accordingly, the memory channel is reduced to a memoryless one defined on the larger
Hilbert spaces ${{\cal H}_{Q}}^{\otimes l+m}$, hence allowing one to extend the
coding theorems for memoryless channels.
Remarkably, the double-block strategy can be applied even if the channel is forgetful
although not strictly forgetful.
Therefore, the memoryless expressions for classical and quantum channel capacities in Eqs.\ (\ref{C-from-C1})-(\ref{qcap})
can be applied also in the memory setting for forgetful maps, and the entropic upper bounds
in Eqs.\ (\ref{cbounds}) and (\ref{qbounds}) are exactly achieved (the same holds true for entanglement-assisted capacity).

Forgetful channels have been proven to constitute a dense set
with the topology induced by the cb-norm distance~\cite{KW05}.
That implies that any non-forgetful channel can be approximated by a
forgetful one. Notwithstanding, their capacities may be different.
An example can be given in the context of Markovian channels.
A long-term memory channel, Sec.\ \ref{sec:long}, can be
approximated by a forgetful Markovian channel, Sec.\ \ref{sec:markov}.
However, according to the coding theorem for long-term memory channel
discussed in the following section, the capacity of the latter does not
approximate the capacity of the former \cite{DD09}.

\subsubsection{Long-term memory channels}

An example of memory channels for which the bounds~(\ref{cbounds}) and (\ref{qbounds})
are not  tight, is provided by the  long-term quantum memory channels of the form (\ref{longmemory}) -- see Sec.~\ref{sec:long}.
In that context it is worth noticing that if the set $\{\Phi_i\}_i$ contains a finite number of elements, 
the determination of capacities of averaged channel is equivalent to the determination of capacities of 
the associated compound channel (see Sec.\ \ref{sec:compound}), since for finite sums one can always bound the error probability of the 
individual (memoryless) branches by the error probability of the averaged channel and vice-versa. 
Then, under the circumstance of  $\{\Phi_i\}_{i=1}^{N<+\infty}$, it has been shown that the product-state 
classical capacity is given by the expression
\begin{equation}\label{eq:long_psc}
C_1= \sup_{{\cal E}} \left[ \min_{i} \chi( {\cal E}; \Phi_i) \right] \, ,
\end{equation}
where the supremum is taken over all finite ensembles ${\cal E}$ of input states~\cite{DD07}.
This result has been derived by employing a quantum version of Feinstein's
Fundamental Lemma \cite{Feinstein, Khinchin} and a generalization of Helstrom's theorem \cite{helstrom}. 
The basic idea is to allow Alice and Bob to use the first channels uses to determine which, 
among the various possible channels $\Phi_i$, happens to be assigned by the statistical process
that defines the communication line via Eq.~(\ref{longmemory}).
After that, Alice and Bob can use a proper HSW encoding to optimize the communication rate. 
Accordingly it is clear that the maximum rate for which reliable transmission can be  guaranteed is the {\it lowest} one among those allowed by the $\Phi_i$. Indeed by operating
the channel to the highest rate allowed by the collection of maps $\{ \Phi_i\}$ will introduce
errors with finite probability.

The product-state capacity can be generalized to give the
classical capacity of the channel in the usual manner, that
is, by considering inputs that are product states over uses of
blocks of $n$ channels, but may be entangled across
different uses within the same block. This yields the value
\begin{equation}
C = \lim_{n \rightarrow \infty} \frac{1}{n} C_1(\Phi^{(n)})\;,
\end{equation}
which, in general, is smaller than the bound (\ref{cbounds}).
Similarly, the entanglement-assisted classical capacity has been proven to be expressed as \cite{DSD08}
\begin{equation}\label{ans}
C_{ea}= \sup_{\rho} \left[ \min_{i} {I}(\rho ; \Phi_i) \right],
\end{equation}
where $I(\rho ; \Phi_i)$ is the quantum mutual information (\ref{q-mutual}).
Finally, \onlinecite{BBN09} provided the expression for the quantum
capacity
\begin{equation}
Q=\lim_{n\to\infty}\frac{1}{n} \max_{\rho} \left[ \inf_{i} J\left( \rho; \Phi_i^{\otimes n}\right) \right] \, .
\end{equation}
Actually it has been shown, by means of a discretization technique based on $\tau$-nets, that this result holds true for compound channel associated to an arbitrary set $\{ \Phi_i\}$ (not only a finite one).
Finding the best rate for quantum communication over an arbitrary set of channels can be viewed as \emph{universal coding} problem. As such this result looks like a quantum channel counterpart of the universal quantum data compression result discovered in \cite{JHHH98}.

\subsubsection{Ergodic cq-channels with decaying input memory}\label{ergodicchannels}

For cq-channels (Sec.\ \ref{CQQC}) $W: A^{\zz}\times \mathcal{B}^{\zz}\to\cc$ which are stationary ergodic (Sec.\ \ref{sec:causal})
and have decaying input memory (Sec.\ \ref{sec:erg}), a coding theorem has been derived \cite{BB08} such that
the classical capacity is given by
\begin{equation}\label{erg-cap-dima}
C(W)=\sup_{p \textrm{ stationary ergodic}} i(p,W),
\end{equation}
where
\begin{equation}
i(p,W):=\lim_{n\to\infty}\frac{1}{n}\left(S(\rho_p^n)+S(\rho_{W}^n)-S(\rho_{p,W}^n)\right),
\end{equation}
with
\begin{eqnarray}
\rho_{p}^n&=&\sum_{x^n\in A^n}p^n(x^n) |x^n\rangle\langle x^n| \, , \\
\rho_{W}^n&=&\sum_{x^n\in A^n}p^n(x^n) \rho_{x^n} \, , \\
\rho_{p,W}^n&=&\sum_{x^n\in A^n}p^n(x^n) |x^n\rangle\langle x^n|\otimes \rho_{x^n} \, .
\end{eqnarray}
Here $\rho_{x^n}$ denotes the density operator of the output state $W^n(x^n,\cdot)$,
$x^n\in A^n$ and $|x^n\rangle = |e_{x_1}\rangle\otimes\ldots\otimes |e_{x_n}\rangle$ for some orthonormal
basis $\{|e_i\rangle\}_{i=1}^{|A|}$ of $\mathbb{C}^{|A|}$.

The $\sup$ in Eq. (\ref{erg-cap-dima}) is calculated over all stationary ergodic probability
measures $p$ on $A^{\mathbb Z}$. That is, consider a shift $T:A^{\mathbb Z}\rightarrow A^{\mathbb Z}$
of double infinite sequences of $A$, then $p$ is stationary if $p(Ta)=p(a)$ for all
$a\in A^{\mathbb Z}$. Moreover, it is ergodic if for all  $a\in A^{\mathbb Z}$ such that  $Ta=a$ it is $p(a)=0$ or $1$.

The above theorem results as an extension of coding theorem for input memoryless cq-channel
whose proof combines Wolfowitz's code construction \cite{Wo57} and a version of the
Feinstein's lemma \cite{BBT58} based on the notion of the joint input output probability distribution.


\section{Solvable Models}\label{sec:solv}

\subsection{Examples of solvable models for memoryless channels}

This section collects examples of discrete and continuous memoryless
quantum channels for which classical or quantum capacities can be analytically calculated.
For most of them the calculation is made feasible by the fact that the Holevo information
or the coherent information is additive. Hence the regularization in the limit of infinite 
$n$ of Eqs.~(\ref{C-from-C1}) and (\ref{qcap}) is not necessary as the capacities equal 
their product-state version.

\subsubsection{Discrete variable memoryless channels} \label{SEC:DISCRETE}

A closed expression for the classical capacity can be obtained for unital qubit channels (mapping
the two-dimensional identity operator into itself -- see Sec. \ref{sec:qubit}) and for the depolarizing channel acting on a finite
dimensional Hilbert space of arbitrary dimension.
For these channels the Holevo information has been proven to be additive \cite{King02,King03}.

Since any unital qubit channel is unitary equivalent to a Pauli channel~\eqref{PAULI} and, 
as discussed at the end of Sec.~\ref{ssec:oper}, capacities are invariant under unitary tranformations, 
it is sufficient to consider the latter.
As anticipated, the classical capacity for these maps equals its product-state version.
The fundamental ingredient to achieve this goal is the inequality~(\ref{upperSmin}), derived in 
Appendix~\ref{BOUNDSCA}  (which for this special channel can be shown to be achievable), and the fact that $S_{\min}(\Phi^{\otimes n})$
happens to be additive, i.e., $S_{\min}(\Phi^{\otimes n}) = n S_{\min}(\Phi)$.
The resulting expression for the classical capacity is then computed as
 \begin{equation}
C(\Phi) =C_1(\Phi)= 1 - h\left( \frac{1+\xi}{2} \right) \, ,
\end{equation}
where $h$ is the binary Shannon entropy~(\ref{binarysh}) and
$\xi$ is the maximum among $|p_0+p_1-p_2-p_3|$, $|p_0-p_1+p_2-p_3|$ and $|p_0-p_1-p_2+p_3|$. 
One may notice that the capacity reaches its maximum value $1$ if and only if  $\xi=1$, i.e., when at least
two of the probabilities $p_i$ are different from zero. 
Vice-versa the capacity nullify for $\xi=0$, i.e., when $p_i=1/4$ which correspond to a fully depolarizing qubit map sending
$\rho$ into the completely mixed state $\openone/2$. In a similar way
the classical capacity of the qudit depolarizing channel of Eq.~\eqref{DEPOL} can be shown to be
\begin{eqnarray}
C(\Phi)=C_1(\Phi)&=&\log_2 d +\left[1-\lambda- \frac{1-\lambda}{d}\right] \log_2\left[\frac{1-\lambda}{d}\right]
\nonumber\\
&+& \left[\lambda+\frac{1-\lambda}{d}\right] \log_2\left[\lambda+\frac{1-\lambda}{d}\right],
\end{eqnarray}
and the maximum in Eq.~(\ref{oneshotunassist}) is achieved by a set of $d$ equiprobable orthogonal pure states \cite{King03}.
Moreover, the entanglement-assisted classical capacity $C_{ea}(\Phi)$ is given by the same expression as for $C(\Phi)$ but
replacing $d$ with $d^2$.
A similar expression can be derived for the transpose depolarizing channel of Eq.~\eqref{DEPOLT}
which has also been proven to have additive Holevo information \cite{Fannes04,Datta06}.

Concerning the large class of qubit maps in Eq.\ (\ref{kraus:qq}), the full quantum capacity,
corresponding to the product-state one for the degradable case (since the coherent information
is additive) and vanishing for the anti-degradable regime, is given by~\cite{GF05,WP05}
\begin{eqnarray}
\label{qqqcap}
Q(\Phi) = \left\{
\begin{array}{ll}
f(\theta,\phi) & \mbox{for $\cos(2\theta)/\cos(2\phi) > 0$} \\
0 & \mbox{for $\cos(2\theta)/\cos(2\phi) \leq 0$\;,}
\end{array}
\right.
\end{eqnarray}
with
\begin{align}
f(\theta,\phi)=\max_{q\in[0,1]}\Big[
&h\big(q \cos^2\theta +(1-q) \sin^2\phi \big)\nonumber\\
&-h\big(q \sin^2\theta +(1-q) \sin^2\phi \big)\Big].
\end{align}

\begin{figure}[t]
\centering
\includegraphics[width=0.49\textwidth]{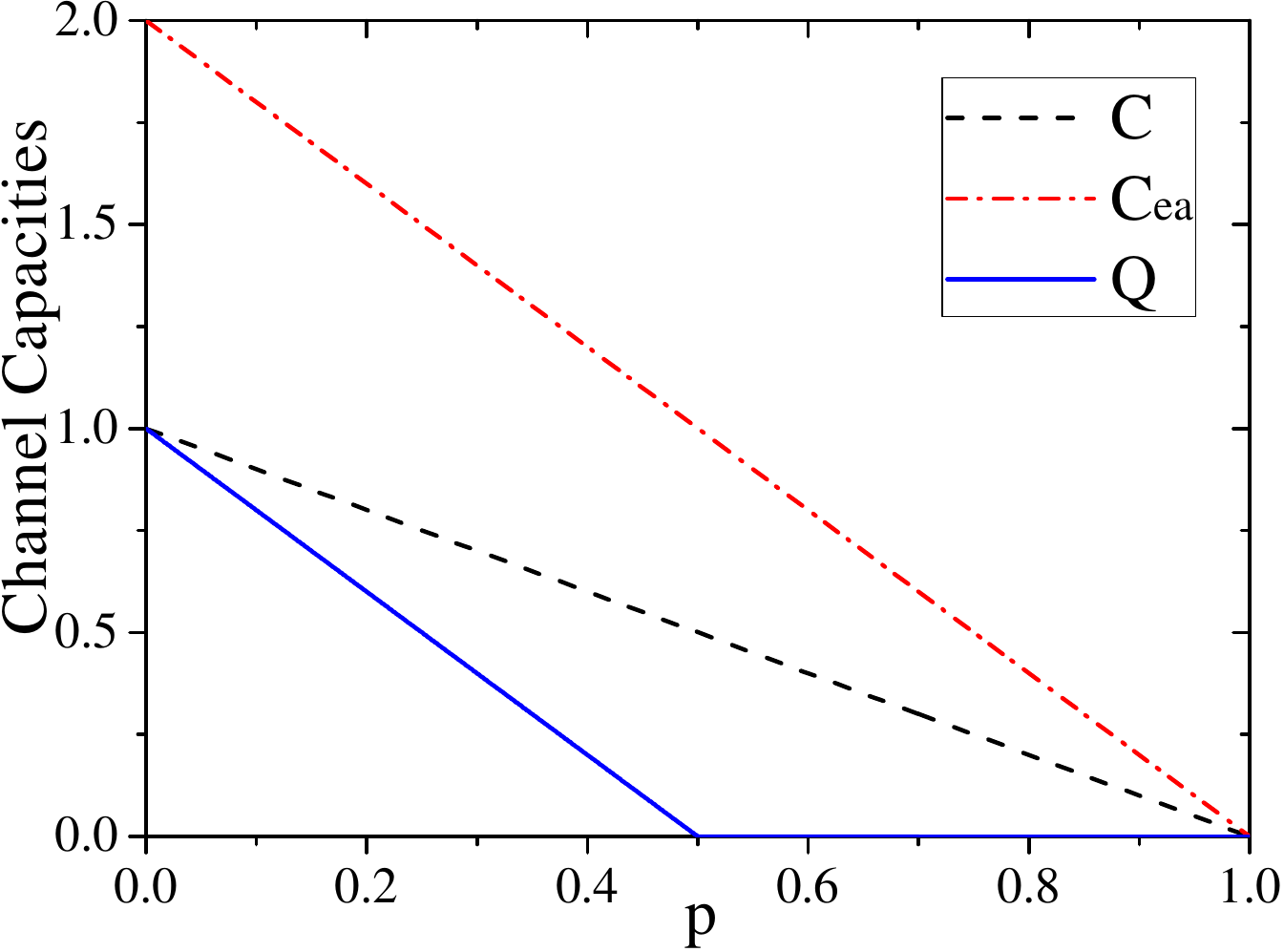}
\caption{Classical capacity ($C$), entanglement-assisted classical capacity ($C_{ea}$) and quantum capacity ($Q$) 
of the erasure channel mapping the input state into itself with probability $p$ and into an orthogonal state 
otherwise -- see Sec.~\ref{SEC:ERASURE}. In the plot the local dimension of the carrier is $d=2$.}
\label{f:fig13}
\end{figure}

Finally, the erasure channel introduced in Sec.~\ref{SEC:ERASURE} is one of the few
examples for which one can compute the whole set of capacities. This map
is degradable for $p \leq 1/2$ and has a quantum capacity $Q(\Phi)=(1-2p)\log_2{d}$ with $d$ being the dimension of the
input carrier;
for $p \geq 1/2$ it is, instead, anti-degradable, hence with vanishing $Q$.
Its Holevo information is also additive, yielding a classical capacity equal to $C(\Phi)=(1-p)\log_2{d}$ which can also be shown to coincide with the two-way classically assisted quantum capacity $Q_2(\Phi)
$. Finally the entanglement-assisted classical capacity is $C_{ea}(\Phi)=2 C(\Phi)=2(1-p)\log_2{d}$ \cite{BDS97,B99} - these quantities are shown in Fig. \ref{f:fig13}.

\subsubsection{Continuous variable memoryless channels}\label{subsec:CVmemless}

The very first  example of  a non trivial CV memoryless channel for which the capacity has
been explicitly computed was provided  by~\onlinecite{HolevoSohmaHirota} who considered  a cq-channel where  classical messages are mapped into
Guassian states obtained by continuously displacing an assigned Gibbs reference state.
This is a special example of one-mode Guassian channels --  see Sec.~\ref{CV-qc}. Under the memoryless condition the latter
can be identified by a triad $(\sf{d}^{(1)},\sf{X}^{(1)},\sf{Y}^{(1)})$,
where $\sf{d}^{(1)}$ is a two-component displacement vector, and $\sf{X}^{(1)}$, $\sf{Y}^{(1)}$ are $2 \times 2$ matrices.
Therefore, a sequence of $n$ consecutive channel uses is described by
a triad $(\sf{d}^{(n)},\sf{X}^{(n)},\sf{Y}^{(n)})$ where $\sf{d}^{(n)} = \bigoplus_{k=1}^n\sf{d}^{(1)}$,
$\sf{X}^{(n)} = \bigoplus_{k=1}^n \sf{X}^{(1)}$, $\sf{Y}^{(n)} = \bigoplus_{k=1}^n \sf{Y}^{(1)}$.


The property of (anti)degradability holds for the single-mode channels
describing the process of linear attenuation and amplification.
These channels are characterized by a single parameter $\eta$, see \ref{CV-qc}, and are known
to be antidegradable (hence having null quantum capacity) for $\eta \leq 1/2$ and degradable
(hence having additive coherent information) otherwise~\cite{CG06,CGH06}.
For $\eta > 1/2$ their quantum capacity reads \cite{WPG07}
\begin{equation}
Q(\Phi) = \log{\eta}-\log{|1-\eta|} \, .
\end{equation}
Analogously, if the mean number of bosonic excitation at the channel input is constrained to be
less than $N$ (see Sec.\ \ref{CV-qc}), for $\eta>1/2$ the constrained quantum capacity reads \cite{HW01,WPG07}
\begin{equation}
Q(\Phi,N) =Q_1(\Phi,N) = g(\eta N)-g(|1-\eta| N) \, ,
\end{equation}
where $g(x) := (x+1)\log{(x+1)} - x\log{x}$ for $x>0$ and $g(x) := 0$ for $x\leq 0$.

Under a constraint of $N$ mean input excitations, the Holevo information 
has been shown to be additive for the lossy bosonic channel ($\eta \in [0,1]$)~\cite{GiovannettiPRL92},
allowing for a single-letter expression for the classical capacity:
\begin{equation}
C(\Phi;N) = C_1(\Phi;N) = g(\eta N) \; ,
\end{equation}
Similarly, the constrained entanglement assisted classical capacity~\cite{GiovannettiPRL91,GiovannettiPRA68} reads
\begin{equation}
C_{ea}(\Phi;N) = g(N) + g(\eta N) - g[(1-\eta)N] \; .
\end{equation}
(Notice that their unconstrained counterparts, differently from the quantum capacity, are unbounded.)
Besides being additive, the Holevo information for the lossy bosonic channel is
maximized for Gaussian inputs.
The same properties have been very recently proven to hold for a broad family of Gaussian
channels, which includes the lossy and noisy channel, the linear amplifier, and the
additive noise channel \cite{Giova2014a,Giova2014b}. 
For all these channels, this result solves and gives a positive answer to a long-standing conjecture
\cite{HW01,GiovannettiPRA04,lloyd09,GHLM10,GPNBLSC12,GES2007,GiovannettiNP13,KONIG1,KONIG2,SEW05,Hir06} 
and proves single-letter expressions for their classical capacities (the latter were summarized in \cite{LupoPS}).
For example, the capacity of the lossy and noisy Gaussian channel reads ($\eta \in [0,1]$)
\begin{equation}
C(\Phi;N) = C_1(\Phi;N) = g(\eta N + (1-\eta)N_\mathrm{th}) - g((1-\eta)N_\mathrm{th}) \; ,
\end{equation}
that of the linear amplifier is ($\eta \geq 1$)
\begin{equation}
C(\Phi;N) = C_1(\Phi;N) = g(\eta N + (\eta-1)) - g((\eta-1)) \; ,
\end{equation}
and for the additive noise channel one has
\begin{equation}
C(\Phi;N) = C_1(\Phi;N) = g(N + N_\mathrm{add}) - g(N_\mathrm{add}) \; .
\end{equation}

\subsection{Examples of solvable models for memory channels}

The main difficulty in the evaluation of the capacities of quantum channels with memory
relies on the requirement of the regularization of the corresponding entropic quantities in the limit of
infinite uses of the channel.
For the case of forgetful channels this gives the exact expression for the capacities, while in general
it provides an upper bound for non-forgetful channels.

Up to date only few models of memory quantum channels have been
fully solved in terms of their capacities.
One is the dephasing channel (in the discrete variable setting),
and the other is the lossy bosonic channel (in the continuous
variable setting), with different types of correlations.

\subsubsection{Discrete memory channels}\label{sec:smdmc}

Referring to the model discussed in Sec.~\ref{sec:memorylessquantum},
consider  a sequence of qubit carriers propagating at rate $\nu$
and interacting each one with a single qubit environment subject
in turn to a relaxation process described by amplitude damping
with a rate $1/\tau$.
Then assume that the carrier-environment interaction is a control-unitary,
such that when the carrier is in $|0\rangle_{q_j}$
nothing happens to the environment, while when $q_j$ is in $|1\rangle_{q_j}$
the environment undergoes the unitary transformation described by the operator $\gamma \sigma_z + \sqrt{1-\gamma^2} \sigma_x$.
One hence has a memory channel whenever the condition $\nu\tau\ll 1$ is not satisfied.
However, it is possible \cite{G05} to trace this model back to a memoryless phase damping
channel $\Phi_{\overline\gamma}$ with $p_z=(1-\overline\gamma)/2$ the probability of $\sigma_z$ error.
Here $\overline\gamma$ is a complicate function of several parameters including $\nu$ and $\tau$ and it reduces to
$\gamma$ for $\nu\tau\ll 1$ (the memoryless limit of Section \ref{sec:fmmqc}), while it can be
$\overline\gamma >\gamma$ for $\nu\tau\ge 1$,
thus making  $\Phi_{\overline\gamma}$ effectively less noisy than $\Phi_\gamma$.

In the case of the phase damping channels (see Sec. \ref{sec:qubit}) the capacities can be explicitly computed.
For instance, since the noise does not affect the populations associated with
the computational basis, the classical capacity of the
phase damping channel ${\Phi}_\gamma$ is $C({\Phi}_\gamma) = 1$.

On the other hand the quantum capacity of a phase
damping channel ${\Phi}_\gamma$ is
$Q({\Phi}_\gamma) = 1-h(p_z)$~\cite{WP05,DS05}
where $h$ is the binary entropy.
One hence has $Q({\Phi}_{\overline{\gamma}}) \geq Q({\Phi}_{\gamma})$ for $\overline{\gamma}>\gamma$, i.e.,
enhanced quantum capacity by memory effects.

\bigskip

Markovian correlated dephasing has been considered by \onlinecite{arrigo07}, where
the degree of correlations is expressed by a correlation parameter $\mu \in [0,1]$
which characterize the Markovian transition probabilities as in Eq.\ (\ref{mem-par}).
Likewise in the above model, the quantum capacity increases when considering an higher degree of memory.
In particular, the memoryless dephasing channel capacity is recovered
for $\mu=0$, while for $\mu=1$ (perfect memory) the channel is asymptotically noiseless, i.e., $Q(\Phi)=1$ \cite{BM04}. \onlinecite{arrigo07} also considered a microscopic model for correlated dephasing defined in terms of a
spin-boson model, where quantum information is encoded in a train of qubits and a single bosonic
mode represents the memory system.
Lower bounds for the quantum capacity of a qubit memory channel with both correlated dephasing and
damping have been evaluated numerically starting from a microscopic spin-boson model with Jaynes-Cummings
interaction in the presence of strong dephasing noise \cite{BDF09,BDF12}.

\bigskip

\onlinecite{PV07,PV08} have considered another model of dephasing memory channel for qubits.
It can be traced back to the scenario introduced in~\cite{GM05} and schematized in Fig.~\ref{fig:memchan2},
where each individual information carrier (a qubit in this case) interacts with a corresponding environment particle,
the correlations being established by the environment multi-particle state.
Specifically they have considered the case where the  two-particle (two-qubit)
interaction is defined by a controlled-phase gate, the environmental particle being the controller qubit that determines which
unitary transformation will be applied to the carrier. As a consequence
the join state of the carriers gets transformed through mixtures of random sequences of identity and $\sigma_z$  operators, each sequence
being characterized by a (correlated) probability which depends upon  the diagonal elements of the environment initial  state.

The interesting feature of this model is that it allows to write  explicit
formulae for the associated capacities for the channel  in terms of properties
of the many-body environment that share a close relationship with thermodynamical quantities.
In particular, the CJ state of their family of correlated channels is a maximally correlated state (i.e., state of the form
$\sum_{i,j} \alpha_{i,j} |ii \rangle \langle jj|$) \cite{R99a,R99b,R01}, and, combining this feature with the forgetfulness of such maps,
one can show \cite{PV07,PV08} that the quantum capacity can be expressed in terms of the regularized diagonal entropy of the system environment, i.e.,
\begin{equation}
Q(\Phi) = 1 - \lim_{n \rightarrow \infty} {S({\rm diag}(\rho_{env})) \over n} \label{qkey}\;,
\end{equation}
where ${\rm diag}(\rho_{env})$ is the environmental state in the computational basis after eliminating all off-diagonal elements
[note that the coding argument used in order to arrive to Eq.~(\ref{qkey}) has been also independently shown by \cite{Ham02}].
For the special case in which the initial state of the environment is described by a classically correlated many-body system
(i.e., diagonal in the computational basis), the last term on the right hand side of Eq.~(\ref{qkey}) coincides with the thermodynamical entropy of the environment.
Hence, the capacity is given by
\begin{equation}
        Q(\Phi) = 1 -
        \left( 1 - \beta {\partial \over
        \partial \beta} \right) \lim_{n \rightarrow \infty}
        {1 \over n} \log_2 Z_n \label{class}\;,
\end{equation}
where $Z_n$ is the partition function for $n$ environment spins, and  $\beta$ is the associated inverse temperature.
In other words, one can exploit results from classical statistical physics in order to compute the capacity, as shown by Eq.~(\ref{class}).

The calculation of the entropy of the associated many-body system, and hence of the quantum capacity of the memory channel,
can be done exactly in certain relevant cases.
One of them is the case of many-body systems described by Matrix Product States (MPS) involving only rank-1 matrices.
For the sake of simplicity, one does focus on a translationally
invariant MPS for a 1D system of 2-level particles, with periodic boundary conditions.
This environmental state is characterized by two matrices ${\sf A}_0$ and ${\sf A}_1$ and
is given by the following expression
$|\psi\rangle = \sum_{i_1\ldots,i_n} {\rm{Tr}}\{{\sf{A}}_{i_1}\ldots {\sf{A}}_{i_n}\} |i_1\ldots i_n\rangle$.
Then, by dephasing each qubit, the resulting unnormalized state is
\begin{equation}
\rho = \sum_{i_1\ldots,i_n} {\rm{Tr}}\left[ \prod_{k=1}^n({\sf A}_{i_k} \otimes \bar{\sf A}_{i_k}) \right] |i_1\ldots i_n\rangle\langle i_1\ldots i_n| ; ,
\end{equation}
where $\bar{\sf A}$ is the complex conjugate matrix of $\sf A$.
It is possible to show that, if $|i_1\ldots i_n\rangle$
has $l$ occurrences of $0$ and $n-l$ of $1$, and $k$
boundaries between $0$s and $1$s blocks,
then the corresponding diagonal elements of $\rho$ are proportional to
$a^{l}b^{n-l} c^k$,
with $a$ (resp. $b$) being the eigenvalue of ${\sf A}_0\otimes \bar{\sf A}_0$ (resp.\ ${\sf A}_1\otimes \bar{\sf A}_1$),
and $c$ being the eigenvalue of $({\sf A}_0\otimes \bar{\sf A}_0)({\sf A}_1\otimes \bar{\sf A}_1)/(ab)$.

Finally, it is worth remarking that \cite{WOV06} showed the existence of Hamiltonians exhibiting quantum phase
transitions and with ground states being Matrix Product States (MPS) involving only matrices of rank-1.
Hence, it can be shown that the diagonal elements of such MPSs are equal to
the probability $\wp$ of microstates in corresponding classical Ising chains.
Therefore, by exploiting this connection, one can easily compute the limit in Eq.\ (\ref{qkey})
by using well known many-body physics methods. Figure \ref{wolfplot} shows the case of the following Hamiltonian
\begin{equation}
\label{HWolf}
\sum_i 2(g^2-1)\sigma_{z,i}\sigma_{z,i+1} - (1+g)^2 \sigma_{x,i} + (g-1)^2 \sigma_{z,i}\sigma_{x,i}\sigma_{z,i+1} \;.
\end{equation}
In this case, one knows that the ground state is a rank-1 MPS which possesses a non-standard `phase transition' at $g=0$,
where indeed some correlation functions are non-differentiable (though continuous) and the ground state energy is analytic \cite{WOV06}.

\begin{figure}[t]
\centering
\includegraphics[width=0.45\textwidth]{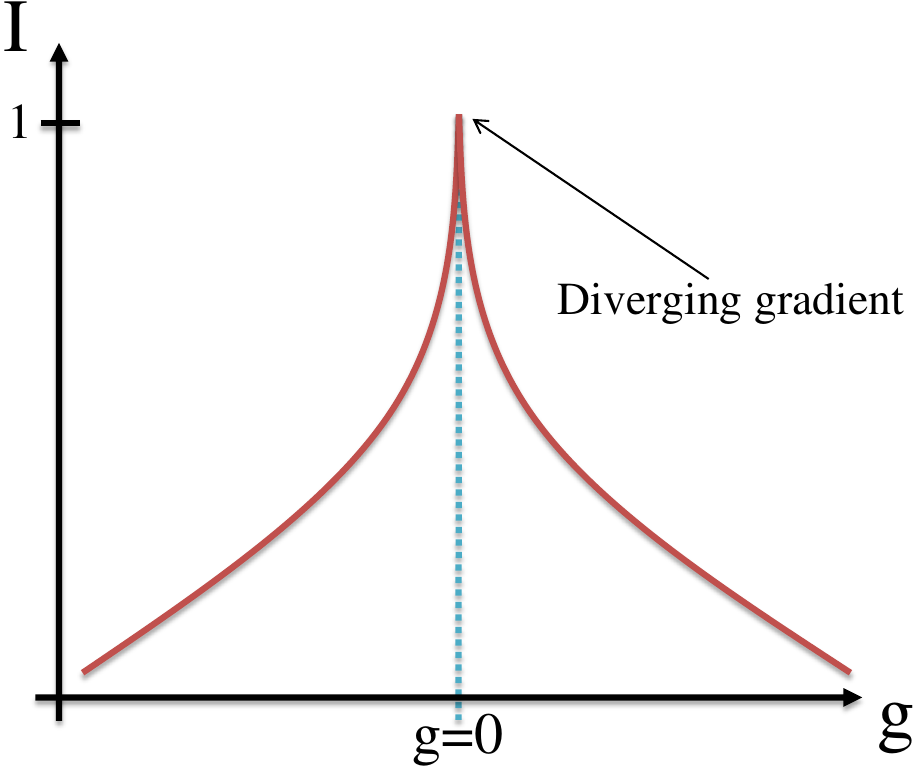}
\caption{Sketch of the capacity behaviour in the case of an environment given by the ground state of the Hamiltonian \eqref{HWolf}. 
Notice the divergent gradient near the `phase transition', i.e., at $g=0$~\cite{PV07}.} \label{wolfplot}
\end{figure}

\subsubsection{Continuous memory channels}\label{sec:smcmc}

Among Gaussian memory channels, one can identify a subclass of
channels for which the memory effects can be {\it unraveled}.
That is, by applying suitable unitary encoding and decoding transformations,
$n$ uses of such channels are mapped to $n$ independent single-mode channels used in parallel.
By applying known results for the memoryless setting one may then compute the
capacities of the memory channel (see Sec.\ \ref{subsec:CVmemless}).

Such a unitary mapping from $n$ uses of a Gaussian memory channel to $n$ parallel
uses of independent single-mode channels was first considered in \cite{CCM+05,CCM+06,GM05},
and then applied for estimating the communication capacities of
Gaussian memory channels in several settings \cite{LMM09,LPM09,LGM10,LGM10b,SKC09}.
A formal definition of the class of memory channels that can unraveled first appeared in \cite{LM10}.

If one takes the one-mode channel as a reference point, representing a single use of the
channel, $n$ uses of the quantum memory channel are characterized by the triads $(\sf{d}^{(n)},\sf{X}^{(n)},\sf{Y}^{(n)})$
such that either $\sf{d}^{(n)}\neq\bigoplus_{k=1}^n\sf{d}^{(1)}$ or
$\sf{X}^{(n)} \neq \bigoplus_{k=1}^n \sf{X}^{(1)}$, $\sf{Y}^{(n)} \neq \bigoplus_{k=1}^n \sf{Y}^{(1)}$ (see Sec.\ \ref{CV-qc}).
A memory channel can be unraveled if there exist unitary transformations
$\Phi_E^{(n)}$, $\Phi_D^{(n)}$, acting on $n$ modes, such that $\Phi_D^{(n)} \phi^{(n)} \Phi_E^{(n)} = \bigotimes_{k=1}^n \phi^{(1)}_k$,
that is, $n$ uses of the memory channel are unitary equivalent to the tensor
product of $n$ independent -- but not necessary identical -- single-mode Gaussian channel
(this mapping is depicted in Fig.\ \ref{unraveling1}).
Since the application of unitary transformations cannot change the capacities of the channel, they can be
equivalently computed for the unraveled channel, in which each input mode
is transformed independently (although in general not-identically).
If one is interested in the calculation of constrained capacities, then one
has to take in account how the constraint changes under the action of the encoding and
decoding unitaries.
A relevant setting is that of encoding and decoding transformations preserving the
constraint. For the case of constrained mean input excitation-number, the constraint is preserved if
$\sum_{k=1}^n a_k^\dag a_k = \Phi_E^{*(n)}\left(\sum_{k=1}^n a_k^\dag a_k\right)$,
a condition which is satisfied when the encoding unitary is a linear passive transformation, e.g. in the case of optical realization, when exploiting a network of beam splitters and phase shifters (see e.g., \cite{FOP05}).

\begin{figure}[t]
\centering
\includegraphics[width=0.48\textwidth]{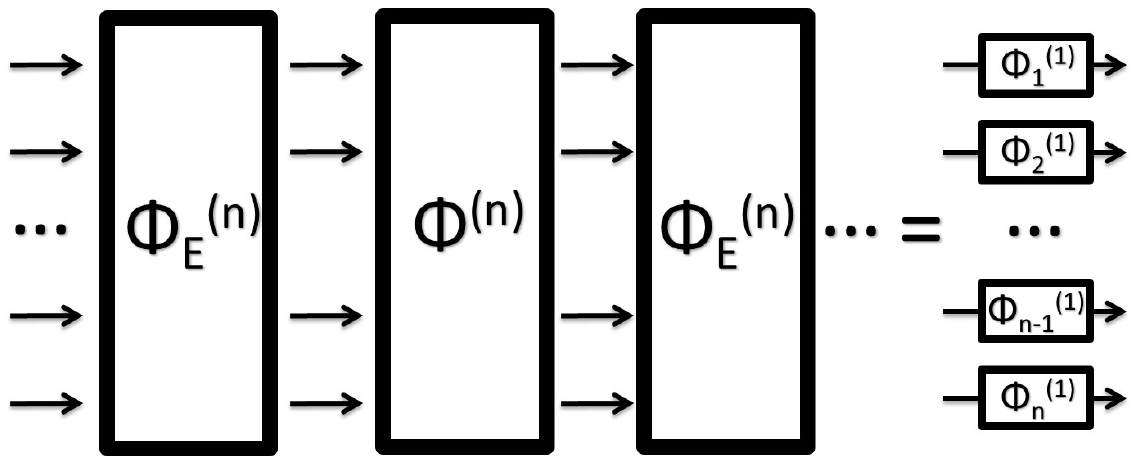}
\caption{Unraveling of $n$ uses of a memory channel.
Each horizontal line indicates one bosonic mode, propagating from the left to the right.
$\phi^{(n)}$ denotes $n$ uses of the memory channel. $E^{(n)}$ and $D^{(n)}$ are
pre-processing and post-processing Gaussian unitaries. $\phi^{(1)}_k$'s are one-mode
Gaussian channels.}
\label{unraveling1}
\end{figure}


If a Gaussian memory channel can be unraveled, then its capacities can be computed upon
reduction to the case of memoryless single-mode Gaussian channel.
This is the case for the model of lossy channel with memory introduced in \cite{LGM10}.
In this model, the action of the channel upon $n$ uses is defined
by the concatenation of $n$ identical unitary transformations coupling the input modes $a_1$, $a_2$, $\dots$, $a_n$
with a collection of local environmental modes $e_1$, $e_2$, $\dots$ $e_n$ and the memory mode $m$.
Specifically the evolution of the $k$-th input mode is obtained by a
concatenation of two beam-splitter transformations, the first with
transmissivity $\epsilon$ and the second with transmissivity $\eta$, see Fig.\ \ref{figsolvable3}.
This results in a non-anticipatory channel with ISI (see Sec.\ \ref{sec:nonant})
having the same structure depicted in Fig.\ \ref{figure01}c.
By varying the transmissivity parameters, the 
model is capable to describe different memory schemes, from the memoryless lossy bosonic channel configuration
(Giovannetti et al., 2004c) (the input $a_k$ influences only the 
output $b_k$), to a channel with perfect memory (all $a_k$
interacts only with the memory mode $m$ ) (see Sec. III.D.3),
to a {\it quantum shift} channel (Bowen and Mancini, 2004) where each input state is replaced by the previous one (this is obtained by setting $\eta =0$, $\varepsilon =1$). 
Extensions of Lupo, Giovannetti, and Mancini (2010a) which encompasses memory effects in linear amplification and thermalization processes are presented in 
 Lupo, Giovannetti, and Mancini (2010b) and De Palma, Mari, and Giovannetti (2014), respectively. All these models can be unraveled into the
tensor product of one-mode lossy or amplifier channels.
The capacities of these memory channels can hence be computed following four steps:
first the memory channel is unraveled into the direct product of the single-mode Gaussian channels;
second the optimization of the relevant entropic function is performed mode-wise under constrained mean input excitation number;
then the distribution of the mean excitation number over the input modes is optimized;
finally the asymptotic limit of infinite channel uses is considered.
The optimization of the distribution of the mean excitation number leads to a {\it quantum water
filling} solution for the capacity of the memory channel, where the way the mean excitation
number is distributed over input modes is analogous to the way water distributes into a vessel \cite{CTbook}.
While algorithms for the optimization were presented in \cite{PLM09,SKC11},
the most delicate point is the consideration of the asymptotic limit \cite{LMM09,LGM10}.

Memory channels with additive noise are characterized by having $\sf{X}^{(n)}=\openone$.
They can be realized by means of multimode CV teleportation
protocol \cite{V94,BK98,BST02}, where the teleportation resource is a multimode state \cite{CGP10}.
The memory channel considered in \cite{CCM+05,CCM+06} belongs to this class.
The latter was defined for two channel uses, represented by two bosonic modes,
which are affected by correlated additive noise.
A generalization of this model to the case of more than two channel uses was first 
introduced in \cite{RM07a} and subsequently in \cite{SKC09,LMM09}, where the additive noise, 
characterized by the matrix ${\sf Y}^{(n)}$, constitutes a Markov process.
It is easy to recognize that these models define SI memory
 channels, which are instances of the general scheme depicted in Fig. 13 and first introduced in  Giovannetti and Mancini (2005).
Here Gaussian memory effects were introduced  by imposing that  the $n$ input modes interact modewise with a joint (possibly
entangled) Guassian state of $n$ environmental modes through  beam-splitter transformations of 
transmissivity
$\eta$ (the associated ${\sf Y}^{(n)}$ matrix of the channel being $(1-\eta){\sf C}^{(n)}$, where ${\sf C}^{(n)}$ 
is the CM of the environmental state).

These Gaussian memory channels can be unraveled whenever the matrix ${\sf Y}^{(n)}$ have
a suitable form, furthermore under certain conditions they can be unraveled with the use of energy-preserving 
unitary pre-processing transformation \cite{LPM09,PZM08} (see also \cite{RSG+05}).

It worth noticing that, differently from the case of discrete-variable memory channels (see Sec.\ \ref{sec:markov}),
there is no transitional behavior in these models of Gaussian memory channels: the optimal input states are either
separable or entangled according to the model symmetries \cite{CCM+05,CCM+06,LM10}.
As entangled states cannot be prepared locally, it is crucial to identify suboptimal
input states that can be prepared efficiently. This issue was considered in \cite{SKC12},
where it was shown that encoding classical information via Gaussian matrix-product
states \cite{GMPS1,GMPS2}, which can be efficiently prepared, may allow to achieve a
reliable communication rate close to the channel capacity.
An analysis of correlated additive Gaussian channels beyond the case of Markovian correlations
was presented in \cite{SKC11}.

Finally, it is worth remarking that the study of Gaussian memory channels has also stimulated and motivated a deep
analysis of the communication capacities of the single-mode memoryless Gaussian channel \cite{LupoPS,PLM09,SKC10}.
In particular \cite{PLM09} and \cite{SKC10} provided a complete characterization of one-mode Gaussian channels,
respectively for the case of lossy channels and additive noise.

\begin{figure}
\centering
\includegraphics[width=0.48\textwidth]{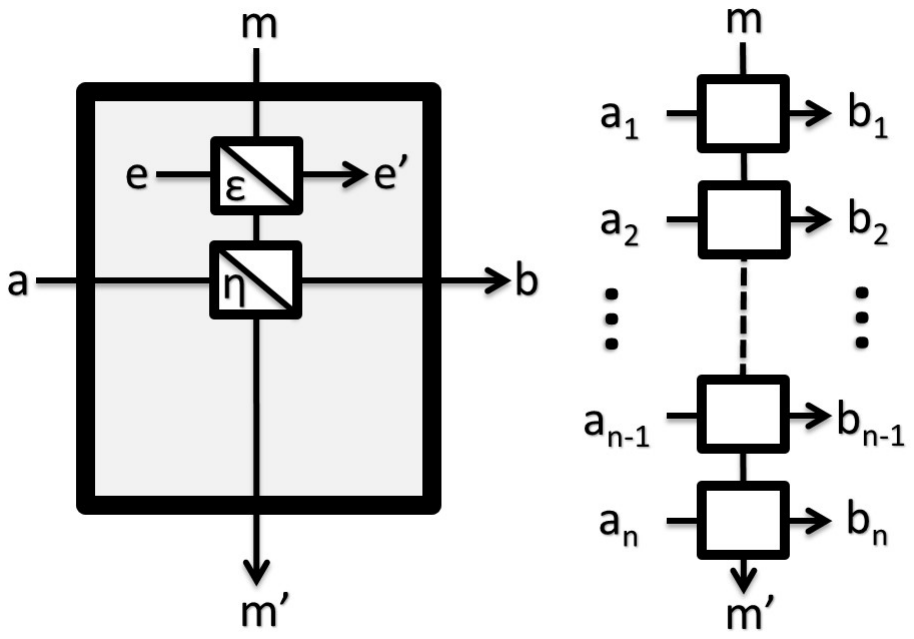}
\caption{(Color online) Left: a single use of the lossy bosonic memory channel \cite{LGM10}.
Right: the $n$-fold concatenation of the memory channel:
photons entering in the $k$-th input mode $a_k$ can only emerge in
the output ports $b_{k'}$ with $k' \geq k$.} \label{figsolvable3}
\end{figure}


\section{Quantum channels divisibility and dynamical maps}\label{sec:memdyn}

In this section we leave the input-output scenario, which 
has characterized all the previous parts of the review,
and focus on the memory effects that may arise when studying the dynamical 
evolution of a system that is evolving in time while interacting with an external environment, see Fig.~\ref{fig001}.

As discussed in Sec.~\ref{sec:comp} the concatenation of CPTP maps defines a new quantum channel.
It is also worth considering whether the converse is also true, that is,
under which conditions a quantum channel
$\Phi\in \mathfrak{P}:=\mathfrak{P}(Q\mapsto Q)$  acting on a system $Q$,
can be expressed as a
concatenation of other elements of $\mathfrak{P}$.
This is intimately related to the
semigroup structure of the set of quantum channels, hence with dynamical
maps and master equations.


\subsection{Divisible and indivisible quantum channels}\label{sec:div}

Loosely speaking, by divisibility of a quantum channel $\Lambda\in\mathfrak{P}$ one refers to the possibility of
decomposing it in terms of concatenation of other channels, i.e., to the possibility of writing
$\Lambda=\Lambda_1\circ \Lambda_2$, with $\Lambda_i\in\mathfrak{P}$.
Obviously, every channel $\Lambda\in\mathfrak{P}$  is divisible in the
following way: $\Lambda=(\Lambda\circ {\cal U}^{-1}) \circ {\cal U}$, with ${\cal U}$ any unitary map.
A non trivial definition of (in)divisibility has been introduced by \onlinecite{WC08}.
According to that, a quantum channel $\Lambda \in \mathfrak{P}$ is indivisible if every decomposition of
the form $\Lambda=\Lambda_1\circ \Lambda_2$, with $\Lambda_i\in\mathfrak{P}$, implies
that either $\Lambda_1$ or $\Lambda_2$ is a unitary conjugation.
Otherwise $\Lambda$ is said to be divisible.
It happens that quantum channels with maximal Kraus rank ($d^2$) are divisible \cite{WC08}.

Hereafter the subset of $\mathfrak{P}$ of divisible channels is denoted as $\mathfrak{D}$.
The notion of divisibility can then be refined by considering
different kinds of divisible quantum channels.
First, one introduces a notion of Markovianity for quantum channels
related to their decomposability, rather than to their composability as done in Sec.\ref{sec:tqmc}.
According to \cite{WC08} a quantum channel is called Markovian if it is an element of a
continuous one-parameter semigroup of CPTP maps.

In such a case there exists a (Liouvillian) generator $\mathcal{L}$
 such that the quantum channel can be written as
  $\Lambda(t)=e^{t \mathcal{L}} \in \mathfrak{P}$ for all $t\geq 0$.
A standard form for such generators was derived in \cite{GKS76,L76}:
 \beq
 \mathcal{L}\rho = i [\rho,H] + \sum_{\alpha,\beta}
 G_{\alpha,\beta} \left(F_\alpha\rho F_\beta^\dagger -\frac{1}{2}\{F_\beta^\dagger
 F_\alpha,\rho\}\right)\label{GENERATOR}
 \eeq
 where  $G\geq 0$, $\{\;,\;\}$ denotes the anti-commutator and the operators $H$ and $F_\alpha$ respectively describe the Hamiltonian and non-Hamiltonian dynamical terms.

Through the (Liouvillian) generator $\mathcal{L}$ one can write down
the dynamical (master) equation for the system density operator $\rho$ \cite{BPbook}
\begin{equation}\label{ME}
\frac{d}{dt}\rho(t)=\mathcal{L}\rho(t).
\end{equation}
Its solution, for given initial condition $\rho(t_0)$,
reads $\rho(t) = \Lambda(t-t_0)\, \rho(t_0)$
with $\Lambda(t-t_0) = e^{(t-t_0){\cal L}}$
obeying the \emph{homogeneous} composition law
\begin{equation}\label{COMP}
    \Lambda(t_1) \circ \Lambda(t_2) = \Lambda(t_1+t_2) \ ,
\end{equation}
for $t_1,t_2 \geq 0$, hence defining a one-parameter semigroup of CPTP maps.
As consequence, Eq.\ \eqref{ME} is called Markovian master equation.

A class of Markovian master equations of this kind can be obtained as the
continuous-time limit of a concatenation of identical system-bath interactions.
These models, known as {\it collision models} \cite{AL87,TDV00,SZSGB02,ZB05,Rau63,Zimanetal02,Zimanetal05}, are
defined by the iterated unitary interactions of the system $Q$ with $n$ identical
reservoirs $E=(e_1, \dots, e_n)$.
This {\it cascade} process, depicted in Fig.~\ref{cascade}, defines a quantum channel of the form
\begin{equation}\label{eqn:cascade}
\Phi^n(\rho_Q) = \mathrm{Tr}_{E} \left[ U_{Q e_1} \cdots U_{Q e_n}
\left( \rho_Q \otimes \omega_E^{\otimes n} \right)
U_{Q e_1}^{\dag} \cdots U_{Q e_n}^{\dag} \right] \, ,
\end{equation}
where $U_{Q e_j}$'s are $n$ instances of a unitary transformation coupling the
system $Q$ with the environmental systems.
A comparison  with Fig.\ \ref{figure01}a  
is useful to 
enlighten the  relations  between  this model and the unitary dilation of  memory channels introduced in  Eq.\ (\ref{eqn:memory_model_10}): basically,
 in passing
from  the latter to Eq.~(\ref{eqn:cascade})
the  environment and the  carriers have
exchanged their roles transforming the spatial
correlations of Eq.~(\ref{eqn:memory_model_10}) into
temporal correlations. 
An hybrid approach which includes both effects has 
been recently introduced in \cite{GP12}: as shown in Fig.\ \ref{GP12fig}
 the scheme has the same structure of Fig.\ \ref{cascade} for each row, and the same of
Fig.\ \ref{figure01}a for each column.
This model provides a link between memory channels and time-continuous dynamical evolutions.

\begin{figure}[t]
\centering
\includegraphics[width=0.48\textwidth]{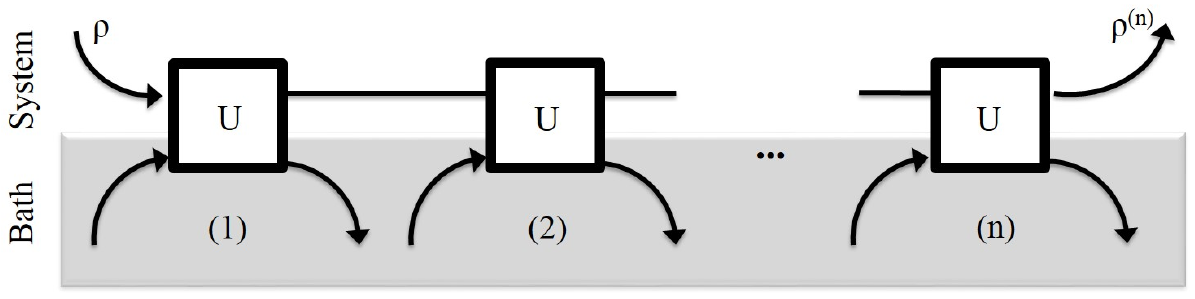}
\caption{The cascade structure of a collision model, defined by the concatenation
of identical unitaries \cite{SZSGB02}.}\label{cascade}
\end{figure}

The set of Markovian quantum channels is denoted below by $\mathfrak{M}$.
Clearly $\mathfrak{M}\subset\mathfrak{D}$ because
any Markovian quantum channel can be
divided into a large number of infinitesimal channels
being it the solution of the (time-independent) master equation \eqref{ME}.

Then one can attempt to single out the class of quantum channels
that can be split into \emph{infinitesimal} pieces,
i.e., into channels arbitrary close to the identity.
Clearly it would contain $\mathfrak{M}$. Actually
the set  $\mathfrak{I}$ of infinitesimal divisible quantum channels can be defined \cite{WC08}
as the closure of the set of all
families $\{\Lambda(t_2,t_1)\in\mathfrak{P}|t_1,t_2\in[0,t]\}$ of quantum channels for which there exists a continuous mapping $[0,t]\times [0,t]\rightarrow\mathfrak{P}$ onto $\{\Lambda(t_2,t_1)\}$ such that
\begin{enumerate}
    \item[1)] $\Lambda(t_3,t_2)\circ\Lambda(t_2,t_1)=\Lambda(t_3,t_1)$, for all  $0\leq t_1\leq t_2\leq t_3\leq
    t$,
    \item[2)]
    $\lim_{\epsilon\rightarrow 0}|||\Lambda_{\tau+\epsilon,\tau}-{\rm id}|||_2=0$, for all  $\tau\in[0,t)$,
\end{enumerate}
where $||| \cdots |||_2$ is the superoperator norm defined in Eq.~(\ref{fromtoNORM}) of Appendix~\ref{sec:dist2} -- the closure being intended
with respect to the associated distance.
For a given family it is like to say that there is a continuous path in $\mathfrak{P}$
(where one can move by concatenating quantum channels)
connecting any element of the family with the identity.

Actually one could consider in the above definition
a set $\mathfrak{I}_M$ analogous to $\mathfrak{I}$ with the restriction
$\Lambda\in\mathfrak{M}$, i.e., of the form $\Lambda(t)=e^{t\mathcal{L}}$.
It will obviously be $\mathfrak{I}_M\subseteq\mathfrak{I}$.
Intuitively also the converse should be true since any quantum channel
close to the identity is `almost Markovian' according to the definition of Markovian quantum channel.
In fact, it has been proven in \cite{WC08} that
any infinitesimal divisible quantum channel can be (arbitrary well) approximated
by a product of Markovian quantum channels.

\begin{figure}[t]
\centering
\includegraphics[width=0.3\textwidth]{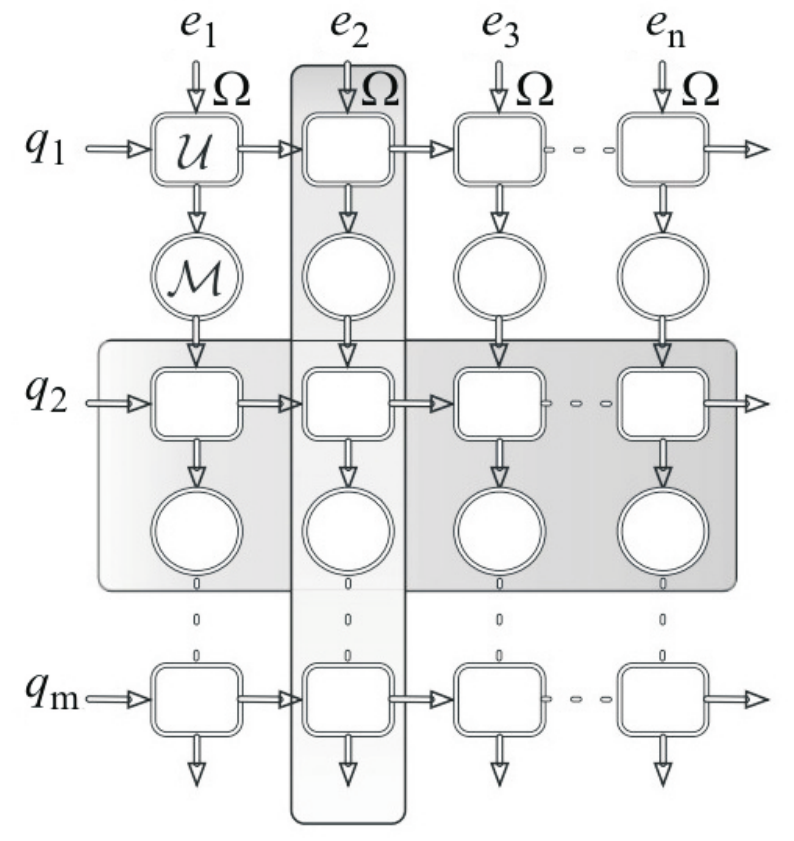}
\caption{The cascade structure leading to the master equation for correlated
quantum channels discussed in \cite{GP12}, described by Eq.\ (\ref{eqn:cascade}).
Each row corresponds to a single collision model (see Fig.\ \ref{cascade}), and
each column correspond to a memory channel (see Fig.\ \ref{figure01}a).}\label{GP12fig}
\end{figure}

In summary one has the following chain of inclusion $\mathfrak{M}\subset\mathfrak{I}\subset\mathfrak{D}\subset\mathfrak{P}$.
The complement of  $\mathfrak{D}$ to $\mathfrak{P}$ is given by the indivisible quantum channels.


\subsection{Non-Markovian master equations}

The simplest generalization of the dynamical equation (\ref{ME})
is obtained by introducing a time-dependent Liouvillian $\mathcal{L}(t)$ admitting
the representation (\ref{GENERATOR}), but with time-dependent operators, $H(t)$ and $F_\alpha(t)$.
Hence, the time-dependent equation for the dynamical map $\Lambda(t,t_0)$
\begin{equation}\label{D-M-t}
    \frac{d}{dt}\Lambda_{t,t_0} = {\cal L}(t) \circ \Lambda(t,t_0)  \ , \ \ \ \  \Lambda(t_0,t_0)=\mathrm{id}\ ,
\end{equation}
has formal solution
\begin{equation}\label{T-local}
    \Lambda(t,t_0) = \mathbb{T}\, \exp\left(\int_{t_0}^t{\cal L}(\tau)d\tau\right) \, ,
\end{equation}
where $\mathbb{T}$ denotes time-ordering.
Differently from the time-homogeneous case \eqref{COMP}, the explicit dependence on time implies that the
dynamical map $\Lambda(t,t_0)$ is no more a function of `$t-t_0$' only.
Notwithstanding, it still satisfies the {\it inhomogeneous} composition law
\begin{equation}\label{COMP-t}
    \Lambda(t,s) \circ \Lambda(s,t_0) = \Lambda(t,t_0) \, ,
\end{equation}
for any $t \geq s \geq t_0$.
The Markovian character is hence preserved by the time-dependent dynamical equation (\ref{D-M-t}) and it implies the infinitesimal divisibility discussed in Section \ref{sec:div}.
This is obviously true if one intends the Markovian character simply expressed by an associative binary operation like \eqref{COMP-t} (a quantum version of the Chapman-Kolmogorov equation). 
However it results that the Chapman-Kolmogorov equation is a necessary but not sufficient condition for having
Markov chains (processes) \cite{VSLPB11}.

On the other hand, from the fact that any infinitesimal divisible quantum channel can be 
(arbitrary well) approximated by a product of Markovian quantum channels (as discussed at 
the end of Sec.\ \ref{sec:div}), it follows that every infinitesimally divisible quantum channel
can be written as a solution of a time-dependent master equation (\onlinecite{WC08} proved this fact 
for $d=2$ and argued the same for $d>2$).
Hence, loosely speaking one can say that the class of infinitesimal divisible channels 
corresponds to the set of solutions of time-dependent master equations.


A more general dynamical equation comes from the Nakajima-Zwanzig projection operator technique \cite{N58,Z60,BPbook} and reads as follows:
\begin{equation}\label{NZ}
\frac{d}{dt}\, \rho(t) = \int_{t_0}^t \mathcal{K}(t - u)\,\rho(u) \, du\ , \ \ \ \rho(t_0)=\rho_0\ \, .
\end{equation}
Here one has memory effects modeled by the {\it memory kernel}
super-operator $\mathcal{K}(t)$.
Hence, the rate of change of the state at time also depends on its history, and
the Markovian setting (\ref{ME}) is recovered when $\mathcal{K}(\tau) =
2\delta(\tau)\mathcal{L}$.

The dynamical map $\Lambda(t,t_0)$ associated to the non-Markovian evolution (\ref{NZ})
is a solution of
\begin{equation}\label{D-NM}
    \frac{d}{dt}\Lambda(t,t_0) = \int_{t_0}^t d\tau\, \mathcal{K}(t - \tau) \circ
    \Lambda(\tau,t_0)  \ , \ \   \Lambda(t_0,t_0)=\mathrm{id}\ .
\end{equation}
It appears to be a function of both $t_0$ and $t$.
However, one can notice that the dynamics of an open quantum system can be always
understood as the reduced dynamics of its unitary dilation (see Sec.\ \ref{sec:qcbo})
which includes the environment.
Being the unitary dynamics of an isolated system homogeneous in time, it follows that,
once the degrees of freedom of the environment are taken into account, the dynamical map
will be only a function of the difference `$t-t_0$', that is, $\Lambda(t,t_0) \equiv \Lambda(t-t_0)$.
This mirrors the fact that any solution of (\ref{D-NM}) is also a solution of the time-dependent equation \cite{CK10a}
\begin{equation}\label{local-NM}
     \frac{d}{dt}\Lambda(t-t_0) = \mathcal{L}(t,t_0) \circ \Lambda(t,t_0) \
     , \ \ \Lambda(t_0,t_0) = \mathrm{id}\ ,
\end{equation}
with a time-dependent Liouvillian defined by the logarithmic derivative of the dynamical map
$\mathcal{L}(t-t_0) :=\left( \frac{d}{dt}\Lambda(t-t_0)\right) \circ \Lambda^{-1}(t-t_0)$.
Nevertheless, the explicit dependence of the generator on the initial time `$t_0$'
implies that $\mathcal{L}$ is effectively non-local in time.
Although the formal solution of (\ref{local-NM}) is analogous to (\ref{T-local}),
it does not satisfy the composition law (\ref{COMP-t}),
a fact which represents a signature of memory effects.

Then, a fundamental problem is to find those conditions on the memory kernel $\mathcal{K}(t)$ that
ensure that the time evolution map $\Lambda(t,t_0)$ is CPTP, i.e., a quantum channel.
Contrary to the Markovian case, a full characterization of legitimate memory kernels is still missing.

In \cite{CK12} a class of memory kernels giving
rise to legitimate quantum dynamics (quantum channels) has been provided.
The construction is based on a simple idea of normalization: starting from a family of
(possibly non-trace-preserving) CPTP maps satisfying a certain additional condition one
is able to `normalize' it in order to obtain a legitimate dynamics, i.e., a CPTP map.
Non-Markovian master equations have been also described in~Ref.~\cite{CICC13,Rybar02} by generalizing
the collision models discussed in the previous section
and in~\cite{SL05} exploiting adaptive strategies
that involve the measurements of the system
environment followed by local transformations.


\subsection{Markovian vs non-Markovian dynamics}

Given a CPTP map, the problem of determining whether or not it admits an infinitesimal
generator of the form (\ref{GENERATOR}), has been proven to be computationally hard \cite{WECC08,CEW12}.

For CPTP maps that do not belong to $\mathfrak{M}$, a measure of non-Markovianity
has been introduced in \cite{WECC08} in terms of the minimal amount of white noise $\mathcal{L}_\mu$
that has to be added in order to make $\log\Lambda + \mathcal{L}_\mu$ of the
 form (\ref{GENERATOR}).

Besides the Markovianity definition given in Section \ref{sec:div} and the above mentioned quantifier of (non)Markovianity other proposals have been put forward, see e.g. Refs.~\cite{RHP10,LuoFuSo12,LuWaSu10,BLP09}.

On one hand, \onlinecite{RHP10}
considered the equivalence between Markovian dynamics and infinitesimal divisibility
and introduced a measure of deviation from it.
Given a maximally entangled state $|\beta\rangle$ of the system of interest and a suitable ancillary system,
due to the Choi-Jamiolkowski isomorphism \eqref{CJiso}, $\Lambda(t+\epsilon,t) $ is a CPTP map iff
$ \left(\Lambda(t+\epsilon,t) \otimes {\rm id}\right) |\beta\rangle\langle\beta|\ge 0$. Then, one can consider
$\|\left(\Lambda(t+\epsilon,t) \otimes {\rm id}\right) |\phi\rangle\langle\phi|\|_1$
as a measure of the non-CPTP character of $ \Lambda(t+\epsilon,t) $. In fact, due to the trace preserving property,
this quantity equals 1 iff $ \Lambda(t+\epsilon,t) $ is CPTP, otherwise it is greater than 1.
Actually, the derivative of this quantity has been considered
\begin{equation}\label{g}
g(t) := \lim_{\epsilon \rightarrow 0}
\frac{\left\|
\left( \Lambda(t+\epsilon,t) \otimes \mathrm{id}\right) |\beta\rangle\langle\beta|
\right\|_1
- 1}{\epsilon} \, .
\end{equation}
It happens that $g(t) > 0$ iff the map $\Lambda$ is indivisible.

On the other hand \cite{BLP09} used
a fundamental property of CPTP maps, namely the fact that they cannot increase the trace-distance
\begin{equation}
D( \Lambda(t,0)(\rho_1), \Lambda(t,0)(\rho_1) ) \leq D( \rho_1, \rho_2 ) \, ,
\end{equation}
for any pair of states $\rho_1$, $\rho_2$.
If a family of CPTP maps is infinitesimally divisible, the monotonicity of the
trace distance holds true locally, that is,
\begin{equation}
 \frac{d}{dt} \, D( \Lambda(t,0)(\rho_1), \Lambda(t,0)(\rho_1) ) \leq 0 \, .
    \label{sig}
\end{equation}
According to that the dynamical map $\Lambda(t,0)$ is said to be non-Markovian
if there exists a value of $t$ such that Eq. (\ref{sig}) is violated, for some initial states $\rho_1$, $\rho_2$.
Physically, this implies a temporal increase in the distinguishability of the two
quantum states, a consequence the backflow of information from the surrounding environment.

The criteria relying on \eqref{g} and \eqref{sig} allow one to define computable measure of non-Markovianity.
A natural quantifier derived from the criterion of \cite{RHP10} reads
\begin{equation}\label{NRHP}
    \mathcal{N}_{\rm RHP}(\Lambda) = \frac{\int_0^\infty g(t)dt}{1+\int_0^\infty g(t)dt} \ ,
\end{equation}
where $g(t)$ is as in Eq.\ (\ref{g}).
From the criterion of \cite{BLP09} one defines the non-Markovianity quantifier
\begin{equation}\label{NBLP}
    \mathcal{N}_{\rm BLP}(\Lambda) =
    \sup_{\rho_1,\rho_2}\, \int  \frac{d}{dt'} \, D( \Lambda(t',0)(\rho_1), \Lambda(t',0)(\rho_1) ) \Big\vert_{t'=t} dt ,
\end{equation}
where the integral is performed only for those $t$ such that Eq. (\ref{sig}) is violated.

It has been pointed out that the relation between these two criteria
resembles that between separable and PPT states in entanglement
theory \cite{CKR11}. Indeed, any family of CPTP maps which is Markovian according to the
first criterion is as well Markovian according to the second one, that is,
$\mathcal{N}_{\rm RHP}(\Lambda)=0$ implies $\mathcal{N}_{\rm BLP}(\Lambda)=0$,
while the converse is in general not true.
An example comparing non-divisibility and non-Markovianity, for the case of Gaussian channels,
has been recently discussed in Ref.\ \cite{BFO12} while a test of  non-Markovianity for these
maps has been discussed in Ref.~\cite{Vasile11}.

One of the few example of non-Markovian dynamics that are exactly solvable
for their communication capacities is a single qubit coupled to an environment
of non interacting qubits in a star configuration giving rise to dephasing channel \cite{ATL10}.
Its quantum capacity behavior as function of time is strongly dependent
on the couplings parameters and on the temperature of the bath.
For generic values of these parameters, recurrence in the quantum capacity as function of time is of small
amplitude and quickly vanishes.
On the contrary, for commensurable values of these parameters the quantum capacity
becomes a periodic function of time.
This feature indicates the backflow of information from the environment to the central spin:
a signature of non-Markovian dynamics.
This is also related to the increased distinguishability of states pointed out by the
non-Markovianity criterion introduced by \cite{BLP09}.


\section{Summary and Outlook}\label{sec:so}

In the last decades the subject of quantum channels has become prominent
for its usefulness in foundational issues \cite{Kraus83}
as well as in technological applications (see the latest striking experiments in quantum communication
\cite{Ma12,Y13}).
Here, this subject has been addressed using a broad approach that embraces memory effects.
This is because the consideration of spatial and temporal memory effects is becoming increasingly
pressing with the continuing miniaturization of devices and with increasing communication rates.
In this scenario defining general properties and determining communication performance become daunting tasks.
Hence, we have mainly touched topics relevant to and witnessing progresses towards these ends.

At beginning (Sec.\ref{sec:qcbo}) we have reviewed basic features of quantum channel maps and tools for their characterization. 
Some physical examples of temporal and spatial evolutions of quantum systems, 
that the general framework of quantum (input-output) channels can describe, are also discussed. Then we have focussed on multiple channel uses by addressing their structural properties in
Sec.\ref{sec:fmmqc}. There, several quantum memory channels models have been devised and their taxonomy presented. 
However, it is worth noticing that the latter is based on channel representations that consider input and (initial)
memory systems mapped onto output and (final) memory systems.
In a black-box description accounting only for input to output mapping some of these models could result equivalent.
Loosely speaking this could be analogous to the possibility of having different Kraus representations of the same quantum channel.
Hence, a property of quantum channels (fixed point, indecomposability, etc.) should be defined in a more general way,
that is, the channel has the property if there exists at least one of its memory representations that satisfies it.

Reliable communication through quantum channels can be achieved by employing error correcting codes as discussed in Sec.\ref{sec:codes}. Standard quantum codes are designed to counteract independent errors affecting multiple uses of a noisy quantum channel. On the other hand, memory channels produce correlated errors. An extreme case is represented by collective errors affecting a certain number of information carriers at once. In such a case the symmetries of the noise usually allow for the existence of decoherence free subspaces. In the intermediate situations one has to design new codes to counteract errors that are neither independent nor completely correlated. A relevant strategy to produce such codes is by concatenation of standard codes and decoherence free subspaces. 
Another strategy consists in exploiting cyclic property of some codes. Finally a relation between convolutional codes and memory channels has been highlighted.

The general definitions of classical and quantum capacities, unassisted as well as assisted by entanglement, have given in Sec.\ref{sec:ctc} followed by the definition of constrained capacities suitable for continuous channels. For such definitions we have remarked that a super-operator norm approach can be used as well.
Then we have sketched coding theorems. Actually, 
for what concern the capacities evaluation, the main obstacle is the restricted class of channels for which coding theorems are available. Hopefully this can be enlarged by resorting to stationary or ergodic properties of the quantum channels as outlined by \cite{BB09}.
Still within such a class, those channels leaving hopes for an exact capacities computation are the forgetful channels (see Sec.\ \ref{coding:forgetful}).
For this reason it is of utmost importance to derive general criteria to decide whether or not a given channel is forgetful.
Beyond that it would be extremely interesting to establish when memory effects increase the capacity of a quantum channel.
It is also worth noticing that the effects of correlations among errors are in close connection with the property of
superadditivity of the minimum output entropy \cite{Has09}.
The possible memory induced enhancement of capacity of a quantum channel, looked through the dynamical memory model sketched in Sec. \ref{sec:smdmc}, can be seen as due to a sort of Zeno effect \cite{MS77}.
In fact, by frequently inserting information carriers through the channel one prevents the environment to come back
to its stationary state after the passage of each of them, thus less affecting the carriers themselves.
Whereas in the case of quantum channels arising from non-Markovian dynamics like that of Sec. \ref{sec:memdyn},
the increment of capacity can be explained by the back flow of information from environment to system.

Known solvable (in terms of capacities) models have been discussed in Sec.\ref{sec:solv}. 
Among them, dephasing memory channels possess features related to many-body physics and lossy 
bosonic memory channels  show water filling phenomena similar to fluid mechanics.

Finally, in Sec.\ref{sec:memdyn}, we have shown the conditions under which a quantum channel 
can be `divided' into the concatenation of other quantum channels, i.e., its action results as the 
composition of other quantum channels. This possibility is closely related to quantum channels 
intended as dynamical maps. Then one can distinguish between Markovian and Non-Markovian dynamics, 
the latter showing memory effects in time. As a consequence we have briefly accounted for some measures 
quantifying deviation from Markovian dynamics, although a general consensus on that subject is not yet reached.

All in all examples of quantum channels showing memory effects are abundant in quantum information processing.
An unmodulated spin chain has been proposed as a model for short distance quantum communication \cite{B03}.
In such a scheme, the state to be communicated over the channel is placed on one of the spins of the chain,
propagates for a specific amount of time, and is then received at a distant spin of the chain.
When viewed as a model for quantum communication, it is generally assumed that a reset of the spin chain occurs
after each signal, for instance by applying an external magnetic field, resulting in a memoryless channel.
However, a continuous operation without resetting corresponds to a quantum channel with memory \cite{BBM+08}.
Another model of a quantum channel with memory is the so-called {\em one-atom maser} or {\em micromaser} \cite{BDF09}.
In such a device, excited atoms interact with the photon field inside a high-quality optical
cavity. If the photons inside the cavity have sufficiently long
lifetime, atoms entering the cavity will feel the effect of the
preceding atoms, introducing ISI correlations (see Sec.\ \ref{sec:nonant})
among consecutive signal states.

Another source of correlated noise in the propagation of the electromagnetic field
is due to atmospheric turbulence, whose effects on the signal propagation can be modeled
as random changes of the channel's characteristics \cite{SV09,SV10}.
Moreover, the decoherence induced by atmospheric turbulence introduces cross talks \cite{TB09,B11},
i.e., ISI correlations (see Sec.\ \ref{sec:nonant}), when information
is encoded in the transverse degrees of freedom of the electromagnetic field, e.g.,
the orbital angular momentum.
Furthermore, the propagation of the quantum electromagnetic field in linear dispersive
media, including the free-space propagation and through linear optical systems,
can be described by a quantum channel with memory \cite{Shapiro2009,MM10,GiovannettiQIC,LUPOpraR2011}
where wave diffraction introduces memory effects. 

Memory effects also arise in the context of quantum cryptography. 
Quite generally, one can categorize the collective attacks 
within the framework of memoryless channels, while the coherent attacks 
within the memory channels framework \cite{SBCDLP09,GRTZ02}. 
However, this link has been subjected to limited attention and probably 
needs further explorations. Actually, in one-way quantum key distribution 
memory effects that introduce correlations among transmitted symbols can advantage the eavesdropper \cite{RM07b}.
Only if the legitimate users have the control of the noise correlations, by properly
tuning them, they can reduce eavesdropper information \cite{VOPM11}.
Instead, in two-way quantum key distribution checking the presence
or absence of noise correlations can help in counteracting eavesdropper attacks \cite{PMBL08}.
Then, an analysis of memory effects in other channel uses-configurations, like
zero error channel capacity, channels with feedback, channels with unknown parameters
and multiuser channels, should be pursued.

Finally, moving to the framework of time-continuous quantum evolution, non-Markovian effects
are relevant in several physical systems characterized by the interaction with a structured environment.
Examples are in the framework of solid state physics, as quantum dots in photonic crystals \cite{VJB02,MALLRFL11},
and in the soft matter framework as the case of exciton dynamics surrounded by their protein
environment \cite{PH08,CCDHP09,RCA09,TERNW09,CHP10}.

In summary, more efforts are needed to gain a full understanding of quantum
channels, however the presented work constitutes a rather general frame where the
still missing dowels of the puzzle could be settled.

\acknowledgments

The authors would like to thank
K. Banaszek,
N. J. Cerf,
D. Chruscinski,
N. Datta,
S. F. Huelga,
C. Macchiavello,
O. Pilyavets,
M. B. Plenio,
A. Shabani,
B. Vacchini,
 S. Virmani, and M. Wilde for
for useful discussions and for careful reading the manuscript.
S.M. acknowledges fruitful and enjoyable collaboration with Garry Andrew Bowen though whom he became acquainted with the subject of quantum channels.
The work of F.C. has been supported by a Marie Curie Intra European Individual Fellowship and a Marie Curie Career Integration Grant within the 7th European Community Framework Programme, under the grant agreements MemoryQuantumICT No. 235086 and QuantumBioTech No. 293449, respectively, and by the Future in Research (FIRB) Programme of the Italian Ministry of Education, University and Research (MIUR), under the FIRB-MIUR grant agreement No. RBFR10M3SB.
The work of C.L. and S.M. has been supported by the Future and Emerging Technologies (FET) programme within the Seventh Framework Programme for Research of the European Commission, under the FET-Open grant agreement CORNER, No. FP7-ICT- 213681.

\appendix

\section{Distance measures}\label{sec:dist}

A proper way to measure the distance between  two states $\rho_1$, $\rho_2 \in \mathfrak{S} ({\cal H}_{Q})$
of a quantum system $Q$,  is provided by the \emph{trace distance} defined as:
\begin{eqnarray} \label{TRACEDISTANCE}
D(\rho_1,\rho_2) :=\frac{1}{2}\| \rho_1-\rho_2 \|_1 \;,
\end{eqnarray}
with $\| O \|_1 := \tr{\sqrt{O^\dag O}}$ being the trace-norm of the operator $O$~\cite{NC00,Wbook13}. While fulfilling all the conditions of a
regular distance (i.e., positivity, symmetry and triangular inequality)
the trace distance possesses other interesting properties which makes it operationally well defined.
For instance it is bounded between $0$ and $1$ (reaching the latter value only when $\rho_1$ and $\rho_2$ have orthogonal support). 
Furthermore the trace distance is preserved under unitary transformations, i.e., $D( U \rho_1 U^{\dagger}, U \rho_2 U^{\dagger})=D(\rho_1,\rho_2)$ (implying that the distance between physical states does not
depend upon the coordinate system used to describe them)
but  it is contractive under CPTP maps $\Phi$, i.e.,
\begin{eqnarray}
D(\Phi(\rho_1),\Phi(\rho_2)) \leq D(\rho_1,\rho_2)  \; ,
\end{eqnarray}
(implying that the action of noise tends to {\it blur} the difference among states).
Finally  $D(\rho_1,\rho_2)$ can be identified with the maximum distance between
the statistical distributions $\{ p_x(\rho_1)=\mbox{Tr}[ E_x \rho_1]\}_{x\in X}$ and  $\{ p_x(\rho_2)=\mbox{Tr}[ E_x \rho_2]\}_{x\in X}$ obtained by performing the  same POVM measurement $\{ E_x\}_{x\in X}$ on $\rho_1$
and $\rho_2$.

Another quantity useful to gauge how close two density matrices $\rho_1$ and $\rho_2$ are,
is the \emph{fidelity} \cite{U76,J94}
\begin{equation}\label{fidelity}
F(\rho_1,\rho_2) := \| \rho_1^{1/2} \rho_2^{1/2} \|_1^2 \;,
\end{equation}
which for $\rho_1$ being rank one, i.e., $\rho_1=|\psi_1\rangle\langle\psi_1|$, coincides with the
probability of finding $\rho_2$ in the vector $|\psi_1\rangle$, i.e.,
\begin{eqnarray} \label{fid1}
 F(|\psi_1\rangle,\rho_2) = \langle \psi_1 |\rho_2 |\psi_1\rangle\;.
 \end{eqnarray}
The function $F(\rho_1,\rho_2)$  is symmetric (i.e., $F(\rho_1,\rho_2)=F(\rho_2,\rho_1)$) and always
 in the range $[0,1]$ (equal to $1$ if and only if $\rho_1=\rho_2$ and vanishing
 for density operators with orthogonal supports, e.g., for orthogonal pure states).
Furthermore, $F$ is invariant under the action of a unitary evolution,
$F( U \rho_1 U^{\dagger}, U \rho_2 U^{\dagger})=F(\rho_1,\rho_2)$, and increasing under CPTP map,
\begin{equation}
F(\Phi(\rho_1),\Phi(\rho_2)) \geq F(\rho_1,\rho_2)  \; .
\end{equation}
While not a distance itself, the fidelity is directly linked to the Bures distance~\cite{bures69} via the identity  $D_B(\rho_1,\rho_2):=[2-2\sqrt{F(\rho_1,\rho_2)}]^{1/2}$.
Trace distance and fidelity are related by the following inequalities
\begin{eqnarray} \label{DistFidIn}
1-\sqrt{F(\rho_1,\rho_2)}\leq  D(\rho_1,\rho_2) \leq \sqrt{1- F(\rho_1,\rho_2)}\;,
\end{eqnarray}
therefore states which have high values of fidelity are also close in trace distance, and vice-versa.

\section{Quasi-Local Algebras}\label{app:algebras}

Quasi-local algebras are the proper mathematical tools to describe infinitely extended quantum lattice systems \cite{BR79}.
For the sake of simplicity let us consider a chain of infinitely many qubits (spins) placed
in a one dimensional lattice $\Z$.
Then, to each lattice site $j\in\Z$ attach a Hilbert space $\hr_j\simeq\cc^2$ and consider the associated algebra of bounded operators $\A_j=\B(\hr_j)$. If one restricts to a finite part of the chain, say a set $\Lambda\subset\Z$, it is possible to define the following tensor products
\begin{equation}
\hr_{\Lambda}=\otimes_{j\in\Lambda} \hr_j,\qquad
\A^{\Lambda}=\B(\hr_{\Lambda})=\otimes_{j\in\Lambda}\A_j.
\end{equation}
Operators $a\in\A^{\Lambda}$ are called \emph{local operators} as they are operators `localized' in $\Lambda$. Clear examples of local observables are the Pauli operators $\sigma_{x,j}$, $\sigma_{y,j}$, $\sigma_{z,j}$,
attached to the site $j\in\Z$. However, if the region $\Lambda$ is infinite the situation becomes tricky.
In fact, although it is still possible to attach an Hilbert space $\hr$ to the whole chain such that local operators
$\sigma_{j}$ act on $\hr$, this cannot be done in a unique way.
It is then preferable to proceed by considering the algebraic properties of local operators not requiring a global Hilbert space $\hr$.
To this end, let us first notice that having two finite regions $\Lambda,\Lambda'$ such that
$\Lambda\subset\Lambda'\subset\Z$, the operators
\begin{equation}
a\in\A^{\Lambda}\quad {\rm and}\quad a\otimes \openone_{\Lambda'\setminus\Lambda}\in \A^{\Lambda'},
\end{equation}
describe the same physical object.
 Therefore, one can identify $\A^{\Lambda}$ with the sub-algebra $\A^{\Lambda}\otimes\openone_{\Lambda' \backslash \Lambda}$  of  $\mathcal{A}^{\Lambda'}$
 through the map
 \begin{equation}
\mathcal{A}^{\Lambda}\ni a\mapsto a\otimes \openone_{\Lambda'\setminus\Lambda}\in \mathcal{A}^{\Lambda'}.
\end{equation}
Doing so one gets a system of matrix algebras $\A^{\Lambda}$ which are ordered by inclusion
\begin{equation}
\Lambda\subset\Lambda'\Rightarrow \A^{\Lambda}\subset\A^{\Lambda'},\quad
\forall \;{\rm finite}\; \Lambda,\Lambda'\subset\Z.
\end{equation}
This construction leads to the possibility of defining, for two arbitrary local operators $a_1,a_2$,
\begin{itemize}
\item[i)]
their linear combination $\mu_1a_1+\mu_2a_2$ with $\mu_1,\mu_2\in\cc$;
\item[ii)]
their product $a_1a_2$;
\item[iii)]
their adjoints $a_1^{\dag}$ (resp. $a_2^{\dag}$).
\end{itemize}
To this end one only needs to find a region $\Lambda$ such that the matrix algebra $\A^{\Lambda}$ contains both operators $a_1,a_2$.

More precisely one can introduce the space of all local operators by the union
\begin{equation}
\mathcal{A}^{loc}:=\bigcup_{{\rm finite}\;\Lambda\subset \Z}\mathcal{A}^{\Lambda},
\end{equation}
and then equip this space with a vector space structure, a product (associative and bilinear), a ${}^{\dag}$-operation and a unit element $\openone$. In such a way $\A^{loc}$ becomes the \emph{algebra of local observables}.

Going further on, one can associate to each $a\in\A^{loc}$ its norm $\|a\|$ by
finding the region $\Lambda$ such that $a\in\A^{\Lambda}$ and then using the standard operator norm.
A problem is given by the fact that $\A^{loc}$ is not complete with respect to such a norm, i.e., not all Cauchy sequences converge. This problem can be overcame by taking the norm-closure of $\A^{loc}$ and get
the \emph{algebra of quasi-local observables}
\begin{equation}
\A^\Z=\overline{\A^{loc}}^{\|\cdot\|}.
\end{equation}
Here \emph{quasi-local} stands for the fact that $\A^\Z$ besides all local observables, also contains non-local observables which can be approximated in norm by local ones.
$\A^\Z$ results a $C^*$-algebra and its elements can be regarded in many respects as bounded operators \cite{BR79}.

It is also useful to consider methods to transform abstract elements $a\in \A^\Z$ into operators $\pi(a)$ acting on a Hilbert space $\hr$, i.e., representations of quasi-local algebra.
 A representation $\pi$ on Hilbert space $\hr$ is a  homomorphism $\pi:\A^\Z\to\B(\hr)$, i.e., a linear map satisfying $\pi(ab)=\pi(a)\pi(b)$ and $\pi(a^{\dag})=\pi(a)^{\dag}$.
Unfortunately for spin chains there is not a unique representation that can be used for all purposes, but  one has to choose the representation which is most appropriate to the given physical context.
This ambiguity also reflects on the definition of \emph{states}. In fact one might be tempted to use density operators $\rho$ on the Hilbert space $\hr$ which carry the representation $\pi$. However since different representations correspond to different physical contexts one should use all possible representations (in fact each density operator in any representation can describe a state). Clearly, it would be much better to describe states in a way independent from the representation. Thus
a state of $\A^\Z$ is defined as a linear functional $\psi:\A^\Z\to\cc$ which is positive ($\psi(a^{\dag}a)\ge 0$, $\forall a\in\A$) and normalized ($\psi(\openone)=1$).
This means that given a representation $\pi$ and a density operator $\rho$ on $\hr$, the corresponding state is the functional $\psi_\rho(a)={\rm tr}\{\pi(a)\rho\}$.

The possibility to find for each state $\psi$ a Hilbert space $\hr$ carrying a representation $\pi$ and a 
density operator $\rho$ such that $\psi=\psi_\rho$ is guaranteed by the Gelfand-Naimark-Segal theorem \cite{BR79}.
It states that each state $\psi$ can be represented by a state \emph{vector} $|v_\psi\rangle$ on a suitable
Hilbert space. In other words, it is like to say that it is always possible to provide a `purification' of the state $\psi$.

One can introduce into the quasi-local algebra ${\cal A}^{\Z}$
a shift operation $T:{\cal A}^{\Z} \to {\cal A}^{\Z}$ by the following action:
\begin{equation}
\mathcal{A}^{\Lambda}\ni a\simeq a\otimes\openone_\A \mapsto T(a):=\openone_\A \otimes a\simeq a\in \mathcal{A}^{\Lambda+1} \, ,
\end{equation}
where $ A \otimes \openone_\A $ (resp. $ \openone_\A \otimes A $) stands for the tensor product
between $A$ belonging to ${\cal A}^{\Lambda}$ and identity of ${\cal A}$ on the
site to the right of $ \Lambda $ (resp. between identity of ${\cal A}$ on the site
to the left of $ \Lambda $ and $a$ belonging to ${\cal A}^{\Lambda}$).

Moving from the action of the shift $T$, it is possible to introduce the notion of
\emph{stationary} state $\psi$ on $\mathcal{A}^{\Z}$ when $\psi\circ T=\psi$ holds true.
The set of stationary states on $\mathcal{A}^{\Z}$ turns out to be convex.
Then a state $\psi$ on $\mathcal{A}^{\Z}$ is called \emph{ergodic} (with respect to the shift)
if it is extremal on this set.


\section{Decomposition for non-anticipatory quantum channels}\label{sec:appendix}

This section provides an explicit derivation of the decomposition of non-anticipatory channels
in Eq.~(\ref{eqn:memory_model_10}) based on a generalization of analysis presented in~\cite{PHHH06,ESW01,KW05,BGNP01}.

Let then $\{ \Phi^{(n)}; n=1,2, \dots \}$ be a family of CPTP maps describing a non-antipatory quantum channel.
Adopting the unitary representation in Eq.~(\ref{dia}), for each $n$ one can define a unitary transformation $W^{(n)}_{q_n, q_{n-1}, \cdots, q_1,M}$
coupling the first $n$ carriers to a common environment $M$ which allows one to write
\begin{align}
\Phi^{(n)}( \rho_Q^{(n)}) = \mbox{Tr}_M [ & W^{(n)}_{q_n, q_{n-1}, \cdots, q_1,M} \nonumber \\
&(\rho_Q^{(n)} \otimes \omega_M^{(n)}) {W^{(n)}}^\dag_{q_n, q_{n-1}, \cdots, q_1,M} ] \; . \label{appendix1}
\end{align}
The environmental state $\omega_M^{(n)}$ is in general a function of $n$ but is independent on the input state $\rho_Q^{(n)}$.
Without loss of generality here it is assumed to be a pure state, $\omega_M^{(n)} = |\omega^{(n)}\rangle_M\langle \omega^{(n)}|$.

In general the unitary couplings $W^{(n)}_{q_n, q_{n-1}, \cdots, q_1,M}$ can have a complicated dependence upon~$n$:
however since the channel is non-anticipatory they must obey the following rule
\begin{align}
& \mbox{Tr}_{q_n,M} [ W^{(n)}_{q_n, \cdots, q_1,M}
( \rho_Q^{(n)} \otimes \omega_M^{(n)} )
{W^{(n)}}^\dag_{q_n, \cdots, q_1,M} ] = \nonumber \\
& \mbox{Tr}_{M} [ W^{(n-1)}_{q_{n-1}, \cdots, q_1,M}
( \rho_Q^{(n-1)} \otimes \omega_M^{(n-1)} )
{W^{(n-1)}}^\dag_{q_{n-1}, \cdots, q_1,M} ] \;, \nonumber
\end{align}
where $\rho_Q^{(n-1)} = \mbox{Tr}_{q_n} [\rho_Q^{(n)} ]$ is the reduced density operator of $\rho_Q^{(n)}$ associated with the
first $n-1$ carriers.
Applying this relation to a pure input state $\rho_Q^{(n)}$ of the form $|\psi\rangle_{q_n} \otimes |\phi\rangle_{q_{n-1},\cdots, q_1}$
one notices that the vectors $W^{(n)}_{q_n, q_{n-1}, \cdots, q_1,M} |\psi\rangle_{q_n} \otimes |\phi\rangle_{q_{n-1},\cdots, q_1}\otimes |\omega^{(n)}\rangle_M$ and
$W^{(n-1)}_{q_{n-1}, \cdots, q_1,M} |\phi\rangle_{q_{n-1},\cdots, q_1} \otimes |\omega^{(n-1)}\rangle_M$ are both purifications
of the state $\Phi^{(n-1)}( \rho_Q^{(n-1)})$.
Therefore there must exist a unitary transformation $U_{q_n M}$ acting on the latter which satisfies the identity \cite{NC00}
\begin{equation}
W^{(n)}_{q_n, q_{n-1}, \cdots, q_1,M} |\omega^{(n)}\rangle_M = U_{q_nM} W^{(n-1)}_{q_{n-1}, \cdots, q_1,M} |\omega^{(n-1)}\rangle_M \;.
\end{equation}
Iterating this $n$ times yields
\begin{equation}
W^{(n)}_{q_n, q_{n-1}, \cdots, q_1,M} |\omega^{(n)}\rangle_M = U_{q_nM} U_{q_{n-1}M} \cdots U_{q_1M} |\omega^{(0)}\rangle_M \;.
\end{equation}
which replaced into Eq.~(\ref{appendix1}) implies Eq.~(\ref{eqn:memory_model_10}).

\section{Explicit derivation of capacity upper bounds}
  \label{a:DERIVO}
Here we present an explicit derivation of the upper bounds
(\ref{cbounds}) and (\ref{qbounds})
 for the classical and quantum capacities defined in Eq.~(\ref{clascap}) of a (non necessarily memoryless)  quantum channel $\Phi^{(n)}$.

 \subsection{Upper bound for the classical capacity} \label{BOUNDSCA}

The derivation of Eq.~(\ref{cbounds}) follows by merging
 the Holevo Bound~\cite{Holevo73a,Holevo73b}
 with the
 {\it classical} Fano inequality~\cite{CTbook}.
For this purpose one reminds that given two random variables $X$ and $Y$
 connected by conditional probability distribution $p(y|x)$
 and a correspondence rule which assign values of  $X$ to each of the values of
  $Y$,  the Fano inquality
  allows one to lower bound the  mutual information $I(X:Y)$ as
 \begin{eqnarray} \label{FANOC}
 I(X:Y) \geq  H(X) - h(P_e) - P_e\; \log_2( |X|-1) \;,
 \end{eqnarray}
where
 $|X|$ is the number of elements of
 the variable $X$,  $h(p)$ is the binary Shannon entropy~(\ref{binarysh}),
and finally  $P_e$ is the average error probability
that the correspondence rule is violated by the conditional probability $p(y|x)$.
Specifically, assuming for simplicity that the $X$ and $Y$ span the same alphabet of symbols and that correspondence rule assign to $Y$ the same symbol on $X$, we have
  $P_e = 1- \sum_{x} p(x) p(y=x|x)$.
   Identify then $X$ with the messages $m$ that Alice is mapping from ${\cal M}$
 to the $n$ carriers  via
 the coding channel $\Phi_E^{(k \to n)}$, and with
$Y$ the elements of ${\cal M}$ which Bob is retrieving  via his decoding   mapping
  $\Phi_D^{(n \to k)}$. Under the assumption that the probability that Alice is selecting  the messages
  from ${\cal M}$ with uniform probability, and reminding that ${\cal M}$ contains $2^k$ elements,
  Eq.~(\ref{FANOC}) yields
  \begin{eqnarray}
 I(X:Y) &\geq&  k  - h(P_e) - P_e\; \log_2(2^k-1)\nonumber \\
 & \geq& k -  h(\epsilon) - \epsilon  k  \;,  \label{FANOC1}
\end{eqnarray}
where we used the fact that $P_e\leq 1-F_{\min}< \epsilon$, with $F_{\min}$ being the minimum fidelity
achieved by the selected  code.  By reorganizing the various terms and by dividing by $n$ we then get
   \begin{eqnarray} \label{FANOC11}
   \frac{k}{n} \leq \frac{I(X:Y)}{(1-\epsilon) n}  + \frac{h(\epsilon)}{(1-\epsilon)n}
 \;.
 \end{eqnarray}
Remind next that $I(X:Y)$ is the mutual information associated to the ensemble of
codewords ${\cal E}= \{ p_m = 2^{-k}; \Phi_E^{(k\rightarrow n)} (m)\}$ generated by Alice
and received by Bob trough the channel $\Phi^{(n)}$.
The Holevo Bound~(\ref{HOLEVOBOUND}) implies then
\begin{eqnarray}
\frac{k}{n} &\leq&  \frac{\chi({\cal E}; \Phi^{(n)})}{(1-\epsilon)n}  + \frac{h(\epsilon)}{(1-\epsilon)n}  \nonumber \\
 &\leq&
  \frac{\max_{{\cal E}}\chi({\cal E}; \Phi^{(n)})}{(1-\epsilon)n}  + \frac{h(\epsilon)}{(1-\epsilon)n}
 \;.
\end{eqnarray}
Since this inequality holds for all encoding/decoding strategies  entering in the capacity definition~(\ref{clascap}), by taking the limit on $k\rightarrow \infty$ and $\epsilon \rightarrow 0$ we finally get the
inequality~(\ref{cbounds}).

The same derivation detailed above can be used to prove the converse part of the HSW theorem
for the one-shot classical capacity $C_1(\Phi)$ of a  memoryless channel $\Phi$, i.e.,
\begin{eqnarray}\label{Acbounds1}
            C_1(\Phi)  \leq \max_{\cal E}
            \chi({\cal E}; \Phi) \, .
\end{eqnarray}
Indeed, exploiting the fact that the channel is memoryless and the  coding
procedure uses only separable codewords $\Phi_E^{(k \to n)}(m)$ one can apply the
sub-additivity of the classical mutual information~\cite{CTbook,Gbook} to replace Eq.~(\ref{FANOC1})
as
\begin{eqnarray}
 \sum_{i=1}^n I(X_i:Y_i) &\geq&  k  - h(P_e) - P_e\; \log_2(2^k-1)\nonumber \\
 & \geq& k -  h(\epsilon) - \epsilon  k  \;,  \label{FANOC2}
\end{eqnarray}
where, for $i=1, \cdots,n$,  $I(X_i:Y_i)$ is the mutual information associated with the
classical input of the $i$-th channel use. Applying  the Holevo Bound to all these terms we get
 \begin{eqnarray}
 \frac{k}{n}
 &\leq&
  \frac{\max_{{\cal E}}\chi({\cal E}; \Phi)}{(1-\epsilon)}  + \frac{h(\epsilon)}{(1-\epsilon)n}
 \;,
\end{eqnarray}
that finally leads to Eq.~(\ref{Acbounds1}) when taking  $\epsilon \rightarrow 0$.

In conclusion, it is worth remarking that Eq.~(\ref{cbounds}) can be used to prove an inequality
which in some cases happens to be useful for deriving explicit expression for $C$ -- see e.g., Sec.~\ref{SEC:DISCRETE}.
This is obtained by noticing that
\begin{eqnarray}
\max_{\cal E}
            \chi({\cal E}; \Phi^{(n)}) &\leq&  \max_{\rho} S(\Phi^{(n)} (\rho)) - \min_{\rho} S(\Phi^{(n)}(\rho)) \nonumber \\
&\leq&  n \log_2 d - S_{\min}(\Phi^{(n)})\;.
\end{eqnarray}
with $S_{\min}(\Phi^{(n)}) =  \min_{\rho} S(\Phi^{(n)}(\rho))$ being the minimum entropy
one can reach at the output of the channel.
In the above derivation the first inequality follows directly from the definition of the Holevo information,
while the second from the fact that the entropy of a state of ${\cal H}_{Q}^{\otimes n}$ is
not larger than $\log_2 d^n$ ($d$ being the dimension of ${\cal H}_{Q}$).
Replacing this into Eq.~(\ref{cbounds})  we then get
\begin{eqnarray} \label{upperSmin}
C \leq \log_2 d - \lim_{n\rightarrow \infty} \frac{S_{\min}(\Phi^{(n)})}{n}\;.
\end{eqnarray}

\subsection{Upper bound for the quantum capacity}\label{BOUNDSQA}

The derivation which follows is an adaptation to the memory channel scenario
of the proof of \cite{BNS98}.
The starting point in this case is the
the quantum Fano inequality presented in Eq.~(\ref{QFANO}), the data-processing inequality~(\ref{DATAPROCJ}) and the bounds~(\ref{superDIS}). From them one can easily verify that
given a generic density matrix $\tau$ of the reference set~${\cal M}$
 the following relation holds
 \begin{eqnarray}  \frac{S(\tau)}{n} &\leq&  \frac{J(\tau; \Phi_D^{(n \to k)} \circ {\Phi}^{(n)} \circ \Phi_E^{(k \to n)})}{n} \nonumber \\
 &+&
  \frac{2}{n} h(F_e(\tau; \Phi_D^{(n \to k)} \circ {\Phi}^{(n)} \circ \Phi_E^{(k \to n)}) \nonumber \\
  &+&  \frac{4}{n} [ 1 -F_e(\tau; \Phi_D^{(n \to k)} \circ {\Phi}^{(n)} \circ \Phi_E^{(k \to n)}) ] \log_2 (4^k -1)
   \nonumber  \\
   &\leq&  \frac{\max_{\rho} J(\rho ; \Phi^{(n)})}{n} +
  \frac{2}{n}h(1-3\epsilon/ 2) + \frac{6}{n}  \epsilon k \;, \label{QFANO2}
 \end{eqnarray}
 where the second inequality has been obtained by using again the data-processing
 inequality ~(\ref{DATAPROCJ})  and  maximizing $J$ over all possible inputs states
 $\rho$ of the $n$ channels uses,  and by using
 the fact that the minimum fidelity of the code is lower bounded by $1-\epsilon$ and the fact that
 this  implies that the corresponding entanglement fidelity fulfills
   $F_e(\tau; \Phi_D^{(n \to k)} \circ {\Phi}^{(n)} \circ \Phi_E^{(k \to n)}) > 1 -3\epsilon/2$, -- see Eq.~(\ref{FeFin2}).
   Specifying Eq.~(\ref{QFANO2})   for the maximally mixed state state
  of  ${\cal M}$, we then get
   \begin{eqnarray}  \frac{k}{n}    &\leq&  \frac{\max_{\rho} J(\rho ; \Phi^{(n)})}{n} +
  \frac{2}{n} h(1-3\epsilon/ 2)  + \frac{6}{n} \epsilon k  \;, \nonumber
 \end{eqnarray}
 which in the limit of $k\rightarrow \infty$ and $\epsilon\rightarrow 0$ yields finally~(\ref{qbounds}).


\bibliographystyle{apsrmp}

\end{document}